\documentstyle[epsfig,natbib2,natbibmnfix]{mn}

\newcommand{\be}{\begin{equation}}
\newcommand{\ee}{\end{equation}}
\newcommand{\bea}{\begin{eqnarray}}
\newcommand{\eea}{\end{eqnarray}}
\newcommand{\bc}{\begin{center}}
\newcommand{\ec}{\end{center}}
\newcommand{\ol}[1]{ {\overline{#1}}}
\newcommand{\lu}{\,h^{-1}{\rm kpc}}
\newcommand{\kms}{\,{\rm km}\, {\rm s}^{-1}}
\newcommand{\msunh}{\,h^{-1}{\rm M}_{\odot}}
\newcommand{\lmu}{\,h^{-1}{\rm Mpc}}

\newcommand{\dd}{{\rm d}}

\newcommand{\gadget}{{\small GADGET}}
\newcommand{\subfind}{{\small SUBFIND\/}}

\bibpunct[,]{(}{)}{;}{a}{}{,}

\title{Populating a cluster of galaxies -- I. Results at $z=0$}

\author[V. Springel, S.\ D.\ M.\ White, G. Tormen and G. Kauffmann]
{\parbox{18cm}{Volker Springel$^{1,2}$\footnotemark, Simon D. M. White$^1$, 
Giuseppe Tormen$^3$\newline and Guinevere Kauffmann$^1$}\vspace{0.3cm}\\
$^1$Max-Planck-Institut f\"{u}r Astrophysik, Karl-Schwarzschild-Stra\ss{}e 1, 
85740 Garching bei M\"{u}nchen, Germany\\
$^2$Harvard-Smithsonian Center for Astrophysics, 60 Garden Street, Cambridge, MA 02138, USA\\
$^3$Dipartimento di Astronomia, Universita di Padova, vicolo dell'Osservatorio 5, 1-35122 Padova, Italy}

\begin{document}

\maketitle
\begin{abstract}
We simulate the assembly of a massive rich cluster and the formation
of its constituent galaxies in a flat, low-density universe. Our most
accurate model follows the collapse, the star-formation history and
the orbital motion of all galaxies more luminous than the Fornax dwarf
spheroidal, while dark halo structure is tracked consistently
throughout the cluster for all galaxies more luminous than the
SMC. Within its virial radius this model contains about $2\times 10^7$
dark matter particles and almost 5000 distinct dynamically resolved
galaxies. Simulations of this same cluster at a variety of resolutions
allow us to check explicitly for numerical convergence both of the
dark matter structures produced by our new parallel N-body and
substructure identification codes, and of the galaxy populations 
produced by the phenomenological models we use to follow cooling, star
formation, feedback and stellar aging. This baryonic modelling is tuned 
so that our simulations reproduce the observed properties of isolated 
spirals outside clusters. Without further parameter adjustment our 
simulations then produce a luminosity function, a mass-to-light ratio,
luminosity, number and velocity dispersion profiles, and a 
morphology-radius relation which are similar to those observed in real
clusters. In particular, since our simulations follow galaxy merging 
explicitly, we can demonstrate that it accounts quantitatively for the
observed cluster population of bulges and elliptical galaxies.
\end{abstract}
\footnotetext{E-mail: vspringel@mpa-garching.mpg.de}

\section{Introduction}

The last two decades have witnessed substantial progress towards an
understanding of hierarchical galaxy formation within
the framework of a universe dominated by cold dark matter (CDM). For
an appropriate choice of the cosmological parameters, the CDM theory
provides a remarkably successful description of large-scale structure
formation, and it is in good agreement with a large variety of
observational data. Much of this progress has been achieved by
detailed analytical and numerical studies of the collisionless
dynamics of the dark matter. As a result, this part of cosmic
evolution is now quite well understood. However, the actual formation
of the luminous parts of galaxies within CDM universes involves many
complex physical processes in addition to gravity, for example,
shocking and cooling of gas, and star formation with its attendant 
regulation and feedback mechanisms. The theoretical modeling of 
important aspects of these processes is still highly uncertain.

Not surprisingly, the lack of precise specifications for treating
the relevant physics has also hampered direct numerical studies of
galaxy formation. In addition, such studies are confronted with the huge
range of scales over which these physical processes interact.
Difficulties in finding robust and appropriate algorithms for
handling ``subgrid'' physics have so far prevented hydrodynamical 
simulations from reproducing many basic properties of galaxies, although
more recent work is beginning to achieve some notable successes both for
individual galaxies \citep[e.g.][]{Ka91,Nav94,St95,Mi96,Wa96,Na97,St99} and
for the distribution of galaxy populations
\citep[e.g.][]{Ce93,Ce00,Ka96,Kat99,We97,Wein2000,St95,Bl99,Pea99,Pea2000}.

Much of our current understanding of galaxy formation has 
been learned through `semi-analytic' techniques, as laid
out originally by \citet{Wh91}, \citet{Col91} and \citet{Lac91},
building on the scenario first sketched by \citet{Whi78}.  In
these models, each of the complicated and interacting physical
processes involved in galaxy formation is approximated using 
a simplified, physically based model. These processes include the
growth through accretion and merging of dark matter haloes, the shock heating
and virialization of gas within these haloes, the radiative cooling of
gas and its settling to a rotationally supported disk, star formation, and the
resulting feedback from supernovae and stellar winds, the evolution
of stellar populations, absorption and reradiation by dust, and galaxy
merging with its accompanying starbursts and morphological
transformations.  At the expense of uncertainties introduced by the 
simplifying assumptions, semi-analytic techniques can access a 
much larger dynamic range than numerical simulations, they
allow a fast exploration of parameter space and of the influence of
the specific simplifying assumptions chosen, and they facilitate
direct comparisons with a wide range of observational data.

Over the last few years, a number of groups have used semi-analytical
models to study galaxy formation, and to interpret observations of 
galaxy populations at low and high redshift
\citep{Lac93b,Kau93a,Kau94,Col94,Hey95,Bau96b,Bau96a,Kau95a,Kau95b,Kau96b,
Kau96a,Gu98,Kau98,Bau98,Bau99,Mo98,Mo99,Dev98,Som99,Kau2000,Cole2000,
vdB00,Boi00,Ha00,Som00}.
Stellar population synthesis modelling allows detailed
photometric comparison with observation, including studies of the
strong evolution apparent in observed high redshift galaxy samples.
With a small number of free parameters, semi-analytic models have been
quite successful in allowing a unified and coherent interpretation
of a broad range of galaxy properties, for example, luminosity
functions, Tully-Fisher and Faber-Jackson relations, number counts, 
the distributions of morphology, color and size, global and individual
star formation histories, background radiation contributions from the
UV to the far IR, clustering strengths, and the observed relations 
between AGN's and their host galaxies. Most studies concentrate on 
a few specific issues. Oversimplifications in the adopted physical 
models often show up as inconsistencies with observation in other
areas. For example, until recently most studies had difficulty
simultaneously to fit the zero-point of the Tully-Fisher relation and the
luminosity function of field galaxies. Recent work has removed some
of the most serious oversimplifications in this area and has
reduced the discrepancy significantly \citep{Som99,vdB00}.

The construction of dark matter merging history trees
\citep{Kau93b,Som98b,Sh99} is an important ingredient in semi-analytic
models.  In most studies, Monte-Carlo realizations of merging
histories for individual objects are generated using the extended
Press-Schechter formalism \citep{Pre74,Bon91,Lac93}. A disadvantage of
this approach is that there is little information about the spatial
distribution of galaxies, although two point correlations can be
estimated using the methods introduced by \citet{MoW96} \citep[see
also][]{Bau99}.  In order to study clustering in more detail
semi-analytic models have been combined with cosmological N-body
simulations; the galaxy population within each of the virialised
haloes present in a given output of the simulation is created by a
Monte Carlo ``semi-analytic'' realisation of its prior history and
these galaxies are then attached to the halo centre and to random
``satellite'' particles within the halo
\citep{Kau97,Gov98,Ben00b,Ben00a,We00}. This allows mock catalogues of
galaxies to be constructed which contain all the spatial and kinematic
information of real redshift surveys.

In a natural extension of this approach, one can use N-body
simulations not only to provide the mass distribution at a given 
time, but also to reconstruct individual halo merging histories, 
a scheme first tried in a crude form by \citet{Wh87}.
This allows one to avoid the uncertainties inherent in
the Press-Schechter formalism, and it provides
spatial and kinematic distributions not just for the final
galaxies but for their progenitors at all earlier times as well. Thus it
is effectively equivalent to a full dynamical simulation of galaxy formation 
and clustering, but with the advantage that the computationally intensive
part of the procedure, the original N-body simulation, does not
need to be repeated every time the assumptions about baryonic processes
are changed. The disadvantages in comparison with the
simpler scheme just discussed are that the resolution of
the merging history trees is limited by that of the N-body simulation
and that particle data must be stored sufficiently often for the
trees to follow adequately the growth of structure.
The method thus requires frequent data dumps from high resolution 
simulations of  cosmologically representative volumes, leading
to a substantial raw data volume.

Recently, \citet{Rou97} studied merging history trees directly from
N-body simulations using scale-free simulations and a rather limited
number of simulation outputs. A much more extensive study has been
published by \citet[hereafter KCDW]{Kau99}, who constructed merging
history trees from two high-resolution N-body simulations using a
total of 51 output times between redshift $z=20$ and $z=0$. KCDW
grafted semi-analytic models of galaxy formation onto the simulated
merger trees to study how the clustering of galaxies is
related to intrinsic properties like luminosity, colour and morphology.
In subsequent papers, they
used this methodology to predict the evolution of clustering to high
redshift \citep{Kau99b}, to construct realistically selected mock
redshift surveys \citep{Dia99}, and to study the spatial and kinematic
distributions of galaxies within clusters \citep{Dia2000}.

Using a somewhat different technique, \citet{Kam99} also made use of
the full merging history of N-body simulations. They modified an
existing N-body code so that heavier `tracer' particles were
introduced during the execution of the simulation. These tracer 
particles, identified with galaxies, replaced locally overdense groups
of the orginal particles. This approach neglects the
internal structure of galaxy/halo systems, for example the fact
that such systems can be stripped of much of their mass as they orbit
within a cluster. In addition it is somewhat inflexible since
any change in the assumptions about how baryonic part of galaxies 
forms and evolves requires the simulation to be repeated.

Our approach follows the methodology of KCDW, extending it to deal
with higher resolution simulations.  In such simulations, substantial
substructure, partially stripped haloes of cluster galaxies, can be
identified within dark matter clumps corresponding to galaxy groups
and clusters. To this end, we study four simulations of the formation
of a rich cluster of galaxies.  These simulations follow the same
object, a Coma-like cluster of mass $8\times 10^{14}\,\msunh$ in a
flat, low-density universe, but use different mass resolutions.  We
resolve the virial region of the final cluster with $0.12$, $0.61$,
$3.5$ and $20$ million particles, respectively, and we sample the
field in the region immediately around the cluster at the same
resolution but with about twice as many particles in each case. The
cosmological tidal field is represented by an additional boundary
region with $\sim 3.1$ million particles extending to a distance of
$70\lmu$ from the cluster. This sequence allows us to test explicitly
how our results depend on numerical resolution.

We develop a new algorithm, \subfind, for identifying substructure
within the clumps formed in these simulations. This algorithm defines
`subhaloes' as locally overdense, self-bound particle groups, and is
able to detect hierarchies of substructure using just a single
simulation output.  As in KCDW, we have stored 51 simulation snapshots
from $z=20$ to the current epoch, and we trace the merging history of
groups and their subhaloes from output to output. We modify the
semi-analytic recipes employed by KCDW to allow the inclusion of
subhaloes, and we analyse the changes resulting from this increase in
fidelity to the physical system modelled.

In particular, we study the luminosity function of the cluster, its
mass-to-light ratio, the Tully-Fisher relation of spirals in the
surrounding field, the Faber-Jackson relation of field and cluster
ellipticals, and the {$B-V$} colors of our model galaxies.  We show
that the new subhalo-scheme gives rise to a pronounced radial
segregation by morphology.  We investigate luminosity segregation by
comparing the radial profiles of galaxies by number, morphology and
luminosity with that of the dark matter. Finally, we study how the
velocity dispersion profile of the galaxies depends on their
luminosity and colour and compare it with the corresponding quantity
for the dark matter. For all these quantities we are able to
demonstrate good numerical convergence for all galaxies brighter than
the SMC.

Our approach also allows a detailed study of the formation history of
the cluster and its galaxies. In particular, we have analysed the star
formation history of field and cluster galaxies, the spatial and
temporal origin of the `first' stars that end up in the cluster
\citep{Whi99}, and the evolution of the merger rate of galaxies. These
results are discussed in a companion paper.

Interestingly, the new subhalo analysis improves the agreement with
observational data, especially with respect to the cluster luminosity
function.  Most of the bright galaxies in the final, highly-resolved
cluster are still connected to well localized subhaloes within the
smooth dark matter background of the cluster.  There is hence no need
to estimate merging timescales within the cluster using dynamical
friction or approximate cross-section arguments. Mergers are treated
automatically in a fully dynamically consistent way.
As we will discuss, inaccuracies in estimated merging times can lead
to a problem of excessively bright first ranked cluster galaxies in
the simpler methodology which does not track subhaloes.
This problem goes away in our refined approach which also
demonstrates explicitly that merging can account for the observed
fractions of elliptical and bulge-dominated galaxies and for their 
distribution within clusters.

This paper is organized as follows. In Section 2 we describe the N-body
simulations, and in Section 3 we review the techniques of
KCDW, and our specific implementation of them. In Section 4 we discuss
our techniques for identifying dark matter substructure within larger
haloes, and our methods of tracing it from output to ouput in the 
simulations. We then describe the implementation of semi-analytic models
including this subhalo information, and we present results obtained
with these prescriptions in Section 5. Finally, we discuss some
aspects of our findings in Section 6.

\section{Cluster simulations}

In this study, we analyse collisionless simulations of clusters of
galaxies that are generated by the technique of `zooming in' on a
region of interest \citep{Tor97}. In a first step, a cosmological
simulation with sufficiently large volume is used to allow the
selection of a suitable target cluster.  For this purpose, we employed
the GIF-$\Lambda$CDM\footnote{The GIF project is a joint effort by
astrophysicists in Germany and Israel.} model carried out by the Virgo
consortium.  It has cosmological parameters $\Omega_0=0.3$,
$\Omega_\Lambda=0.7$, $h=0.7$\footnote{We employ the convention
$H_0=100\, h\, {\rm km\,s^{-1}\,Mpc^{-1}}$.}, spectral shape
$\Gamma=0.21$, and was cluster-normalized to $\sigma_8=0.9$.  The
simulation followed $256^3$ particles of mass $1.4\times
10^{10}h^{-1}{\rm M}_{\odot}$ within a comoving box of size
$141.3\lmu$ on a side. Note that this simulation is one of the models
recently studied by KCDW.  We selected the second most massive cluster
that had formed in the GIF simulation for further study.  This cluster
has a virial mass $8.4\times 10^{14}\msunh$, and it appears to be well
relaxed at the present time.

\begin{table*}  
\bc
\caption{Numerical parameters of our cluster simulations.
All four simulations compute the evolution of the same cluster, assuming a
$\Lambda$CDM universe with cosmological parameters $\Omega_0=0.3$,
$\Omega_\Lambda = 0.7$, $\Gamma=0.21$, $\sigma_8=0.9$, and $h=0.7$.
The simulations follow a sphere of matter with comoving diameter
$141\lmu$.  In the Table, we give the particle mass $m_{\rm p}$ used
in the central high-resolution zone, the starting redshift $z_{\rm
start}$, the gravitational softening $\epsilon$ in the high-resolution
zone, the number $N_{\rm hr}$ of high-resolution particles, the number
$N_{\rm bnd}$ of boundary particles, the total number $N_{\rm tot}$ of
particles, and the number $N_{\rm p}$ of processors used in each of
the simulations S1-S4.  The gravitational softening was kept
fixed at the given values in physical coordinates below redshift
$z=9$, and in comoving coordinates above this redshift.
\label{tabsims}
}
\vspace{0.5cm}
\begin{tabular}{ccccc}
   & S1 & S2 & S3 &S4  \\
\hline
$m_{\rm p}\; [\msunh]$ & $6.87\times 10^{9}$  & $1.36\times 10^{9}$ & $2.38\times 10^{8}$ &
$4.68\times 10^{7}$ \\
$z_{\rm start}$ & 30 & 50 & 80 & 140 \\
$\epsilon \; [\lu]$ &  6.0   & 3.0 & 1.4 & 0.7\\
$N_{\rm hr}$ & 450088  & 1999978 & 12999878 & 66000725\\
$N_{\rm bnd}$ & 3029956  & 3117202 & 3016932  & 3013281 \\
$N_{\rm tot}$ & 3480044  & 5117180 & 16016810 & 69014006\\
$N_{\rm p}$ & 16 & 32 & 128 & 512 \\
\hline
\end{tabular}
\ec
\end{table*}

\begin{figure*}
\bc
\resizebox{8cm}{!}{\includegraphics{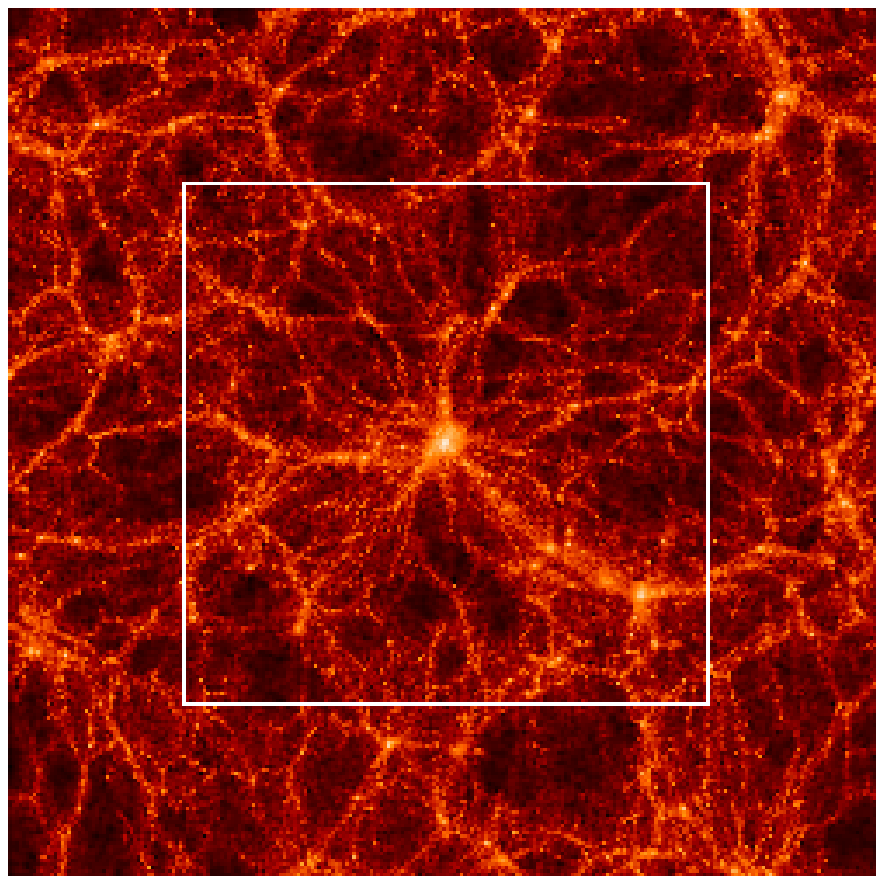}}%
\hspace{0.5cm}\resizebox{8cm}{!}{\includegraphics{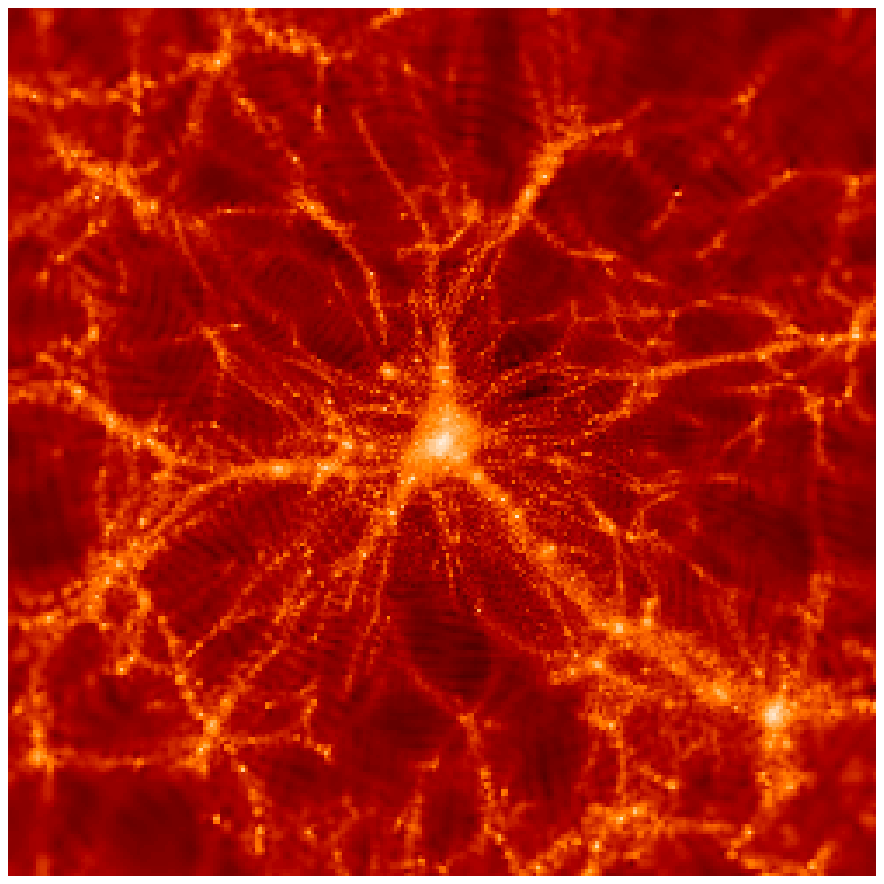}}%
\caption{The projected mass density fields in slices of thickness
10$\lmu$ around the cluster center in the original GIF simulation
(left), and in the S3 resimulation (right).  The left image is
$141\lmu$ on a side, and the white square marks the region ($85\lmu$
on a side) that is displayed in the image of the resimulation on the
right.  In the right panel, you may notice small traces of the
spherical grid used to represent the density field in the boundary
region.  Note that these residuals of the grid structure are just seen
because of projection effects that arise in the visualization
technique.
\label{figSimuls}}
\ec
\end{figure*}

\begin{figure*}
\bc
\resizebox{15cm}{!}{\includegraphics{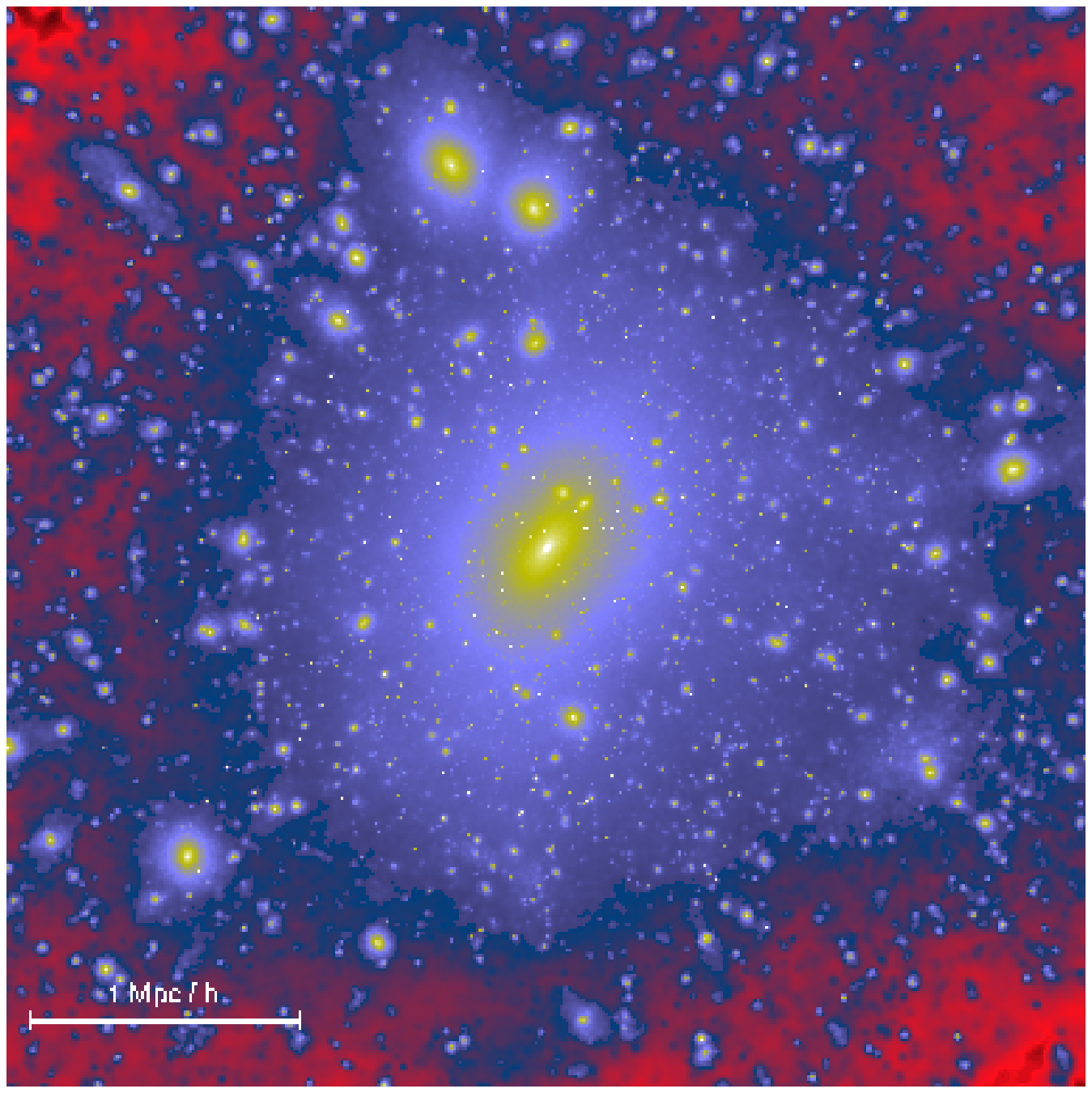}}
\caption{The dark matter distribution of the S4 cluster at $z=0$.  The
image shows all the mass in a box of sidelength $4\lmu$ around the
cluster center. To render the substructure visible, particles have
been weighted by their local density (computed by adaptive kernel
estimation), and a logarithmic color scale has been applied. Note that
the small bright dots that are visible in the cluster should not be
mistaken as noise -- they are in fact self-bound subhaloes and
correspond to surviving cores of haloes that have fallen into the
cluster at some earlier time.
\label{figSimulT4}}
\ec
\end{figure*}

In a second step, we simulated the formation of this cluster again,
using greatly increased mass and force resolution. To this end, the
particles in the final GIF-cluster and in its immediate surroundings
were traced back to their Lagrangian region in the initial
conditions. The corresponding part of the displacement field was then
sampled using a glass-like particle distribution with smaller particle
masses than in the GIF simulation. Due to the increase in resolution,
the fluctuation spectrum could now be extended to smaller scales. We
added a random realization of this additional small-scale power, while
we kept all the waves on larger scales that had been used in the GIF
simulation.

Outside this central high-resolution region, we gradually degraded the
resolution by using particles with masses that grow with distance from
the center.  In this `boundary region', we employed a spherical grid
whose spacing grew with distance from the high resolution zone.  The
spherical boundary region extends to a total diameter of $141.3\lmu$,
which is just the box size of the original GIF simulation. Beyond this
region, we assumed vacuum boundary conditions, i.e.~a vanishing
density fluctuation field. Using comoving coordinates, we then evolved
the simulations to redshift $z=0$ with our new parallel tree-code
\gadget. This code uses individual timesteps for all particles, and
was designed to run on massively parallel supercomputers with
distributed memory. Parallelization is achieved explicitly using the
communication library of the {\em Message Passing Interface} (MPI).  A
full account of the numerical and algorithmic details of \gadget~is
given elsewhere \citep{SprGadget2000}.

To be able to study systematic effects arising from numerical
resolution, we simulated the same cluster several times, increasing
the resolution step by step.  In the first step, refered to as
simulation `S1' from here on, the particle mass was just about two
times smaller than in the original GIF simulation. For the
high-resolution zone of S1, we used a total of 450000 particles and a
gravitational softening of $\epsilon=6.0\lu$. The boundary region was
represented with an additional $\sim 3$ million particles. This
relatively high number of boundary particles was chosen with the
sequence of our planned simulations in mind.  Except for slight
changes at its inner rim, we kept the sampling of the boundary region
fixed for the other simulations, where we populated the central zone
with many more particles.  In simulation `S2', we used 2 million
high-resolution particles, and we reduced the gravitational softening
to $3.0\lu$.  In simulation `S3', we employed a total of 13 million
high-resolution particles with mass $2.4\times 10^{8}\msunh$, and a
softening of $1.4\lu$.  Finally, in simulation `S4' we used 66 million
particles for the high-resolution zone, pushing the particle mass down
to $4.68\times 10^{7}\msunh$ and the spatial resolution down to
$0.7\lu$.  In each case, roughly one third of the particles in the
high-resolution zone ends up in the virialized region of the final
cluster.  This means that S4 resolves a single object with about 20
million particles.

In Table~\ref{tabsims}, we summarize important numerical parameters of
our simulations. Note that we have softened gravity using a spline
kernel. Our cited values for $\epsilon$ are such that the
gravitational potential of a point mass at zero lag is $\Phi=-Gm/
\epsilon$, and that the softened force becomes Newtonian at a distance
$2.8\,\epsilon$.  We have kept the softening length fixed in physical
coordinates below redshift $z=9$, and in comoving coordinates at
higher redshift.  The softening for the boundary particles was set to
much larger values, in an inner shell around the cluster to $15\lu$,
and further outside to $75\lu$.  For all four simulations, we stored
51 outputs, logarithmically spaced in expansion factor between
redshifts $z=20$ and $z=0$.

In Figure~\ref{figSimuls}, we show two images comparing the original
density field of the GIF-simulation with that of our S3 resimulation.
The filaments of dark matter around the cluster are nicely reproduced
by S3, even relatively far away from the cluster, where the resolution
of S3 has already fallen below that of the GIF simulation. It is 
also apparent that S3 has much higher resolution in the cluster
itself.

However, the vast increase in resolution that this new set of
simulations offers is perhaps best appreciated if we zoom in onto the
cluster directly.  In Figure~\ref{figSimulT4}, we show an image of the
dark matter distribution in a box of side-length $4\lmu$, centered on
the cluster that formed in S4, the simulation with our highest
resolution. Note that there is essentially no particle noise visible
in this picture; all the small bright features are genuine
self-gravitating subhaloes. So dark matter haloes forming in CDM
cosmologies exhibit a remarkable richness of substructure, and are
thus quite far from the smooth, over-merged objects suggested by
numerical work on cluster formation until a few years ago.  This new
view of CDM dynamics has only recently been fully established, with
some of the most important work done by \citet{Gh98,Moo99} and
\citet{Kly99}. Our best simulation achieves an even larger dynamic
range than previous work; as we will see, the virialized region of the
S4 cluster contains nearly 5000 self-gravitating subhaloes.

\section{Modeling galaxy formation using N-body merging trees}

In the following, we briefly summarize our specific implementation of
the techniques developed by KCDW to combine semi-analytic models for
galaxy formation with dark matter merging history trees constructed
directly from cosmological N-body simulations. We will later extend
this formalism to include dark matter substructure, and we will be
especially interested in any changes of the results arising from that.

There are essentially two main parts in the modeling: (1) The
measurement of dark matter merging trees from a sequence of simulation
outputs. (2) The implementation of the actual semi-analytic recipes
for galaxy formation on top of these merging trees. Both parts of the
modeling are technically complex and warrant a detailed discussion,
which we provide in the following sections for the sake of
completeness.  Readers who are primarily interested in the results of
our modelling may want to proceed directly to Section 5 upon a first
reading of this paper.

In this Section, we start by describing our implementation of the
techniques of KCDW.  First we treat the construction of the merging
trees, then the physics of galaxy formation. In Section 4, we describe
what we change in the two parts to allow the inclusion of subhalo
information.

\subsection{Following the merging trees}

For each simulation output, we compile a list of dark matter haloes
with the friends-of-friends (FOF) algorithm using a linking length of
$0.2$ in units of the mean interparticle separation. We only include
groups with at least 10 particles in the halo catalogue.  The majority
of such haloes are already stable, i.e. particles found in a
10-particle group at one output time are almost all part of the same
halo in subsequent simulation outputs. For each halo, we also
determine the most-bound particle within the group, where `most-bound'
here refers to the particle with the minimum binding energy.

We then follow the merger tree of the dark matter from output to
output.  A halo $H_{\rm B}$ at redshift $z_{\rm B}$ is defined to be a
{\em progenitor} of a halo $H_{\rm A}$ at redshift $z_{\rm A}<z_{\rm
B}$, if at least half of the particles of $H_{\rm B}$ are contained
within $H_{\rm A}$, and the most bound particle of $H_{\rm B}$ is
contained in $H_{\rm A}$, too. These definitions already suffice to
uniquely define the dark matter merging trees.

\subsubsection{Defining a galaxy population}

So far we are just dealing with catalogues of dark matter haloes. We
now supplement this with the notion of a {\em galaxy population} with
physical properties given by the semi-analytic techniques.  In our
formalism, each dark halo carries exactly one {\em central} galaxy,
and its position is given by the most-bound particle of the halo.
Only the central galaxy is supplied with additional gas that cools
within the halo.

A halo can also have one or several {\em satellite} galaxies, where
the position of each of them is given by one of the particles of the
halo.  Satellites are galaxies that had been central galaxies
themselves in the past, but their haloes have merged at some previous
time with the larger halo they now reside in. Satellite galaxies orbit
in their halo and are assumed to merge with the central galaxy on a
dynamical friction timescale. Note that they are cut off from the
supply of fresh cool gas, so they may only form stars until their
internal reservoir of cold gas is exhausted.

Finally, we define a class of {\em field} galaxies, which are
introduced to keep track of satellites whose particles are currently
not attached to any halo, for example because they have been ejected
out of their parent halo. Usually, these ``lost'' field galaxies are
accreted onto a halo later on.

\subsubsection{Following the galaxy population in time}

At a given output time, we therefore deal with a galaxy population
consisting of {\em central} galaxies, {\em satellite} galaxies, and
{\em field} galaxies, each attached to the position of a simulation
particle.  Starting at the first output time at high redshift (when
the first haloes have formed), we initialize the galaxy population
with a set of central galaxies, one for each halo, with stellar mass,
cold gas mass, and luminosity set to zero.  The physical properties of
these galaxies are then evolved to the next output time, where we
obtain a new galaxy population based on a combination of semi-analytic
prescriptions and the merging history of the dark matter. Propagating
this scheme forward in time from output to output we obtain the galaxy
population at the present time, and at all output times at higher
redshift.

We now describe in more detail our rules and prescriptions for this
evolution. Beginning with the galaxy population at redshift $z_{\rm
B}$, we first generate the galaxies of the new population at redshift
$z_{\rm A}<z_{\rm B}$ based on the merging history of the dark matter.
Using the group catalogues of the corresponding simulation outputs and
the galaxy population at $z_{\rm B}$, we construct an `initial' set of
galaxies at $z_{\rm A}$ as follows:

\begin{enumerate}
\item
Each galaxy at $z_{\rm B}$ is assigned to its new halo at $z_{\rm A}$.
If the particle used to tag a galaxy does not reside in any halo, the
galaxy becomes a field galaxy.
\item 
Each halo at $z_{\rm A}$ selects its central galaxy as the central
galaxy of its most massive progenitor. This central galaxy is
repositioned to the position of the most-bound particle, i.e.~a new
particle tagging the galaxy is selected. The central galaxies of all
other progenitors become satellites of the halo.
\item 
If a halo has no progenitors, a new central galaxy is created at the
position of its most-bound particle.  In the event that the halo
contains one or more galaxies (particles recovered from the field),
the central galaxy is picked as the most massive of these.
\end{enumerate}

Once the new set of galaxies is generated in this way, the properties
of the galaxies are evolved according to the physical prescriptions
described below, resulting finally in the new galaxy population at
redshift $z_{\rm A}$.  Note that some of the satellite galaxies
generated in the initial set for $z_{\rm A}$ will merge with central
galaxies during this evolution, and thus not necessarily be part of
the `final' population at $z_{\rm A}$.

\subsection{Physical evolution of the galaxy population}

We model the following physical processes: (1) Radiative cooling of
hot gas onto central galaxies.  (2) Transformation of cold gas into
luminous stars by star formation.  (3) Reheating of cold gas, or its
ejection out of the halo, by supernova feedback.  (4) Orbital decay of
satellites and their merging with central galaxies.  (5)
Spectrophotometric evolution of the luminous stars. (6) Simplified
morphological evolution of galaxies.  Below we detail the physical
parameterizations adopted for these processes.

\subsubsection{Gas cooling\label{colmodel}}

Gas cooling is modeled as in \citet{Wh91}. We assume that the hot gas
within a dark halo is distributed like an isothermal sphere with
density profile $\rho_{\rm g}(r)$.  Then the local cooling time
$t_{\rm cool}(r)$ can be defined as the ratio of the specific thermal
energy content of the gas, and the cooling rate per unit volume, viz.
\be t_{\rm cool}(r)=\frac{3}{2}\frac{k T \rho_{\rm g}(r)}{ \ol{\mu}
m_{\rm p} n_{\rm e}^2(r)\, \Lambda(T,Z)}.  \ee Here $\ol{\mu} m_{\rm
p}$ is the mean particle mass, $n_{\rm e}(r)$ the electron density,
and $\Lambda(T,Z)$ the cooling rate.  The latter depends quite
strongly on the metallicity $Z$ of the gas, and on the virial
temperature $T=35.9\,(V_{\rm vir}/\kms)^2\,{\rm K}$ of the halo. We
employ the cooling functions computed by \citet{Su93} for collisional
ionisation equilibrium to represent $\Lambda(T,Z)$, but we restrict
ourselves to primordial metallicity in the present study.

We define the cooling radius $r_{\rm cool}$ as the radius for which
$t_{\rm cool}$ is equal to the time for which the halo has been able
to cool `quasi-statically'. We will approximate this time with the
dynamical time ${R_{\rm vir}}/{V_{\rm vir}}$ of the halo.  If the
cooling radius lies well within the virial radius $R_{\rm vir}$ of a
given halo, we then take the cooling rate to be \be \frac{\dd M_{\rm
cool}}{\dd t} =4\pi\rho_{\rm g}(r_{\rm cool})r_{\rm cool}^2 \frac{\dd
r_{\rm cool}}{\dd t} .  \ee Note that we define the virial radius
$R_{\rm vir}$ of a FOF-halo as the radius of a sphere which is
centered on the most-bound particle of the group and has an
overdensity 200 with respect to the critical density. We take the
enclosed mass $M_{\rm vir}= 100H^2R_{\rm vir}^3/G$ as the virial mass,
and we define the virial velocity as $V^2_{\rm vir}=GM_{\rm vir}/R_{\rm
vir}$.

Adopting an isothermal sphere for the distribution of the hot gas of
mass $M_{\rm hot}$ within the halo, i.e.  \be \rho_{\rm
g}(r)=\frac{M_{\rm hot}}{4\pi R_{\rm vir} r^2}, \ee the cooling rate
is then given by \be \frac{\dd M_{\rm cool}}{\dd t} =\frac{M_{\rm
hot}}{R_{\rm vir}}\, \frac{r_{\rm cool}}{2 t_{\rm cool}}.
\label{cool1}
\ee At early times or for low-mass haloes the cooling radius can be
much larger than the virial radius.  In this case, the hot gas is
never expected to be in hydrostatic equilibrium, and the cooling rate
will essentially be limited by the accretion rate onto the central
galaxy. We approximate this rate with \be \frac{\dd M_{\rm accr}}{\dd
t} =\frac{M_{\rm hot}}{2 t_{\rm cool}},
\label{cool2}
\ee which corresponds to $r_{\rm cool} = R_{\rm vir}$ in equation
(\ref{cool1}).  We adopt the minimum of equations (\ref{cool1}) and
(\ref{cool2}) as our actual cooling rate. As in KCDW we find it
necessary to implement an upper limit on the circular velocity
of haloes in which cooling gas is allowed to settle onto a central
galaxy. This accounts for the fact that observed cluster cooling
flows do not form stars at the observed apparent cooling rate (which
corresponds to that estimated above). We follow KCDW in setting
this upper limit at a circular velocity of $350\,{\rm km\, s}^{-1}$.

\subsubsection{Star formation}

In this work, we model the star formation rate of a galaxy as \be
\frac{\dd M_{\star}}{\dd t}= \alpha \frac{M_{\rm cold}}{t_{\rm dyn}},
\ee where $M_{\rm cold}$ is the mass of its cold gas, and $t_{\rm
dyn}$ is the dynamical time of the galaxy.  We approximate the latter
as \be t_{\rm dyn}=\frac{R_{\rm eff}}{V_{\rm vir}}, \ee with $R_{\rm
eff}=0.1 R_{\rm vir}$, i.e.~we set the effective stellar radius to be
a fixed fraction of the virial radius.  Notice that at a fixed
redshift, $R_{\rm vir}$ is proportional to $V_{\rm vir}$, hence
$t_{\rm dyn}$ depends only on redshift. The dimensionless parameter
$\alpha$ regulates the efficiency of star formation and is treated as
a free parameter. In this work, we keep $\alpha$ constant in time, but
we remark that a redshift dependence of $\alpha$ may be required to
provide a better understanding of the rapid evolution of the number
density of luminous quasars \citep{Kau2000} and to match the observed
abundance of Lyman break galaxies at $z\sim 3$ \citep{Som00}.  Once a
galaxy falls into a larger halo and becomes a satellite, the values of
$R_{\rm vir}$ and $V_{\rm vir}$ are not changed any more.  The galaxy
can then continue to form stars until its reservoir of cold gas is
exhausted, but it does not receive new cold gas by cooling processes.

\subsubsection{Feedback}

Assuming a universal initial mass function (IMF), the energy released
by supernovae per formed solar mass is $\eta_{\rm SN}E_{\rm SN}$,
where $\eta_{\rm SN}$ gives the expected number of supernovae per
formed stellar mass, and $E_{\rm SN}$ is the energy released by each
supernova.  The formation of a group of stars with mass $\Delta
M_{\star}$ will thus be accompanied by the release of a feedback
energy of $\eta_{\rm SN}E_{\rm SN}\Delta M_{\star}$, where we adopt
$\eta_{\rm SN}=5.0\times 10^{-3} {\rm M}_{\odot}^{-1}$, based on the
\citet{Sca86} IMF, and $E_{\rm SN}=10^{51}\,{\rm erg}$.

One major uncertainty is how this energy affects the evolution of the
interstellar medium, and how the star formation rate is regulated by
it \citep{Spr99}. We here assume that the feedback energy reheats some
of the cold gas back to the virial temperature of the dark halo. The
amount of gas reheated by this process is then \be \Delta M_{\rm
reheat}=\frac{4}{3}\,\epsilon\,\frac{\eta_{\rm SN}E_{\rm SN}}{V_{\rm
vir}^2}\,\Delta M_{\star}, \ee where the dimensionless parameter
$\epsilon$ describes the efficiency of this process.

Following KCDW, we consider two alternative schemes for the fate of
the reheated gas.  In the {\em retention} scheme, the reheated gas is
simply transferred from the cold phase back to the hot gaseous halo,
and the reheated gas thus stays within the halo.  Alternatively, in
the {\em ejection} scheme we assume that the gas leaves the halo, and
it is only re-incorporated into the halo at some later time.  If
$\Delta M_{\rm ejec}$ is the total gas mass ejected by a galaxy, we
model this reincorporation by decreasing $\Delta M_{\rm ejec}$ to zero
again on the dynamical timescale of the halo.

\subsubsection{Mergers of galaxies}

In CDM universes, large haloes form by mergers of smaller haloes. As a
consequence, mergers of galaxies are an inevitable process. We assume
that the satellite galaxies orbiting within a dark matter halo
experience dynamical friction and will eventually merge with the
central galaxy of the halo. In principle, mergers between two
satellite galaxies are also possible. These events are expected to be
rare, but they do happen occasionally as we show in the companion
paper. For the moment, we neglect these events but we will take them
into account in our subhalo-scheme later on.

N-body simulations by \citet{Na95} suggest that the merging timescale
can be reasonably well approximated by the dynamical friction
timescale \be T_{\rm friction} =
\frac{1}{2}\frac{f(\epsilon)}{C}\frac{V_{\rm c}r_{\rm c}^2}{G M_{\rm
sat} \ln \Lambda } \,.
\label{friceq}
\ee The formula is valid for a small satellite of mass $M_{\rm sat}$
orbiting at a radius $r_{\rm c}$ in an isothermal halo of circular
velocity $V_{\rm c}$.  The function $f(\epsilon)$ describes the
dependence of the decay on the eccentricity of the satellites' orbit,
expressed in terms of $\epsilon=J/J_{\rm c}(E)$, where $J_{\rm c}(E)$
is the angular momentum of a circular orbit with the same energy as
the satellite. The function $f(\epsilon)$ is well approximated by
$f(\epsilon)\simeq \epsilon^{0.78}$, for $\epsilon>0.02$
\citep{Lac93}. $C$ is a constant with value $C\simeq 0.43$, and $\ln
\Lambda$ is the Coulomb logarithm.

We follow KCDW and approximate $r_{\rm c}$ with the virial
radius of the halo when the satellite first falls into it.  To
describe the orbital distribution, we adopt the average value
$\left<f(\epsilon)\right>\simeq 0.5$, computed by \citet{To97}. Note
that this differs slightly from KCDW who drew a random orbit uniformly
from $\epsilon\in[0.02,1]$. We identify the mass of the satellite with
the virial mass of the galaxy at the time when it was last a central
galaxy, and we approximate the Coulomb logarithm with
$\ln\Lambda=\left(1+M_{\rm vir}/M_{\rm sat}\right)$.

When a small satellite merges with a central galaxy, we transfer all
its stellar mass to the bulge component of the central galaxy, and we
update the photometric properties of this galaxy
accordingly. Similarly, the cold gas of the satellite is transferred
to the disk of the central galaxy.  If the mass ratio between the
stellar components of the merging galaxies is larger than some
threshold value (we adopt 0.3 for that), the merger destroys the disk
of the central galaxy completely, and all stars form a single
spheroid, i.e. they generate a bulge.  This is called a major merger.
In addition, we assume that all the cold gas left in the two merging
galaxies is rapidly consumed in a starburst. The stars created in this
burst are also added to the bulge component. Since the central galaxy
is fed by a cooling flow, it can grow a new disk component later on.

\subsubsection{Spectrophotometric evolution}

Photometric properties of our model galaxies can be constructed using
stellar population synthesis models \citep{Bru93}.  In these models,
the number of stars that initially form in each mass range is computed
according to the initial mass function. The stars then evolve along
theoretical evolutionary tracks. In this way, the spectra and colors
of a stellar population formed in a short burst of star formation can
be followed as a function of time. Once the evolution $F_{\nu}(t)$ of
the spectral energy distribution (SED) of a single age population of
stars is known, the SED $S_{\nu}(t)$ of a galaxy can be computed as
\be 
S_{\nu}(t)=\int_0^t F_{\nu}(t-t')\,\dot M_{\star}(t')\, \dd t' 
\ee
from its star formation history $\dot M_{\star}(t)$. Upon convolution
with standard filters, colors and luminosities in the desired bands
can be obtained. In principle, this technique also allows the
redshifting of spectra, and the incorporation of $k$-corrections to
make direct contact with observational photometric data at high
redshift.  In this work we use updated evolutionary synthesis models
by Bruzual \& Charlot (in preparation), which have been computed for
solar metallicity. Note that in this paper we do not attempt to model
the effects of dust. These effects can be be quite substantial in
late-type galaxies and would significantly affect a number of our
results \citep[c.f.][]{Som99}. In particular, when normalising our
models to the observed Tully-Fisher relation, the inclusion of dust
would cause us to assign a higher stellar mass to a galaxy of given
circular velocity

\subsubsection{Morphological evolution}

\citet{Si86} find a good correlation between the $B$-band
bulge-to-disk ratio, and the Hubble-type $T$ of galaxies.  For a
magnitude difference $\Delta M\equiv M_{\rm bulge} - M_{\rm total}$
they find a mean relation \be \left<\Delta M\right>=0.324(T+5) -
0.054(T+5)^2+0.0047(T+5)^3.  \ee Following previous semi-analytic
studies, we assign morphologies based on this
equation.  Specifically, we will usually classify galaxies with
$T<-2.5$ as ellipticals, those with $-2.5<T<0.92$ as S0's, and those
with $T>0.92$ as spirals and irregulars. However, we may allow a shift
in the boundaries between the three classes to obtain a better match
of our models with the observed relative abundances of these three
morphological types (see section 5.3).  Note that galaxies without 
any bulge are classified as type $T=9$.

\subsection{More implementation details}

Our practical implementation of the physical evolution of the
galaxy population includes the following steps.  We first estimate
merger timescales for those satellites that have newly entered a given
halo, i.e.~the galaxies that had not been contained in the largest
progenitor of the halo.  This `merger clock' is then decreased with
time in the subsequent evolution, and the satellite will be merged
with the central galaxy when this time has elapsed. Note that the
merger clock may be reset before the merger happens if the halo
containing the satellite merges with a larger system.

We then compute the total amount of hot gas available for cooling in
each halo.  Assuming on average a universal abundance of baryons equal
to the primordial one, this is simply \be M_{\rm hot}= f_{\rm b}
M_{\rm vir} - \sum_i \left[ M_{\star}^{(i)} + M_{\rm cold}^{(i)} +
M_{\rm ejec}^{(i)} \right], \ee where the sum extends over all
galaxies within the halo.  Here $f_{\rm b}=\Omega_{\rm b}/\Omega_0$
denotes the baryon fraction of the universe. Using the cooling model
of Section~\ref{colmodel}, we then estimate the cooling rate onto each
central galaxy, and we keep this rate constant during the time $\Delta
T$ between the two simulation outputs.

Once these quantities are known, we solve the simple differential
equations describing star formation, cooling and feedback.  We
typically use a number of $N\simeq 50$ small timesteps of size $\Delta
t=\Delta T/N$ for this purpose.  At each of these small steps, new
cold gas is added to the central galaxies.  For each galaxy, we form
some stellar mass $\Delta M_{\star}$ according to its star formation
rate, and we update its current and future photometric properties
accordingly.  The cold gas mass of each galaxy is reduced by the
amount of stars formed, and by the mass of the gas that is reheated or
ejected by supernova feedback.

At the end of each of the small steps, the merger clocks of the
satellites are reduced by $\Delta t$. If a satellite's merging time
falls below zero, it is merged with the central galaxy of its parent
halo. In practice, this means that the luminosity, the stellar mass,
and the gas mass of the satellite are transfered to the central
galaxy, and that the satellite is removed from the list of galaxies.
In addition, in the event of a major merger all cold gas of the
central galaxy is consumed in a short starburst, and all stellar
material is transformed into a spheroid.

\subsection{Choice of model parameters}

Following KCDW, we use the {\em I}-band Tully-Fisher relation to
normalize our models, i.e.~to set the free parameters $\alpha$ and
$\epsilon$ which specify the efficiency of star formation and
feedback, respectively.  For that purpose, we consider the central
galaxies of haloes in the periphery of the cluster, with morphological
types corresponding to Sb/Sc galaxies.  Note that we only use haloes
that are not contaminated by heavier boundary particles. The remaining
number of galaxies is sufficiently large to construct a well defined
Tully-Fisher diagram.  We try to fit the velocity based {\em I}-band
Tully-Fisher relation \be M_I - 5\log h = -21.00 - 7.68 (\log W - 2.5)
\ee measured by \citet{Gio97}.  We set the velocity width $W$ as twice
the circular velocity, and we assume that the circular velocity
$V_{\rm c}$ of a spiral galaxy is $\sim 15\%$ larger than the virial
velocity of that galaxies halo.  This is motivated by detailed models
for the structure of disk galaxies by \citet{Mo98} embedded in cold dark
matter haloes with the universal NFW profile \citep{NFW,NFW2}.

Keeping other parameters fixed, we find that varying $\epsilon$
changes both the slope and the zero-point of the Tully-Fisher relation
strongly. In particular, making feedback stronger tilts the
Tully-Fisher relation towards steeper slopes. On the other hand, the
star formation efficiency $\alpha$ only very weakly affects the
zero-point, but it has a strong effect on the gas mass fraction left
in galaxies at the present time.

In principle, it should be possible to specify the parameters $\alpha$
and $\epsilon$ using the slope and zero-point of the Tully-Fisher
relation {\em alone}.  However, the weak dependence of the
Tully-Fisher relation on $\alpha$ makes this impractical.  As in KCDW,
we instead use an additional criterion and require that the cold gas
(HI plus molecular) mass in a `Milky-Way' galaxy of circular velocity
$V_{\rm c}=220\kms$ is about $8\times 10^{9}\msunh$.

Note that the baryon fraction $f_{\rm b}$ can strongly influence the
cooling rates, and thus the absolute normalization of the models. As
\citet{Wh93} have shown, the baryon content of rich clusters of
galaxies argues for a baryon fraction as high as $f_{\rm
b}=0.1-0.2$. This is inconsistent with big bang nucleosynthesis (BBN)
constraints in a critical density universe, but can be accommodated
within low-density cosmologies, like the one considered in our cluster
models. We will assume $f_{\rm b}=0.15$ in this study, which is
consistent with current BBN constraints.  Our resulting parameter
values are listed in Table~\ref{modparam}.  We expect that slightly
different values of $f_{\rm b}$ will produce similar results when the
parameters $\alpha$ and $\beta$ are adjusted to compensate.

\section{Following halo substructure}

\subsection{Identification of substructure}

A basic step in the analysis of cosmological simulations is the
identification of virialized particle groups, which specify the sites
where luminous galaxies form.  Perhaps the most popular technique
employed for this task is the friends-of-friends (FOF) algorithm. It
places any two particles with a separation less than some linking
length $b$ into the same group.  In this way, particle groups are
formed that correspond to regions approximately enclosed by isodensity
contours with threshold value $\rho\propto 1/b^3$. For an appropriate
choice of $b$, groups are selected that are close to the virial
overdensity predicted by the spherical collapse model. FOF is both
simple and efficient, and its group catalogues agree quite well with
the predictions of Press-Schechter theory.

However, FOF has a tendency to link independent structures across
feeble particle bridges occasionally, and in its standard form with a
linking length of $b\simeq 0.2$ it is not capable of detecting
substructure inside larger virialized objects.  Using sufficiently
high mass resolution, recent studies
\citep{To97,Tor98,Gh98,Kly99,Moo99} were able to demonstrate that
substructure in dense environments like groups or clusters may survive
for a long time. The cores of the dark haloes of galaxies that fall
into a cluster will thus remain intact, and orbit as self-gravitating
objects in the smooth dark matter background of the cluster. In
previous simulations, haloes falling into clusters usually evaporated
quickly, and the clusters exhibited little signs of substructure
\citep[e.g.][]{Fre96}. It now appears, that sufficient numerical force
and mass resolution is enough to resolve this ``overmerging'' problem.

The identification of substructure within dark matter haloes is a
challenging technical problem, and several algorithms to find ``haloes
within haloes'' have been proposed.  In hierarchical
friends-of-friends (HFOF) algorithms \citep{Go98,Kly99} the linking
length of plain FOF is reduced in a sequence of discrete steps,
thus selecting groups of higher and higher overdensity and
eventually capturing true substructure.

Clearly, the need for a well-posed physical definition of
``substructure'' arises early on in such an analysis.  Most authors
have required subhaloes to be locally overdense and self-bound. We
will also adopt this requirement. Note that this implies that any
locally overdense region within a dense background needs to be treated
with an unbinding procedure. This is because a small halo within a
larger system represents only a relatively small fluctuation in
density, and a substantial amount of mass within the overdense region
will just stream through and not be gravitationally bound to the
substructure itself.

Groupfinding techniques that use some criterion of self-boundedness
include the bound density maximum (BDM) algorithm \citep{Kly99}, where
the bound subset of particles is evaluated iteratively in spheres
around a local density maximum.  In the method of \citet{Tor98},
previous simulation outputs are used to track the infall of particle
groups into larger systems.  Once such a particle group from the field
was accreted by a cluster, they simply determined the subset of those
particles that still remained self-bound.

Another approach is followed in DENMAX \citep{Gel94} and its offspring
SKID, where particles are moved along the local gradient in density
towards a local density maximum. Particles ending up in the `same'
maximum are then linked together as a group using FOF. SKID has been
employed by \citet{Gh98} to find substructure in a rich cluster of
galaxies, and to study the statistical properties of the detected
subgroups.

Integrating the gradient of the density field and moving the particles
is not without technical subtleties. For example, a suitable stopping
condition is needed. The new algorithm HOP of \citet{Eis98} tries to
avoid these difficulties by restricting the group search to the set of
original particle positions, just like FOF does. In HOP, one first
obtains an estimate of the local density for each particle, and then
attaches it to its densest neighbour. In this way a set of disjoint
particle groups are formed. However, a number of additional rules are
needed to link and prune some of these groups. For example, HOP may
split up a single virialized clump into several pieces of unphysical
shape, which have to be joined using auxiliary criteria.

It appears that all of these techniques have different strengths and
weaknesses, and that none is completely satisfactory (for example, DENMAX
and HOP do not require the identified substructure to be
gravitationally self-bound).  We have
therefore come up with a new algorithm to detect substructure in dark
matter haloes that incorprates ideas from SKID, HOP, FOF, and IsoDen
\citep{Pfi97}, as well as adding some new ones. For easier reference, we dub
this algorithm \subfind\ (for {\em subhalo finder}).

\subsection{The algorithm \subfind}

Our objective with \subfind\ is to be able to extract substructure,
which we define as locally overdense, self-bound particle groups
within a larger parent group. The parent group will be a particle
group pre-selected with a standard FOF linking length, although
\subfind\ could operate on arbitrary particle groups, or with slight
modifications on all of the particles in a simulation at once. The
use of FOF-groups as input data provides a convenient means to
organize the groups according to a simple two stage hierarchy
consisting of `background group' and `substructure'.

\begin{figure*}
\bc
\resizebox{7.9cm}{!}{\includegraphics{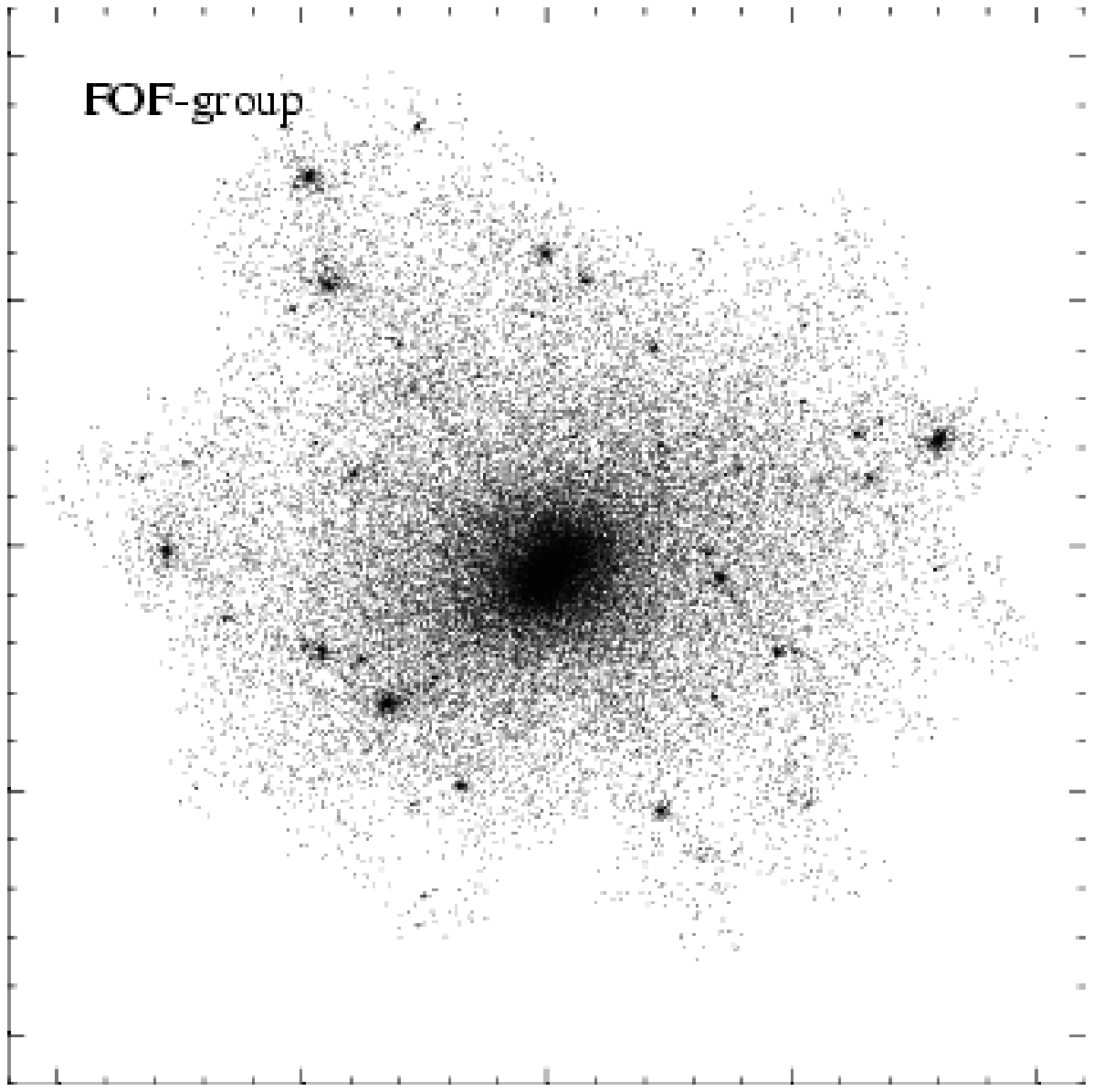}}%
\hspace{0.2cm}\resizebox{7.9cm}{!}{\includegraphics{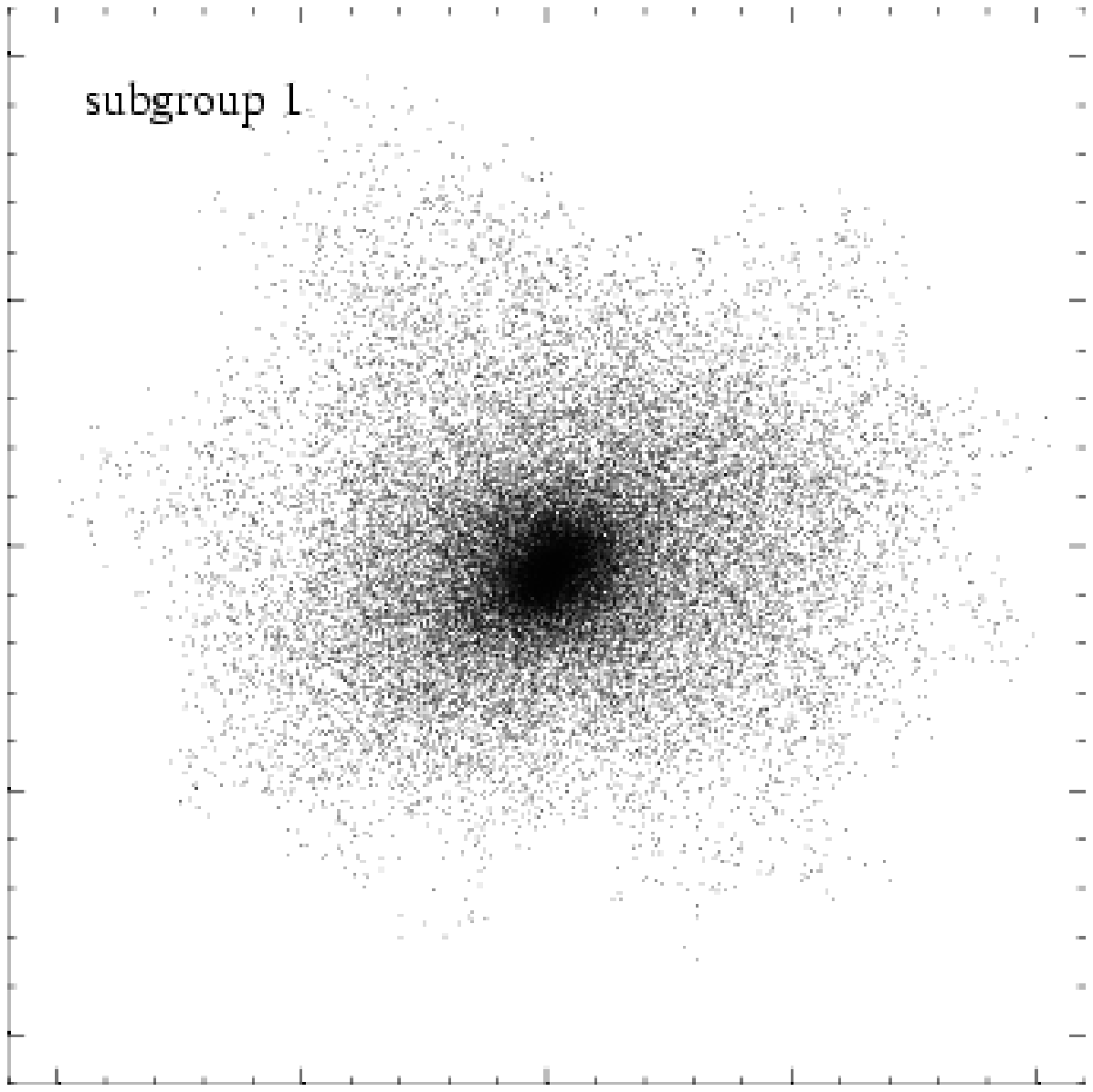}}\\%
\vspace{0.2cm}\resizebox{7.9cm}{!}{\includegraphics{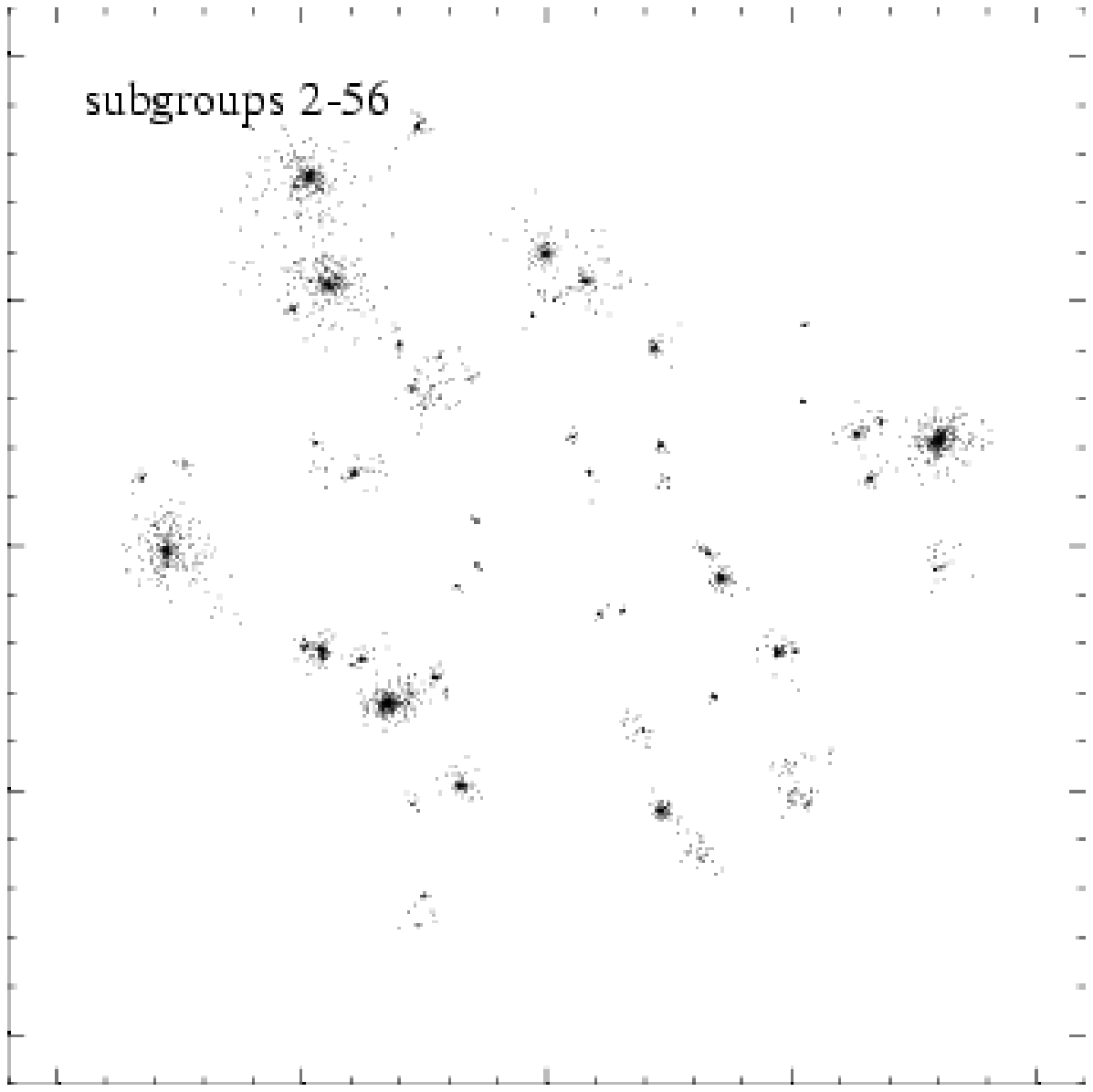}}%
\hspace{0.2cm}\resizebox{7.9cm}{!}{\includegraphics{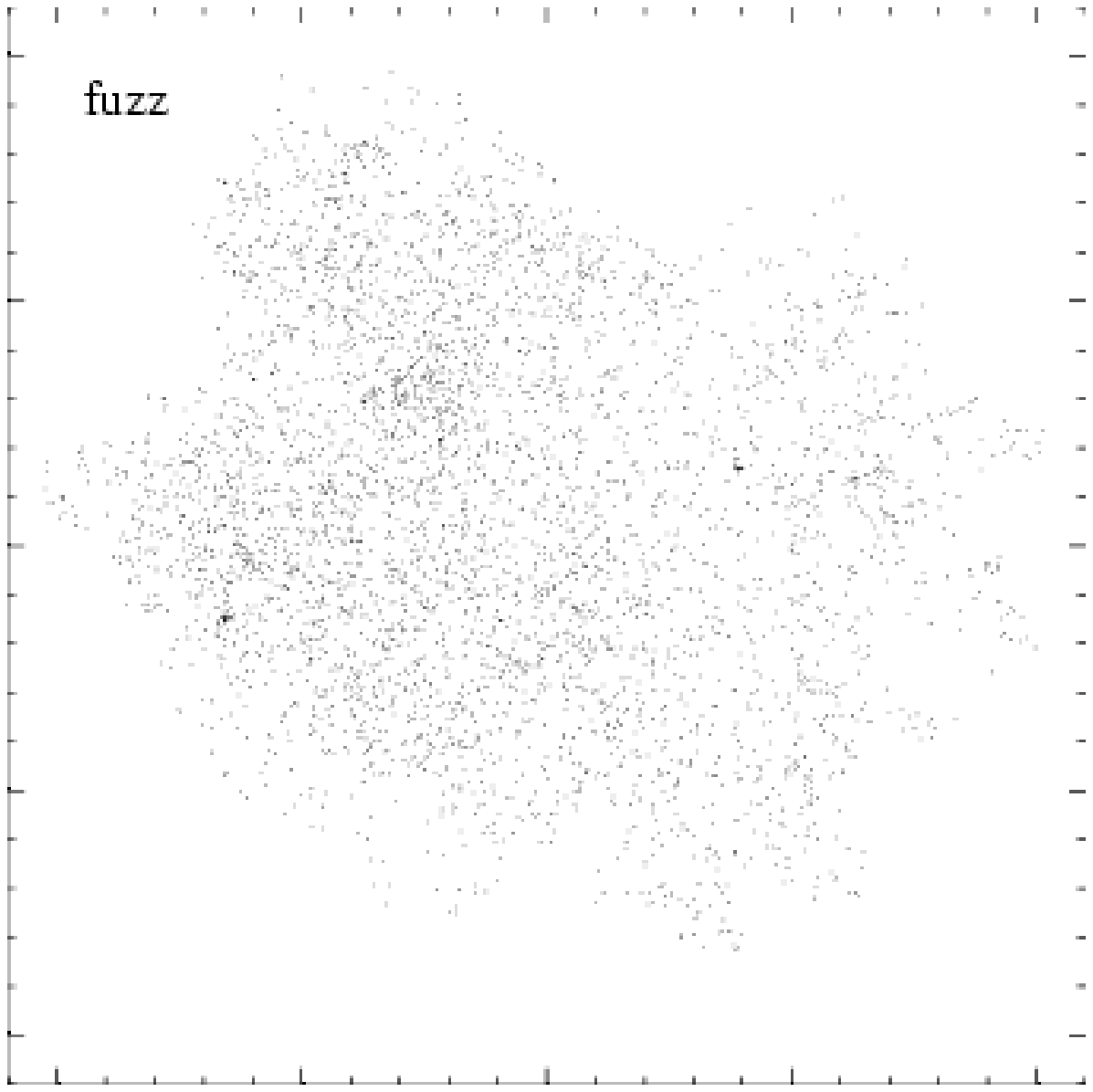}}\\%
\vspace{0.8cm}
\caption{Example for a subhalo identification with \subfind.  The top
left panel shows a small FOF-group (44800 particles), identified at
$z=0$ in the vicinity of the S2 cluster. \subfind\ identifies 56
subhaloes within this group, the largest one forms the background halo
and is shown on the top right, while the other 55 subhaloes are
plotted on a common panel on the lower left. In this example, the
total mass in all the ``true'' subhaloes 2-56 is about 8\% of the
group mass.  Particles not bound to any of the subhaloes form
``fuzz'', and are displayed on the lower right. These particles
primarily lie close to the outer edge of the group. Spatial
coordinates are given in $\lu$.
\label{figExSub2}}
\ec
\end{figure*}

Note that it is unlikely that we lose any substructure by restricting
the search to ordinary FOF-groups. FOF may accidentally link two
structures, a case \subfind\ will be able to deal with, but we rarely
expect FOF (with a linking length of $0.2$) to split a physical
structure into two parts.

In \subfind, we begin by computing a local estimate of the density at
the positions of all particles in the input group. This is done in the
usual SPH-fashion, i.e.~the local smoothing scale is set to the
distance of the $N_{\rm dens}$ nearest neighbour, and the density is
estimated by kernel interpolation over these neighbours.  The
particles may be viewed as tracers of the three-dimensional dark
matter density field. We consider any locally overdense region within
this field to be a {\em substructure candidate}.  More specifically,
we define such a region as being enclosed by an isodensity contour
that traverses a saddle point. How can one find these regions?
Imagine lowering a global density threshold slowly within the density
field.  For most of the time, isolated overdense regions will just
grow in size as the threshold is lowered, except for the moments when
two separate regions coalesce to form a common domain. Note that at
these instances, the contours of two separate regions join at a saddle
point. As a result the topology of the isodensity contour changes as
well.

Our algorithm tries to identify all locally overdense regions by
imitating such a lowering of a global density threshold.  To this end,
we sort the finite number of particles according to their density, and
we `rebuild' the particle distribution by adding them in the order of
decreasing density.  Whenever a new particle $i$ with density $\rho_i$
is considered, we find the $N_{\rm ngb}$ nearest neighbors within the
full particle set. Within this set ${\cal A}_i$ of $N_{\rm ngb}$
particles, we also determine the subset of particles with density
larger than $\rho_i$, and among them we select a set ${\cal B}_i$
containing the two closest particles.  Note that this set may contain
only one particle, or it may be empty.  We now consider three cases:
\begin{enumerate}
\item
The set ${\cal B}_i$ is empty, i.e.~among the $N_{\rm ngb}$ neighbors
is no particle that has a higher density than particle $i$.  In this
case, particle $i$ is considered to mark a local density maximum, and
it starts growing a new subgroup around it.
\item
If ${\cal B}_i$ contains a single particle, or two particles that are
attached to the same subgroup, the particle $i$ is also attached to
this subgroup.
\item
${\cal B}_i$ contains two particles that are currently attached to
different subgroups.  In this case, the particle $i$ is considered to
be a {\em saddle point}, and the two subgroups labeled by the
particles in ${\cal B}_i$ are registered as {\em subhalo
candidates}. Afterwards, the particle $i$ is added by joining the two
subgroups to form a single subgroup.  Note that all subhalo candidates
will be examined for self-boundedness later on in the algorithm.
\end{enumerate}

Working through this scheme results in a list of subhalo candidates,
which can be efficiently stored in a suitably arranged link-list
structure.  Note that a given particle can be member of several
different subhalo candidates, and that the algorithm is in principle
fully capable of detecting arbitrary levels of ``subhaloes within
subhaloes''.

Up to this point, the construction of subhalo candidates has been
based on the spatial distribution of particles alone. A more physical
definition of substructure is obtained by adding the requirement of
self-boundedness. We therefore subject each subhalo candidate to an
unbinding procedure to obtain the ``true'' substructure.  To this end,
we successively eliminate particles with positive total energy, until
only bound particles remain.  We perform the unbinding in physical
coordinates, where we define the subhalo's center as the position of
the most bound particle, and the velocity center as the mean velocity
of the particles in the group.  We then obtain physical velocities
with respect to this group center by adding the Hubble flow to the
peculiar velocities.  Finally, if more than a minimum number of
$N_{\rm ngb}$ particles survive the unbinding, we refer to these
particles as a {\em subhalo}.

An important issue remains of how one should deal with
complications arising from the assignment of particles to several
different subhalo candidates. This does not only occur if one deals
with genuine ``substructure within substructure'', but is actually
quite typical for the algorithm.  For example, imagine a large halo
containing several small subhaloes. Whenever one of the small haloes
`separates' from the main halo, two subhalo candidates are generated
according to case (iii) of the algorithm. Each time the larger of
these groups describes the bulk of the main halo, which will thus
appear several times as a subhalo candidate, although one would like
to consider it only once as an independent physical structure.

We approach this issue by considering only the smaller subhalo
candidate at each branch of the tree generated by the saddle points.
This is based on the notion that we want to examine substructure
within some larger object, and this `background' object is expected to
have larger mass than the actual substructure.  In addition, we
process the subhalo candidates in the inverse sequence as they have
been generated, i.e.~we work through the saddle points from low to
high density.  In this way, a smaller subhalo within a larger subhalo
will always be processed later than its parent subhalo.  As we
consider the subhalo candidates in this order, each particle carries a
label indicating the subhalo it was last detected to reside in.  If in
the process it is found to be contained also within a smaller subhalo,
this label will be overwritten by the new subgroup identifier.

In this way, the complexity of our analysis is reduced by assigning
each particle at most to one subhalo.  We are still able to detect a
hierarchy of small subhaloes within larger subhaloes, albeit at the
expense of reducing the latter by the particles contained at deeper
levels of the hierarchy.  However, this usually does not affect the
corresponding parent subhalo strongly, since the mass of any
substructure within a larger group is usually small compared to that
of the parent group. Nevertheless, it can happen that the extraction
of subhaloes unbinds some additional particles from the parent
subhalo.  For this reason, we check all the disjoint subhaloes at the
end of the process yet again for self-boundedness. Here, we also
assign all particles not yet bound to any subhalo to the ``background
halo'' of the group, which we define as the largest subhalo within the
original FOF input group, and we check whether they are at least bound
to this structure. If not, these particles represent `fuzz',
identified by FOF to belong to the group, but not (yet)
gravitationally bound to it.

In summary, \subfind\ decomposes a given particle group into a set of
disjoint self-bound subhaloes, each identified as a locally overdense
region within the density field of the original structure.  The
algorithm is spatially fully adaptive, and it has only two free
parameters, $N_{\rm dens}$ and $N_{\rm ngb}$. The latter of these
parameters sets the desired mass resolution of structure
identification, and we usually employ $N_{\rm ngb}=10$ for this
purpose.  The results are quite insensitive to the other parameter,
the number $N_{\rm dens}$ of SPH smoothing neighbours, which we
typically set to a value slightly larger than $N_{\rm ngb}$.

Finally, we note that any efficient practical implementation
of the algorithm requires the use of hierarchical tree-data
structures, and fast techniques to find nearest neighbours and
gravitational potentials.  For this purpose, we employ techniques borrowed
from our tree-SPH code \gadget .

In Figure \ref{figExSub2}, we show a typical example of substructure
identified using \subfind.  We selected a small group from
the periphery of the S2-cluster, at $z=0$, for this illustration.  By
eye, one can clearly spot substructure embedded in the FOF-group. The
algorithm \subfind\ finds 56 subhaloes in this case. The largest one
is the `background' halo, shown in the top right panel of
Fig.~\ref{figExSub2}. It represents the backbone of the
group, with all its small substructure removed. This substructure is
made up of 55 subhaloes, which are plotted in a common panel on the
lower left.  While this example shows an isolated and well relaxed
halo, we note that \subfind\ also performs well in cases where FOF
links structures across feeble particle bridges, or when haloes are in
the process of merging. In these cases, the algorithm reliably
decomposes the FOF group into its constituent parts.

\subsection{Subhaloes in the S1, S2, S3, and S4 clusters}

Substructure within dark matter haloes is in itself a highly
interesting subject that merits detailed investigation.  What are the
structural properties of subhaloes within haloes?  What is their
distribution of sizes and masses? What is their ultimate fate?
Answers to these questions are highly relevant for a number of diverse
topics such as galaxy formation, the stability of cold stellar disks
embedded in dark haloes, or the weak lensing of galaxies in
clusters \citep{Gei98,Gei99}.

\begin{figure*}
\bc
\resizebox{7.8cm}{!}{\includegraphics{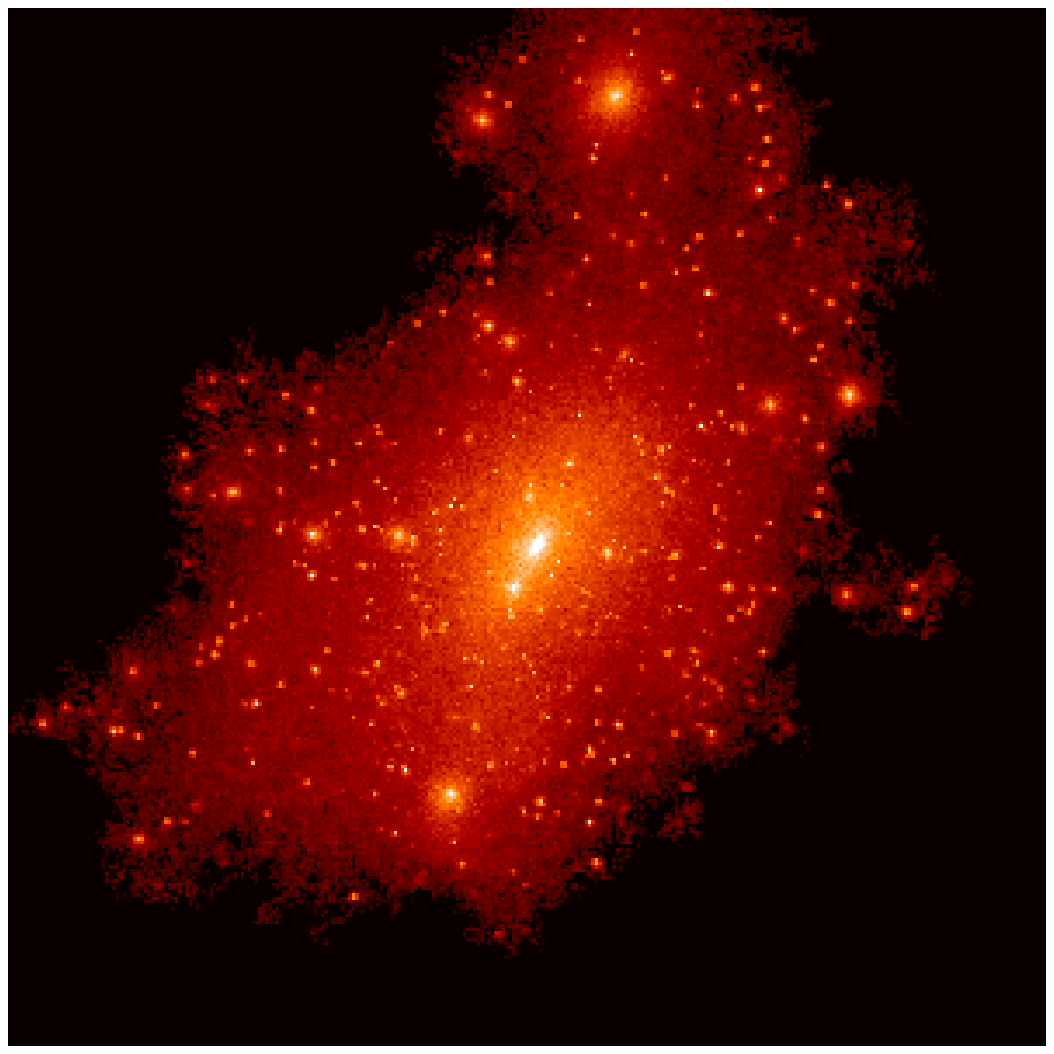}}%
\hspace{0.3cm}\resizebox{7.8cm}{!}{\includegraphics{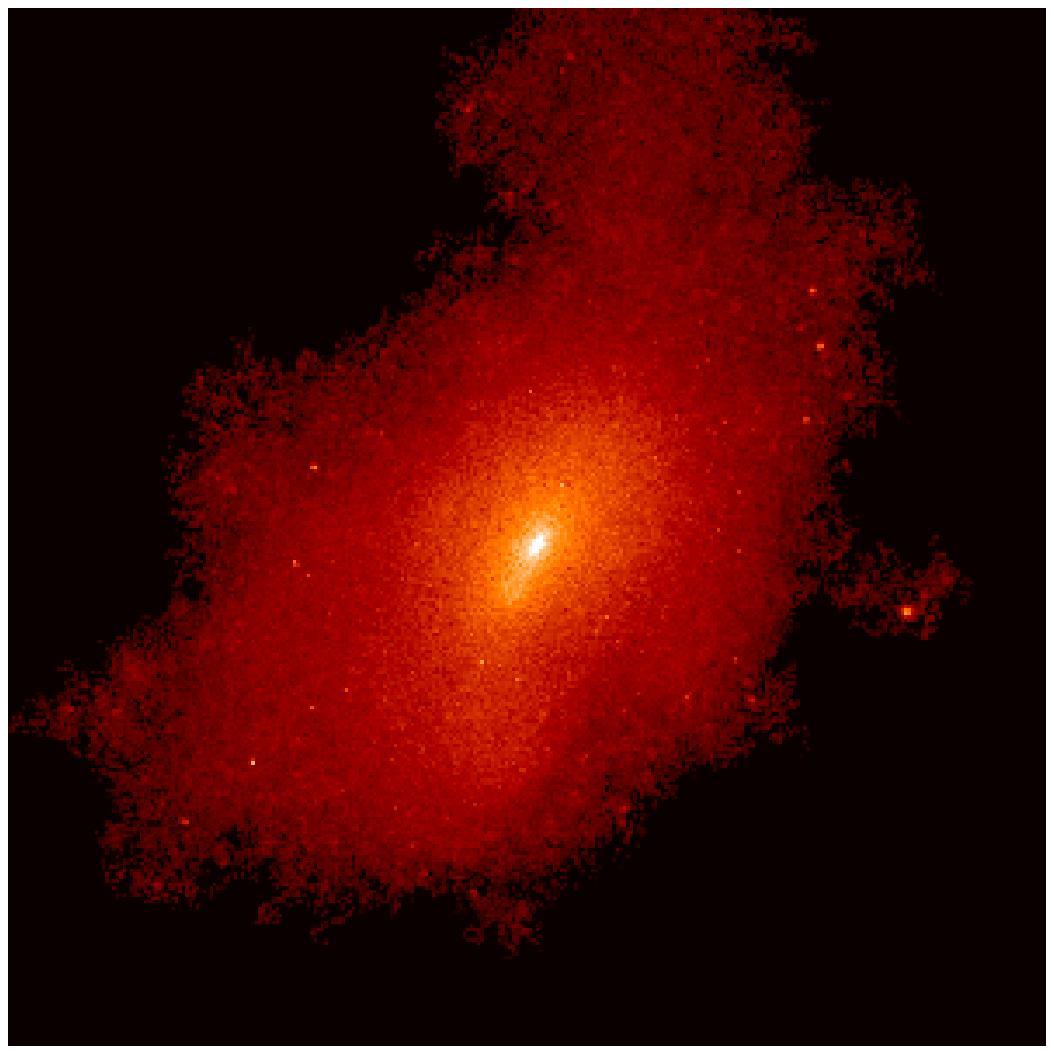}}\\%
\vspace{0.3cm}\resizebox{7.8cm}{!}{\includegraphics{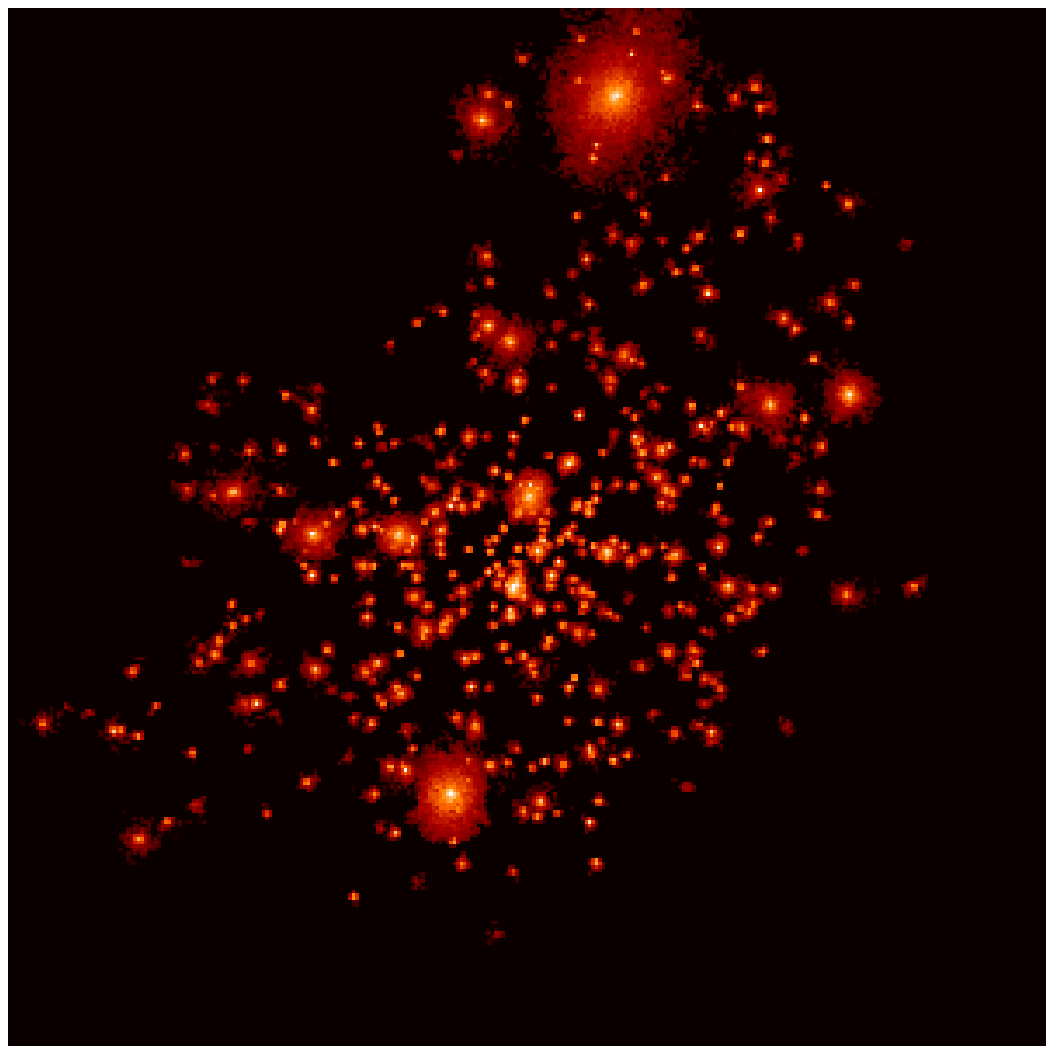}}%
\hspace{0.3cm}\resizebox{7.8cm}{!}{\includegraphics{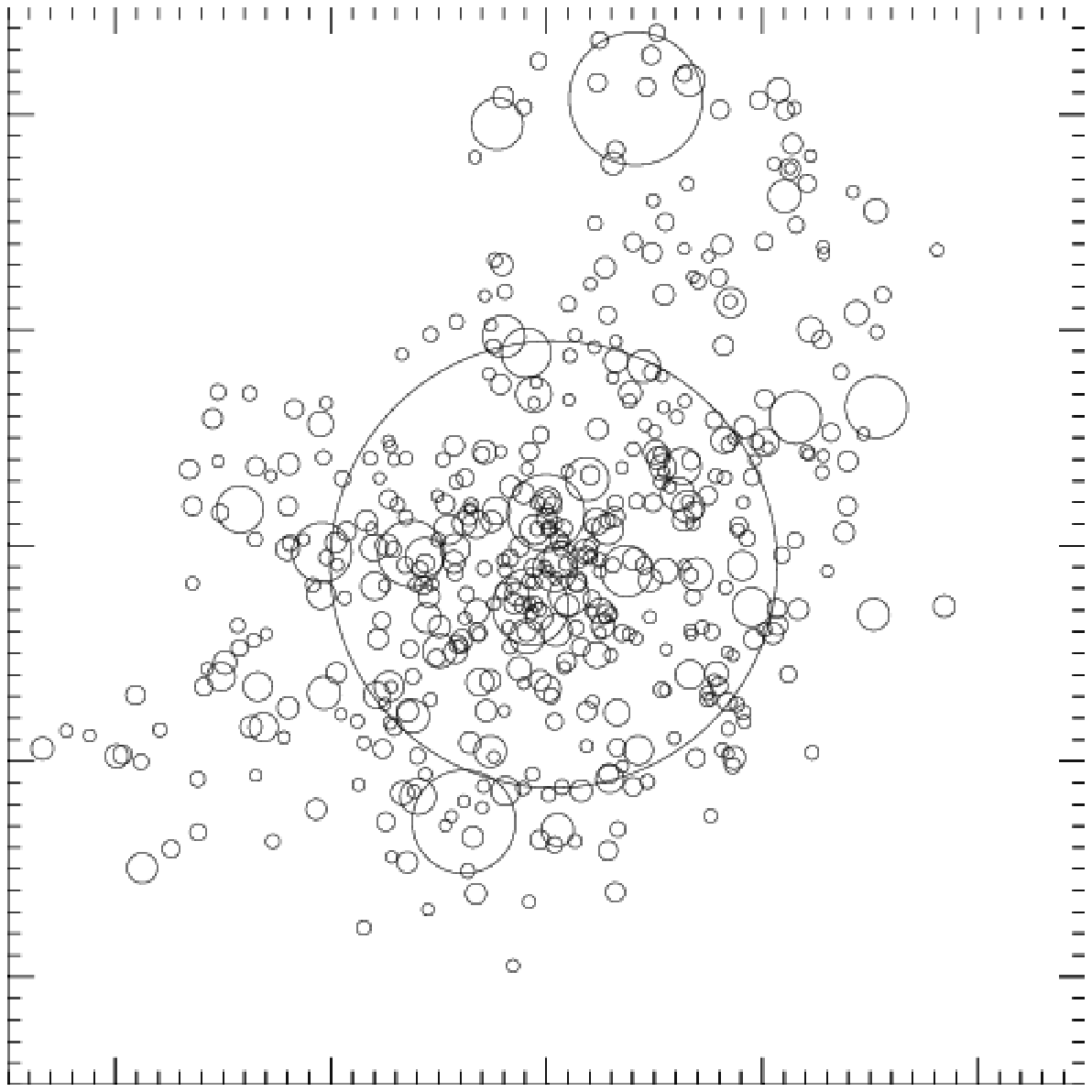}}\\%
\vspace{1cm}
\caption{Substructure in the S2 cluster at $z=0$. The top left panel
shows a color-coded projection of the FOF-group that contains the
cluster.  To highlight the substructure, particles have been given a
weight proportional to the local dark matter density.  In the top
right panel we show the largest subhalo identified by \subfind,
i.e.~the {\em background halo}. The lower right shows the 495 other
subhaloes identified in the object. Finally, on the lower right, we
plot circles at the positions of each identified subhalo, with radius
proportional to the third root of the particle number in the
subhalo. Note that we actually found subhaloes within subhaloes in
this example.
\label{figSubS2}}
\ec
\end{figure*}

\citet{Tor98}, \citet{Gh98} and \citet{Kly99} have addressed some of
these questions, and it will be interesting to supplement their work
with results obtained from our new group-finding technique.  However,
such a study is beyond the scope of the present paper, where we focus
on semi-analytic models for the galaxy population of the cluster.  We
therefore defer a detailed statistical analysis of substructure to
future work, and just report the most basic properties of the
subhaloes identified in the clusters at redshift $z=0$.

\begin{figure}
\bc
\resizebox{8.0cm}{!}{\includegraphics{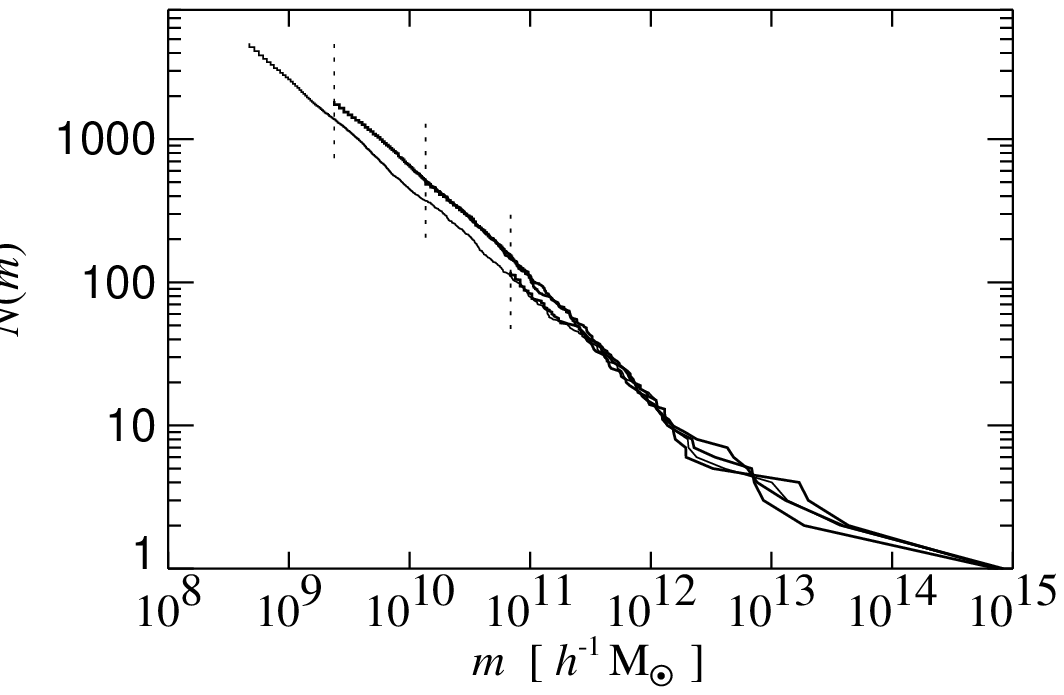}}\\%
\resizebox{8.0cm}{!}{\includegraphics{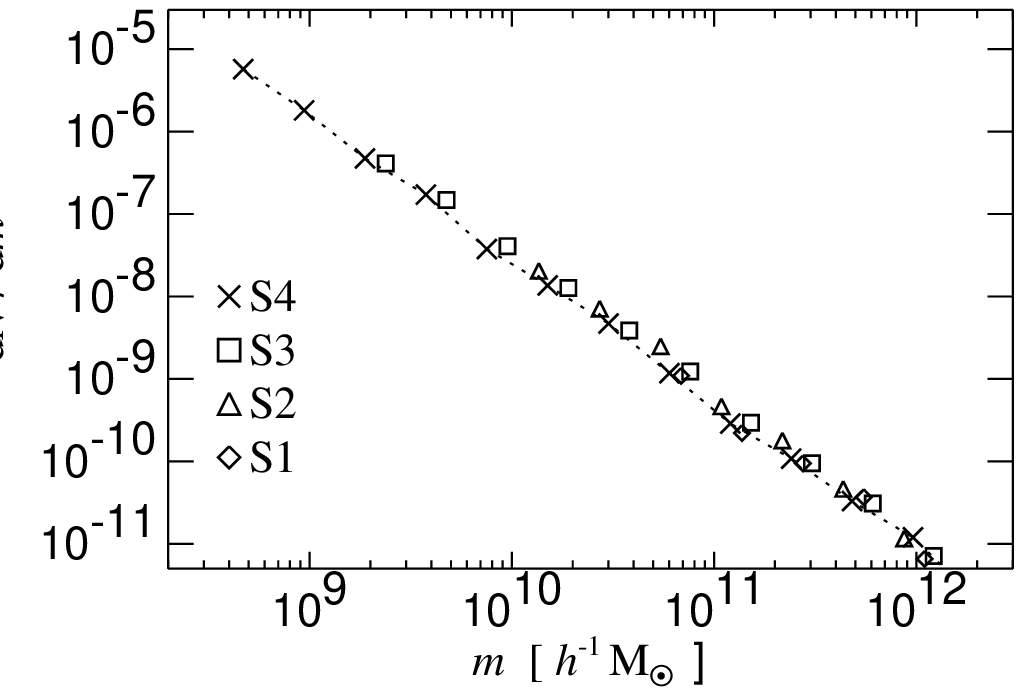}}%
\caption{Subhalo mass functions in the four clusters S1, S2, S3 and S4
at redshift $z=0$.  In the top panel, we plot the cumulative number
$N(m)$ of subhaloes with masses larger than $m$. The short vertical
lines mark the ends of the graphs for the simulations S1 (lowest
resolution), S2, and S3 (second highest resolution).  The agreement
between the four simulations is quite good.  This good agreement
is also present in the differential mass function $\dd N/\dd m$, which
we show in the lower panel.
\label{figmassfunc}}
\ec
\end{figure}

Especially interesting is a comparison of the mass spectra of
subhaloes detected in the four cluster simulations S1, S2, S3 and S4.
In this sequence of simulations, the mass and force resolution
increases substantially, which clearly leads to a larger number of
resolved subhaloes: 118 subhaloes are detect in the
final cluster of S1, 496 in S2, 1848 in S3, and 4667 in S4.

The increase in numerical resolution thus unveils an enormous richness
of structure in the cluster. As an example, we show a plot of the
substructure in the S2-cluster in Figure~\ref{figSubS2}.  Note that
almost all of the additional substructure that becomes visible with
still higher numerical resolution is in objects of smaller mass than
the ones resolved in Figure~\ref{figSubS2}.  This is seen in a
comparison of the cumulative and differential mass spectra, as shown
in Figure~\ref{figmassfunc}.  S1, S2 and S3 are actually well capable
of resolving all the objects above their respective resolution limits.
Even close to their resolution limits they predict the right abundance
of subhaloes of a given mass.

\subsection{Merging trees using substructure}

We now describe our methods to construct merging trees that take the
identified subhaloes into account.  Note that \subfind\ classifies all
particles of a given FOF-group either as lying in a bound subhalo, or as
unbound. Any ordinary FOF-group lacking substructure will also
appear in the list of subhaloes identified by \subfind, albeit only
with its bound subset of particles.  Since the requirement of
self-boundedness is a reasonable physical condition for the definition
of dark haloes, we therefore consider the list of haloes generated by
\subfind\ as the `source list' for our further analysis.  Notice that
we use the FOF-groups as convenient `containers' to establish a
simple two stage hierarchy of haloes. We define the largest subhalo in
a given FOF-group as the {\em main halo} hosting the central galaxy.
All other subhaloes found within the group will be considered to be
substructure of this main halo.

In the following discussion, the term {\em subhalo} will refer to any
self-bound structure identified by \subfind, even if it is just the
self-bound part of an ordinary FOF-group that has no real
substructure.  In addition, we adopt the following definitions:

A subhalo $S_{\rm B}$ at redshift $z_{\rm B}$ is a {\em progenitor of
a subhalo} $S_{\rm A}$ at redshift $z_{\rm A}<z_{\rm B}$, if more than
half of the $N_{\rm link}$ most-bound particles of $S_{\rm B}$ end up
in $S_{\rm A}$.  This definition concentrates on the most-bound core
of each structure, and we have found it to be very robust in tracing
subhaloes between different output times. One can choose $N_{\rm
link}$ as some fraction of the number of particles of subhalo $S_{\rm
B}$, say all or half of them. However, such a condition may fail if
$S_{\rm B}$ is deprived of its outer halo between two outputs, as it
may occasionally happen when a structure of relatively large mass
falls into a cluster. We have found that setting $N_{\rm link}=10$,
equal to our lower particle limit for group identification, can
satisfactorily treat even these cases.  Note that our notion of
`most-bound' refers to the most negative binding energy.  \subfind\
automatically stores each subhalo in the order of increasing binding
energy to facilitate this kind of linking, i.e.~the subhaloes are
effectively stored from inside to outside.

We call a subhalo $S_{\rm B}$ at redshift $z_{\rm B}$ a {\em
progenitor of a FOF-group} $G_{\rm A}$ at redshift $z_{\rm A}<z_{\rm
B}$, if more than half of the particles of $S_{\rm B}$ are present in
$G_{\rm A}$. We also call a FOF-group $G_{\rm B}$ a {\em progenitor of
a subhalo} $S_{\rm A}$, if more than half of the particles of $S_{\rm
A}$ are contained in $G_{\rm B}$. Note that in this latter condition
we deliberately used the particles of $S_{\rm A}$ to define
`membership' in one of the FOF-group at higher redshift.

We expect that once a self-bound structure has formed, most of its
mass will remain in bound structures in the future. However,
occasionally it may happen that a group that was just barely above our
specified minimum particle number at one output time falls below this
limit at the next output time, for example because the group is
evaporated by interactions with other material, or because of noise in
the identification of groups with size close to our identification
threshold.  We call such groups {\em volatile}, and drop them from our
analysis. More precisely: All FOF-groups without any bound subhalo are
considered to be volatile and disregarded. In addition, if a subhalo
is not progenitor to any other subhalo, and not progenitor to any
non-volatile FOF-group, it is considered to be volatile too, and
dropped from further analysis.  Note that basically all subhaloes
eliminated in this way have particle numbers very close to the
detection threshold.

After the elimination of volatile subhaloes, we link subhaloes between
pairs of successive simulation outputs. By construction, every subhalo
may have several progenitors, but itself can only be progenitor for at
most one other subhalo.  In fact, due to the elimination of volatile
subhaloes, a subhalo $S_{\rm B}$ will {\em always} be progenitor to
another subhalo, or at least to a FOF-group $G_{\rm A}$.  If only the
latter is the case, we treat the subhalo $S_{\rm B}$ as a progenitor
of the main halo within the FOF-group.

There remains then the important case that a subhalo $S_{\rm A}$ has
no progenitor. If it also has no progenitor FOF-group, we call this
subhalo a {\em new galaxy}, which is considered to have newly
formed between the two output times. In the galaxy formation scheme,
new galactic `seeds' will be inserted into these subhaloes.

If however the subhalo $S_{\rm A}$ has no progenitor subhalo, but a
progenitor FOF-group $G_{\rm B}$, it is likely that either the subhalo
represents a chance fluctuation, or that it was overlooked in the
identification process at the time $z_{\rm B}$. In the latter case the
corresponding structure will have been merged with the main halo, so
we drop these subhaloes. Their absolute number and the total mass
contained in them is always very small.

In summary, the above identification scheme allows a detailed tracing
of the dark matter merging history tree, from the past to the present.
In particular, the scheme is able to deal with haloes that pop into
existence, that grow in size by accreting additional background
particles, that merge with other haloes of comparable size and lose
their identity, or that fall into a larger halo without being
destroyed completely. The latter case is particularly interesting,
since we expect that a subhalo can survive for some time within the
larger structure. However, its mass can be reduced by tidal stripping,
an effect that can eventually completely dissolve the structure.

\subsection{Inclusion of subhaloes in semi-analytic models}

One of the questions we want to address in this study is how the
inclusion of subhaloes changes the results of semi-analytic models. In
order to highlight such changes we will modify the `standard'
scheme, which is based on the work of KCDW, in a minimum fashion when
subhaloes are included.

\begin{figure*}
\bc
\resizebox{8cm}{!}{\includegraphics{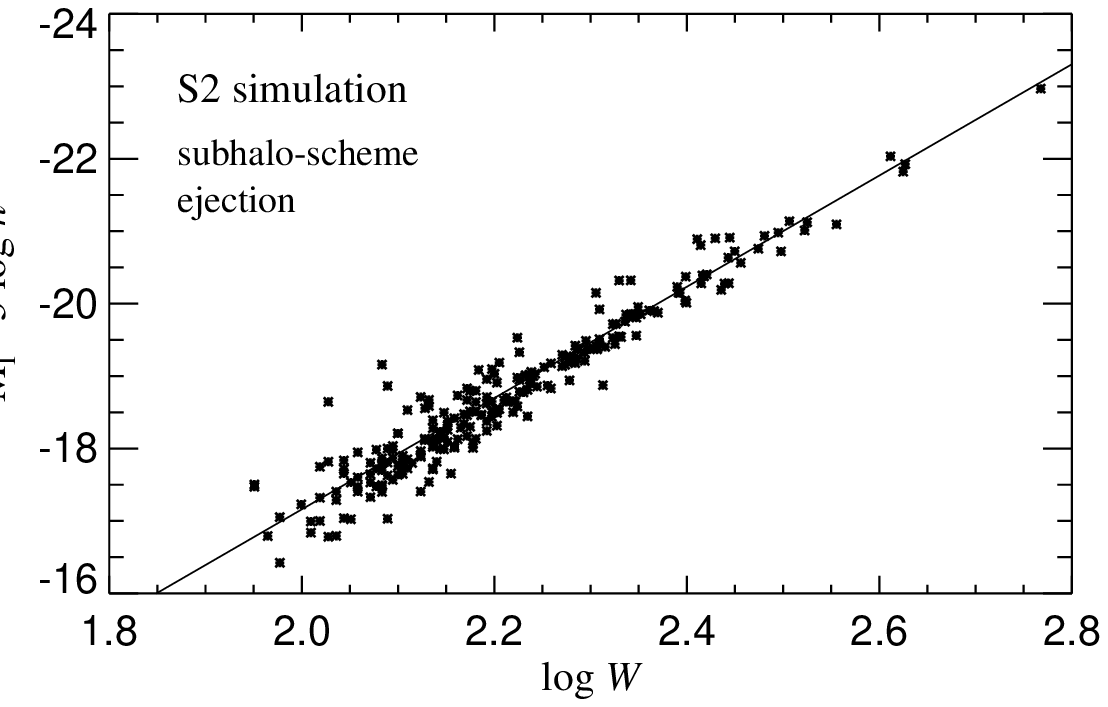}}%
\resizebox{8cm}{!}{\includegraphics{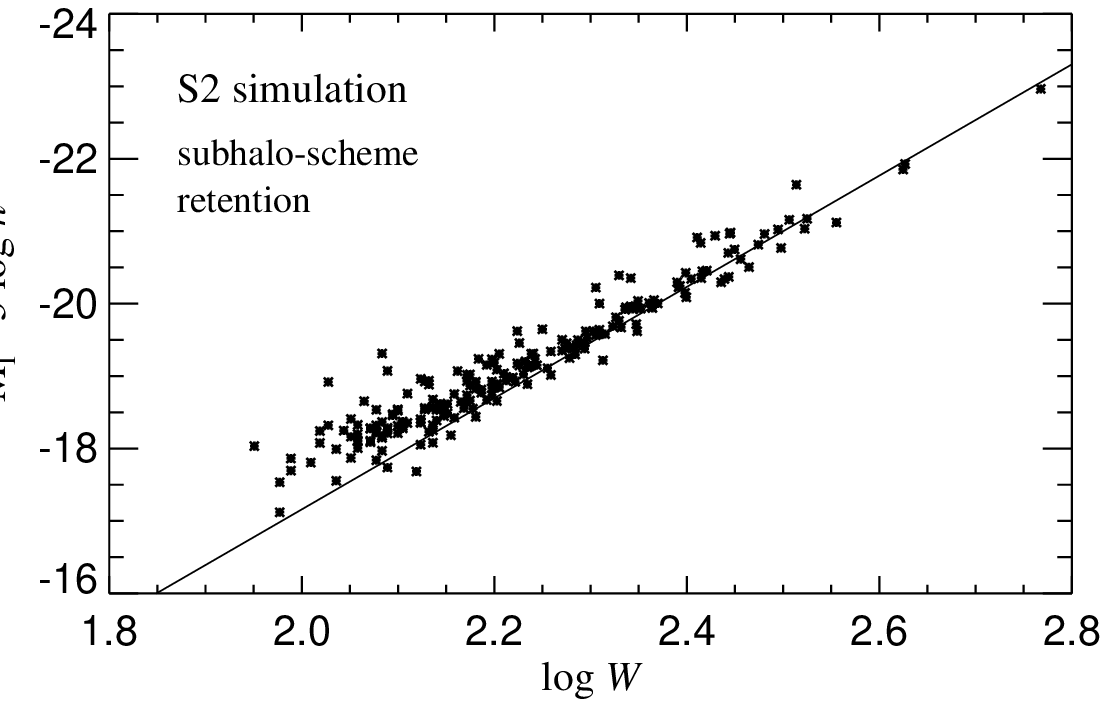}}\\%
\resizebox{8cm}{!}{\includegraphics{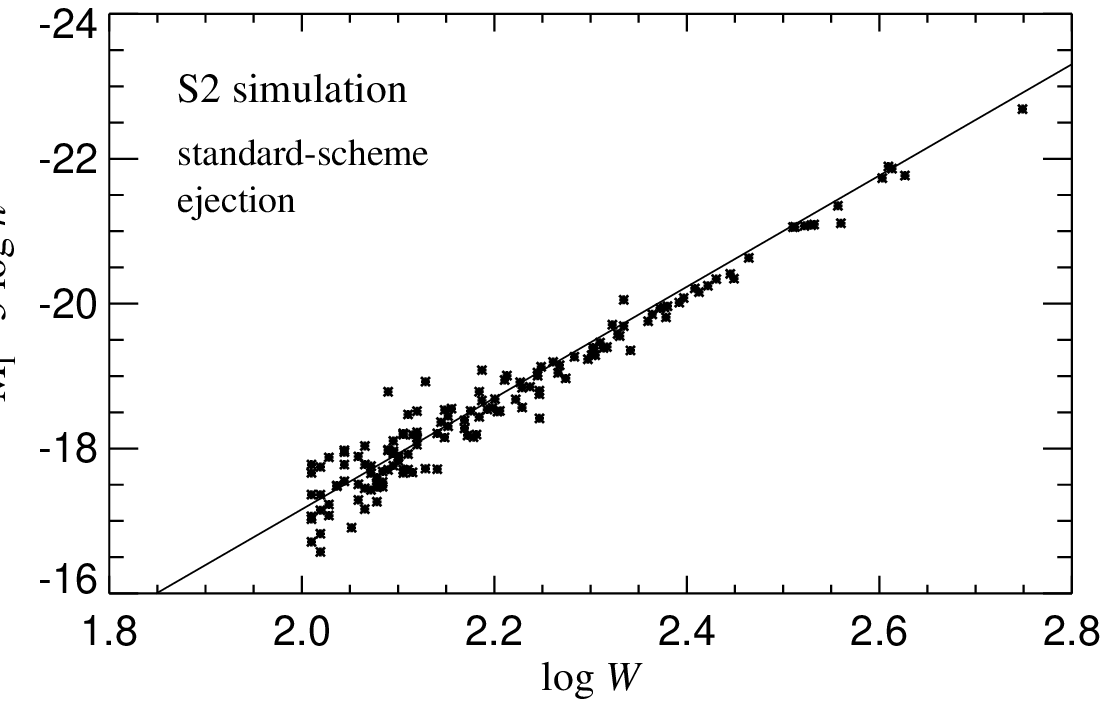}}%
\resizebox{8cm}{!}{\includegraphics{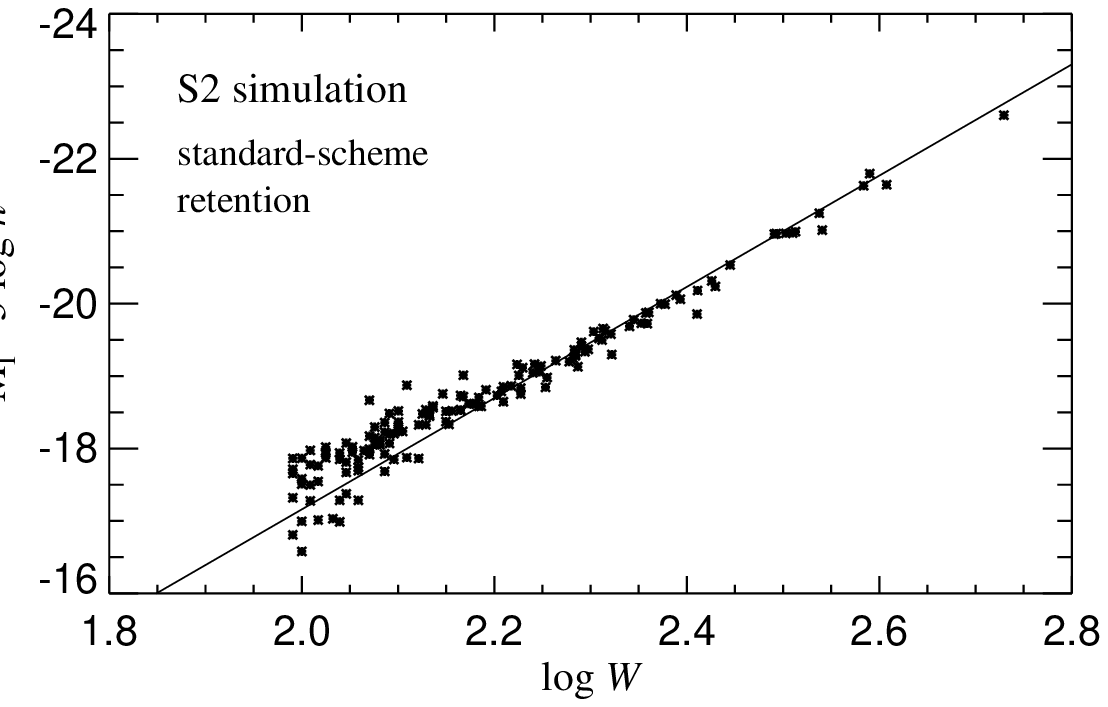}}\\%
\caption{The {\em I}-band Tully-Fisher relation for Sb/Sc galaxies in
the S2 simulation. The two panels on top have been obtained with the
subhalo-scheme using ejection and retention feedback, respectively,
while the two panels on the bottom show the equivalent plots using the
standard-scheme of KCDW.  The galaxies have been selected as central
galaxies of uncontaminated haloes in the periphery of the cluster.
The solid line represents the recent measurement by \citet{Gio97}.
\label{figTF}}
\ec
\end{figure*}

Note that a large variety of semi-analytic ``recipes'' are
conceivable, and that some of the results can depend sensitively on
the specific set of assumptions made, as has been recently highlighted
by KCDW.  Semi-analytic models thus cannot eliminate uncertainties
arising from poorly constrained physics. However, they are ideal tools
to explore the relative importance of various model ingredients, and
hence to constrain their relevance for observed trends in
observational properties.

We now describe the changes in our semi-analytic methodology when
subhaloes are included. From here on, we will refer to the formalism
of KCDW, which only works with FOF groups, as the `standard'-scheme,
and to the analysis that includes substructure as the
`subhalo'-scheme. We note that the word `standard' is not meant to
imply that the corresponding procedure has (yet) the status of a
well-established, widely used method in the field -- it is just used
to refer to the methodology recently developed by KCDW.

Our most important change concerns the definition of the galaxy
population at each output time. We define the largest subhalo in a
FOF-group to host the central galaxy of the group, and this galaxy's
position is given by the most-bound particle in that subhalo. All gas
that cools within a FOF-group is funneled exclusively to the central
galaxy. This definition of `central galaxy' thus corresponds to the
one adopted in the standard analysis.

For the population of galaxies orbiting within a halo, however,
we distinguish between {\em halo-galaxies} and satellite
galaxies.  Here we have coined yet another term; `halo-galaxies' are
attached to the most bound particle of the remaining subhaloes in the
FOF-group. These halo-galaxies were proper central galaxies in the
past, until their halo fell into a larger structure. The
core of this halo is, however, still intact, and thus allows an accurate
determination of the position of the halo-galaxy within the
group. These halo-galaxies may still be viewed as `central galaxies'
of their respective subhaloes, but they are no longer fed by a cooling flow
since their subhalo is not the largest within the FOF group.

Finally, when two (or more) subhaloes merge, the halo-galaxy of the
smaller subhalo becomes a satellite of the remnant subhalo. These
satellites are treated as in the standard analysis. Their position is
tagged by the most-bound particle identified at the last time they
were still a halo-galaxy, and they are assumed to merge on a dynamical
friction timescale with the halo-galaxy of the new subhalo they now
reside in. We need to introduce such satellites in the subhalo-scheme
in order to account for actual mergers between subhaloes, and also
because of the finite numerical resolution of our simulations, which
limits our ability to track the orbits of subhaloes once their mass
has fallen below our resolution limit.  It also allows us to make
direct contact with the standard-scheme in the limit of poor
resolution.  Note that the class of halo-galaxies is absent in the
standard analysis, where all of these galaxies are treated as
satellites.

\begin{table}  
\bc
\caption{Numerical parameters adopted for our semi-analytic models.
$\alpha$ is the star formation efficiency, $\epsilon$ the efficiency
of feedback by supernovae, and $f_{\rm b}$ the baryon fraction. 
\label{modparam}
}
\vspace{0.5cm}
\begin{tabular}{cccc}
Model   & $\alpha$ & $\epsilon$ & $f_{\rm b}$\\ 
\hline
ejection   &   0.05  & 0.10 & 0.15\\ 
retention  &   0.05  & 0.15 & 0.15\\ 
\hline
\end{tabular}
\ec
\end{table}

In the subhalo scheme, we define the virial mass $M_{\rm vir}$ of a
subhalo simply as the total mass of its particles.  For the background
subhalo, we then define virial radius and virial velocity by assuming
that the halo has an overdensity 200 with respect to the critical
density.  For other subhaloes we keep the virial velocity and the
halo's dynamical time at the values they had just before infall.

\section{Results}

\subsection{Tully-Fisher relation}

We use the velocity-based {\em I}-band Tully-Fisher relation \be M_{I}
- 5\log h = -21.00 - 7.68\, (\log W - 2.5)
\label{TFrel}
\ee of \citet{Gio97}, and the requirement of a gas mass of $\sim
8\times 10^{9}\msunh$ in `Milky-Way' haloes, to normalize our
models.  We consider two variants for the implementation of feedback,
the `ejection' model, where gas is blown out of small haloes, and the
`retention' model, where reheated gas is always kept within the halo.

In Figure \ref{figTF}, we show the best-fit Tully-Fisher relations
obtained for these two models, applied to the S2-cluster using the
`subhalo' and the `standard'-schemes.  In the plots, we only
considered central galaxies of haloes that are peripheral to the
cluster, but that are not contaminated by heavier boundary particles.
We also applied a morphological cut, $1.2 \le M_{\rm bulge}-M_{\rm
total}\le2.5$, approximately selecting Sb/Sc galaxies.  In
Table~\ref{modparam}, we list the model parameters thus obtained.  In
the following, we will use the same set of parameter values also for
our other cluster simulations, and for the `standard' semi-analytic
scheme.

The Tully-Fisher relations we obtain exhibit remarkably small
scatter. This is partly a result of the tight coupling we assumed
between the sizes and circular velocities of the disks of spiral
galaxies and the masses of their dark haloes.  Note that additional
scatter can be expected from the distribution of spin parameters of
dark haloes, which gives rise to variations of the disk sizes
associated with a halo of a given mass \citep{Mo98}.

The ejection model fits the slope of the observed TF-relation
relatively easily. However, the retention model is less effective in
suppressing star formation in low mass haloes.  For the same value of
$\epsilon$, the retention scheme therefore produces a shallower
Tully-Fisher relation than the ejection model.  As a result, a larger
value of $\epsilon$ is needed to bring the retention model in
agreement with the observed steepness of the TF-relation. This strong
feedback reduces the overall brightness somewhat, an effect that could
be easily compensated for by slightly larger values for $\alpha$ or
$f_{\rm b}$.  Note that a feedback efficiency of $\epsilon=0.1$ means
that feedback will be quite efficient in haloes of virial velocity
below $V_{\rm SN}= 183 \kms (\epsilon/0.1)^{1/2}$.  In such haloes,
one will have $\Delta M_{\rm reheat}\ge\Delta M_{\star}$, i.e.~the
mass of reheated gas exceeds that of newly formed stars.

\begin{figure*}
\bc
\resizebox{8cm}{!}{\includegraphics{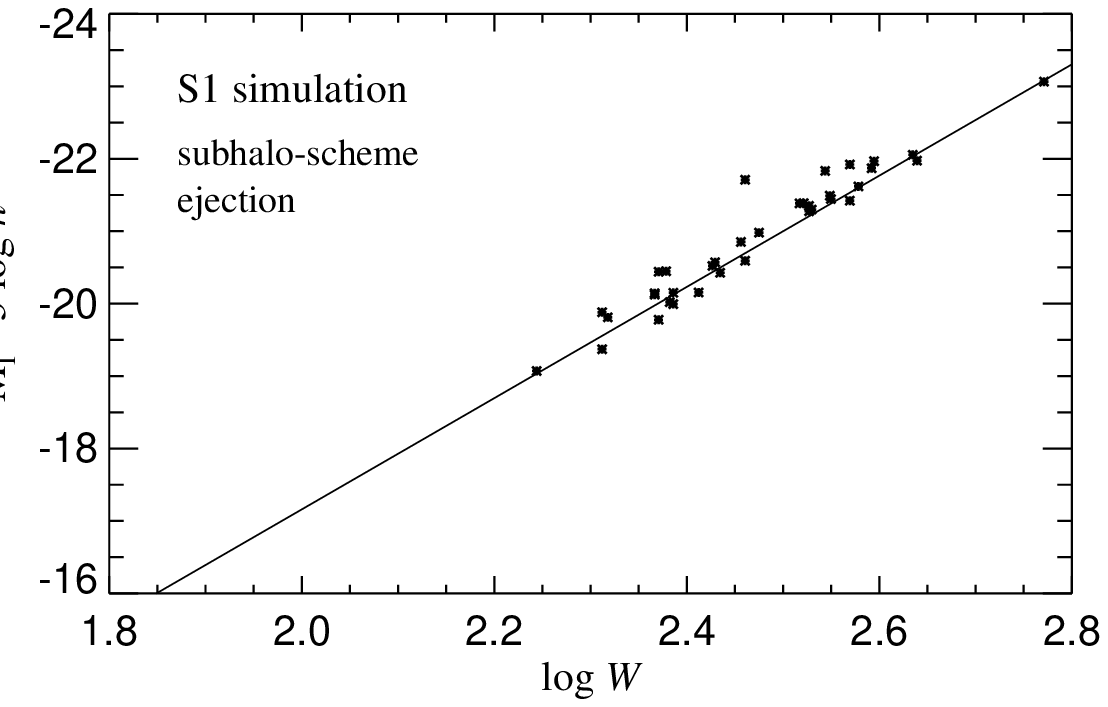}}%
\resizebox{8cm}{!}{\includegraphics{tf_S2_sub_ejt.eps}}\\%
\resizebox{8cm}{!}{\includegraphics{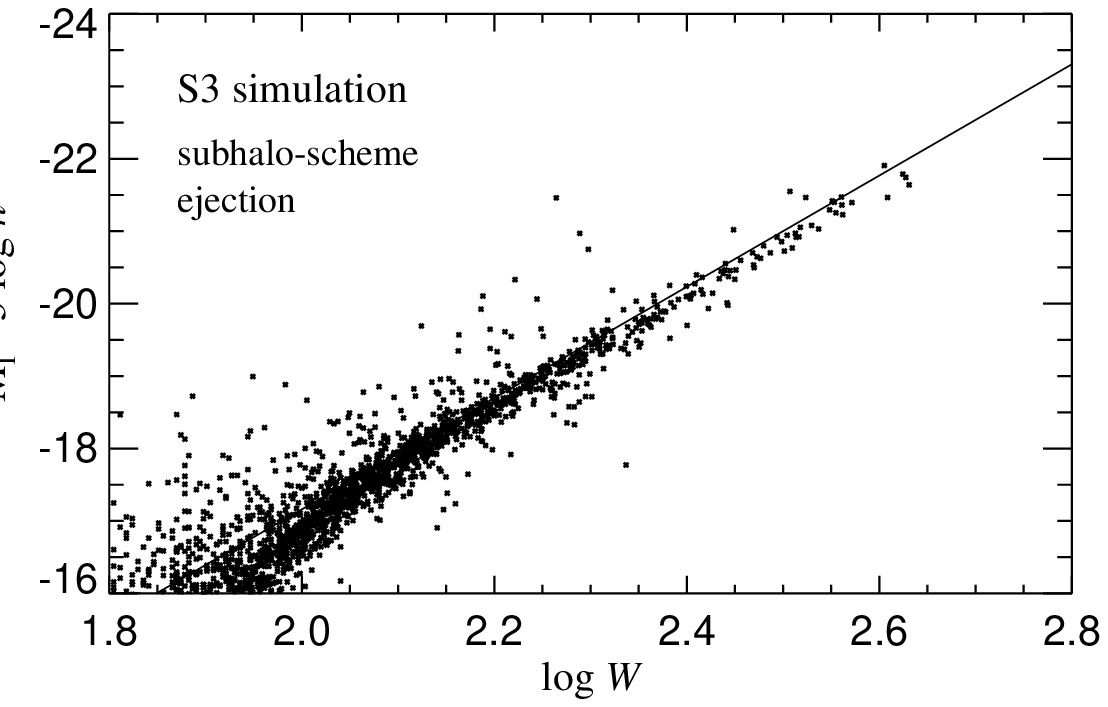}}%
\resizebox{8cm}{!}{\includegraphics{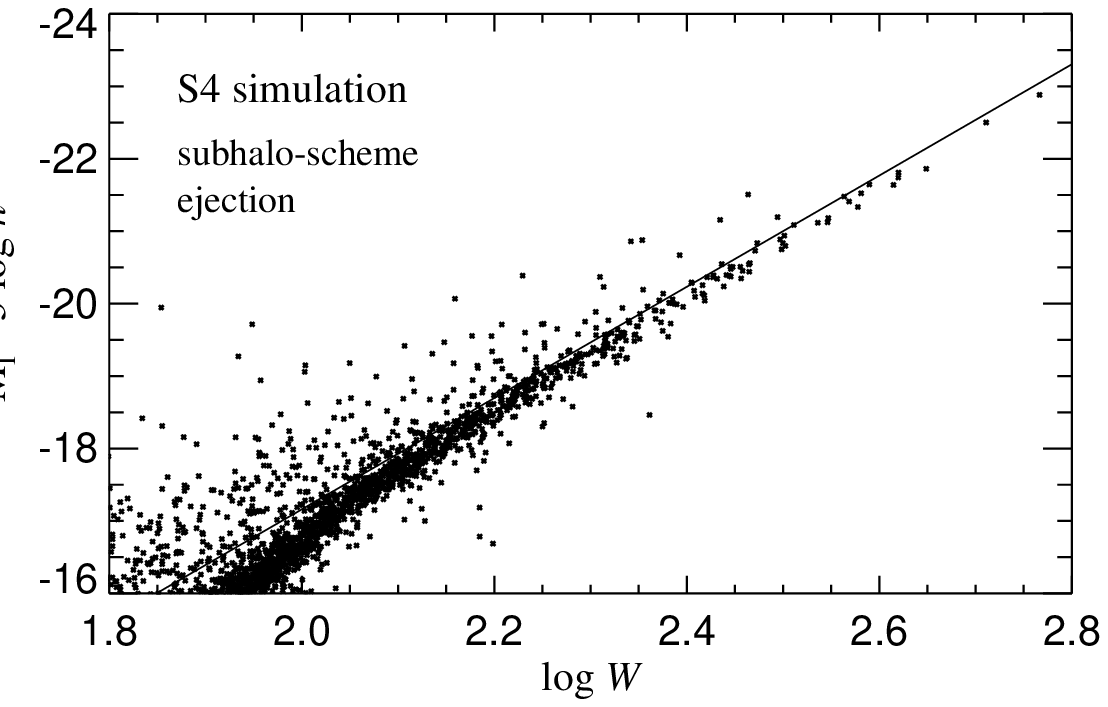}}\\%
\caption{The four panels compare the {\em I}-band Tully-Fisher
relations obtained for the S1, S2, S3, and S4 simulations using the
subhalo-scheme and ejection feedback. The models with retention
feedback, and the models using the `standard' scheme, are able to
match the observed Tully-Fisher relation about equally well.
\label{figTF2}}
\ec
\end{figure*}

It is also interesting to compare the Tully-Fisher relations obtained
for the different cluster simulations (see Figure~\ref{figTF2}).  For
the sake of brevity, we restrict the comparison to the subhalo scheme
with ejection feedback.  In general, there is good agreement between
the two variants of semi-analytic modeling, and between the four
different simulations, despite their large differences in numerical
resolution. For the same choice of free parameters, the slopes and
zero-points of the Tully-Fisher relations agree quite well.  There may
be a weak trend towards fainter zero-points in the sequence S1 to
S4. Note that the scatter in the Tully-Fisher relation increases at
low velocity widths because feedback has a stronger effect on galaxies
with low $V_{\rm c}$. A similar trend in the observed scatter has been
found by \citet{Matt98}.

\subsection{Cluster luminosity function}

\begin{figure*}
\bc
\resizebox{8cm}{!}{\includegraphics{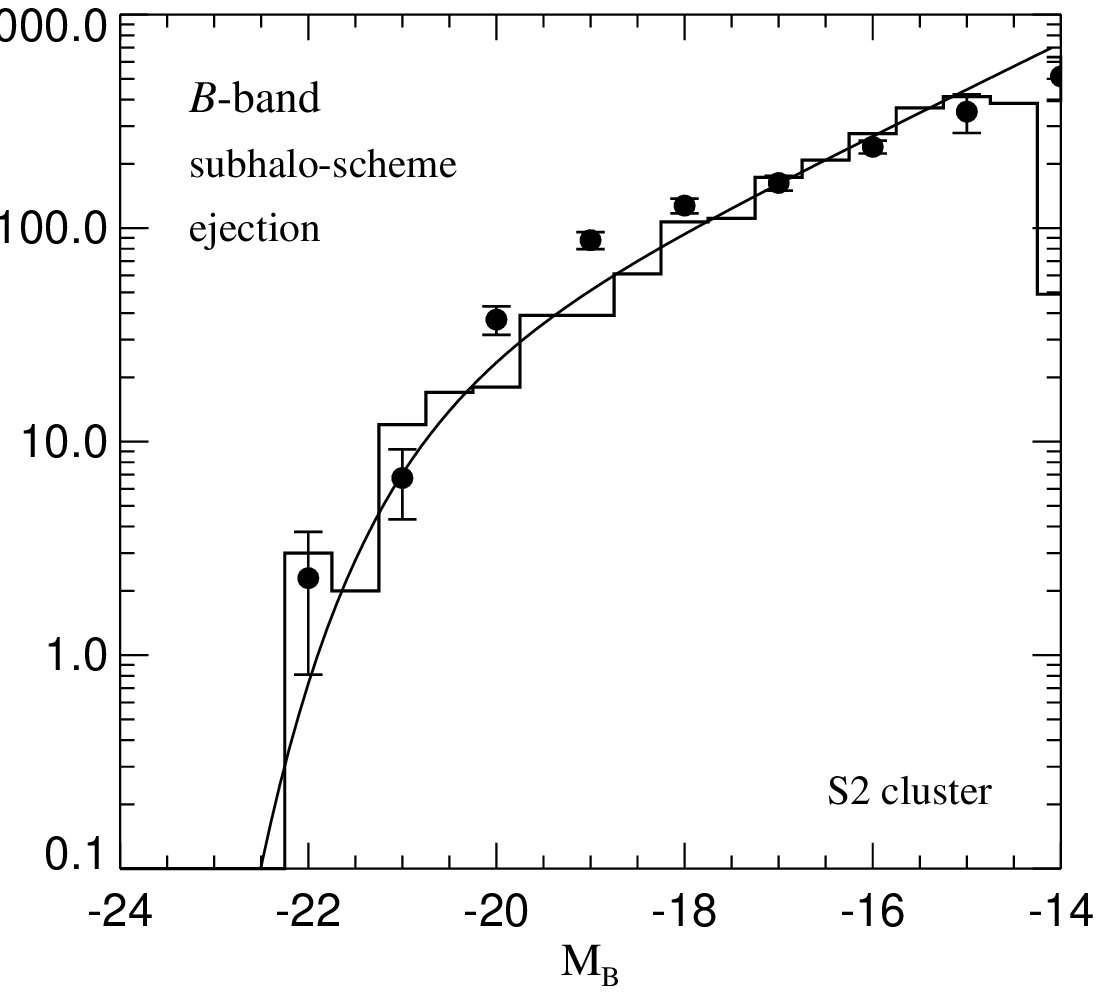}}%
\resizebox{8cm}{!}{\includegraphics{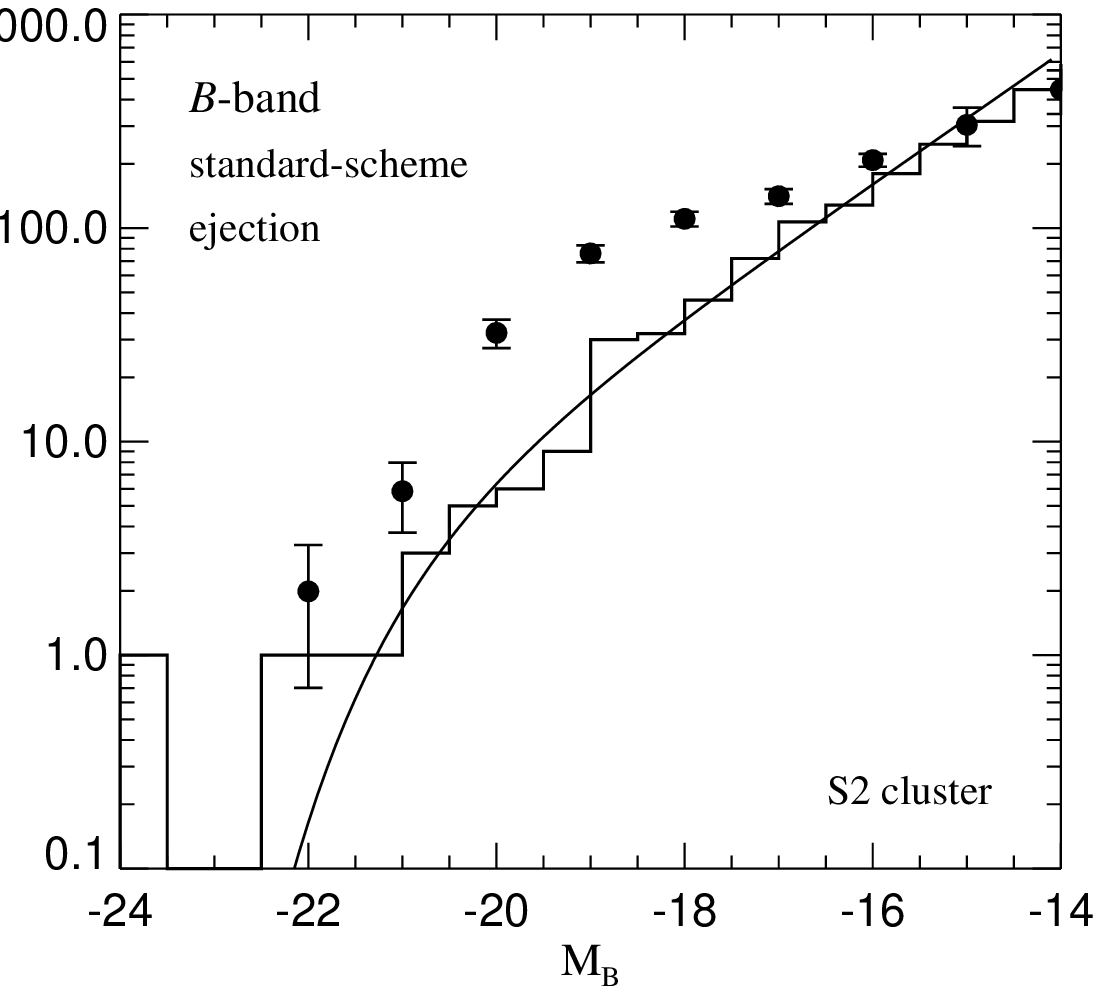}}\\%
\caption{The histograms show the {\em B}-band cluster luminosity
function obtained for the S2 simulation using the subhalo scheme
(left) and the standard scheme (right). We here show results for the
ejection feedback, and note that the results for the retention model
are very similar.  The symbols and error bars give the composite
luminosity function found by \citet{Tre98} as a weighted mean of 9
clusters, all having luminosity functions that are individually
consistent with the composite one. Note that we adjusted the vertical
normalization of the composite function (which is arbitrary) to 
obtain a good fit to the total luminosity of the cluster. The solid
lines are Schechter function fits to the histograms of bin-size 0.5
mag.  For the subhalo model, the faint end slope is $\alpha=-1.21$,
and the turn-off is at $M_{\star}\simeq -21.8$. The standard scheme
results in a steeper slope of $\alpha=-1.31$, and the characteristic
magnitude is not well defined.  Note that the standard model produces
a central galaxy of magnitude -23.9, which appears too bright even comparing
to the brightest cD galaxies.
\label{figLumB}}
\ec
\end{figure*}

In Figure \ref{figLumB}, we show the {\em B}-band luminosity function
of the S2-cluster obtained with the new `subhalo' methodology, and we
compare it to the result of the `standard' semi-analytic recipe of
KCDW.  We plot the number of cluster-galaxies in bins of size 0.5 mag,
and we fit Schechter functions to the counts.  The standard
prescription results in a relatively steep slope of $-1.31$ at the
faint-end, and the ``knee'' of the Schechter function is not well
defined. This is just another reincarnation of a well-known problem in
many previous semi-analytic studies, which predicted too many galaxies
both at the faint and the bright end of the the luminosity function,
i.e.~the shape was not curved enough.  These deficiencies can be
partly cured by invoking additional physics like dust obscuration, or
by employing models with very strong feedback processes.  However,
most semi-analytic studies have not been successful in simultaneously
achieving good agreement with the Tully-Fisher relation and the
luminosity function.

\begin{figure*}
\bc
\resizebox{8cm}{!}{\includegraphics{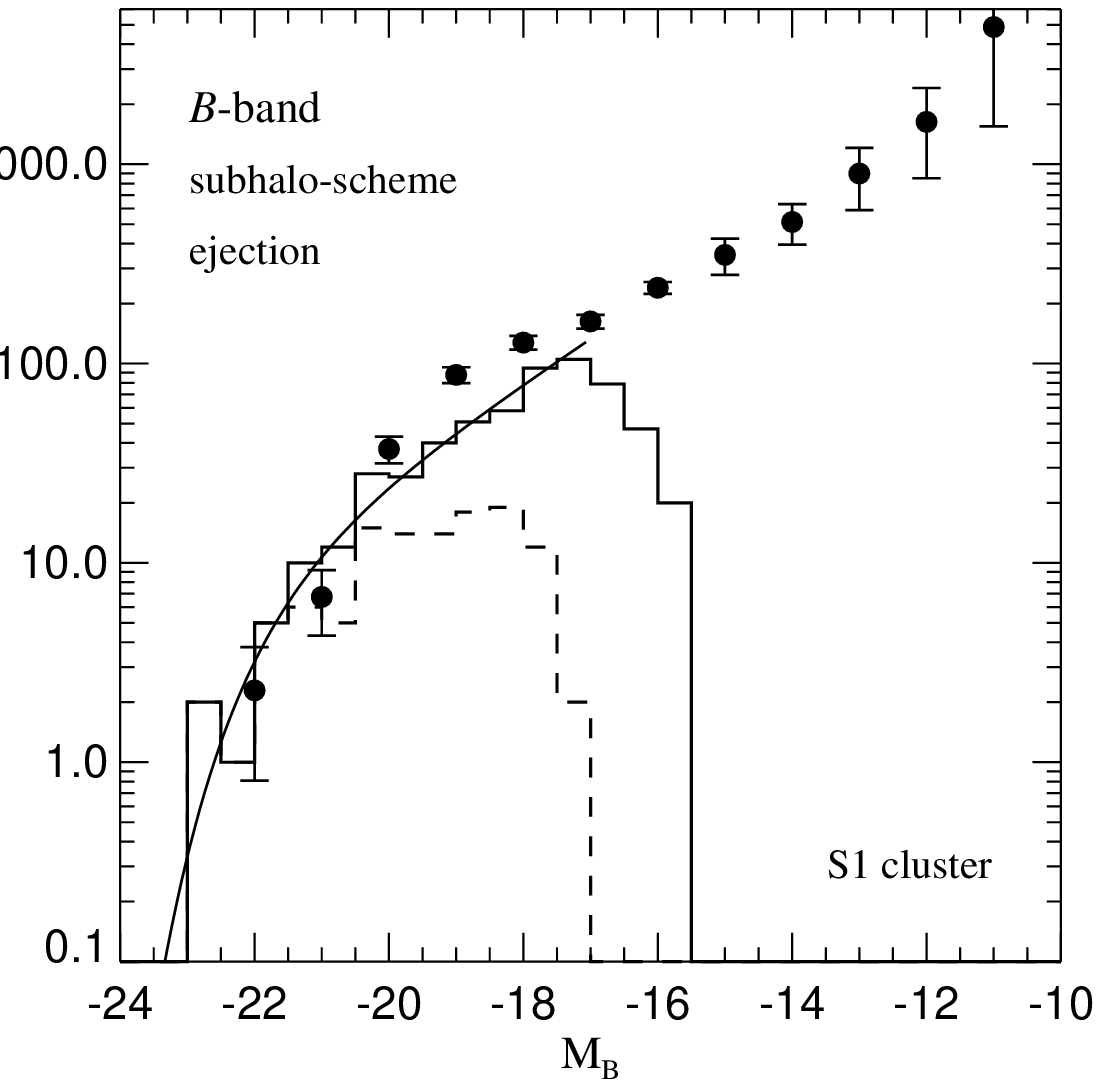}}%
\resizebox{8cm}{!}{\includegraphics{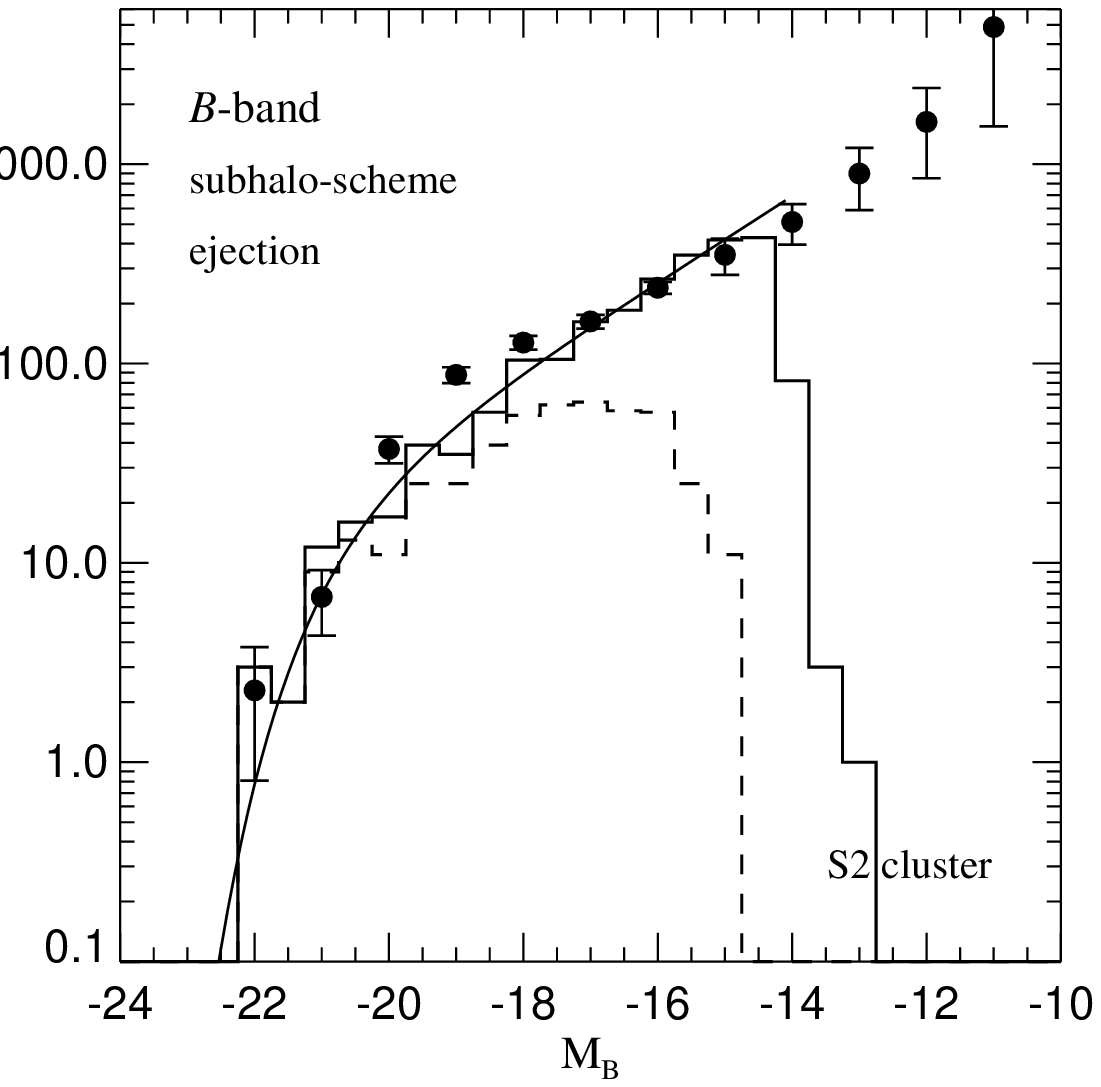}}\\%
\resizebox{8cm}{!}{\includegraphics{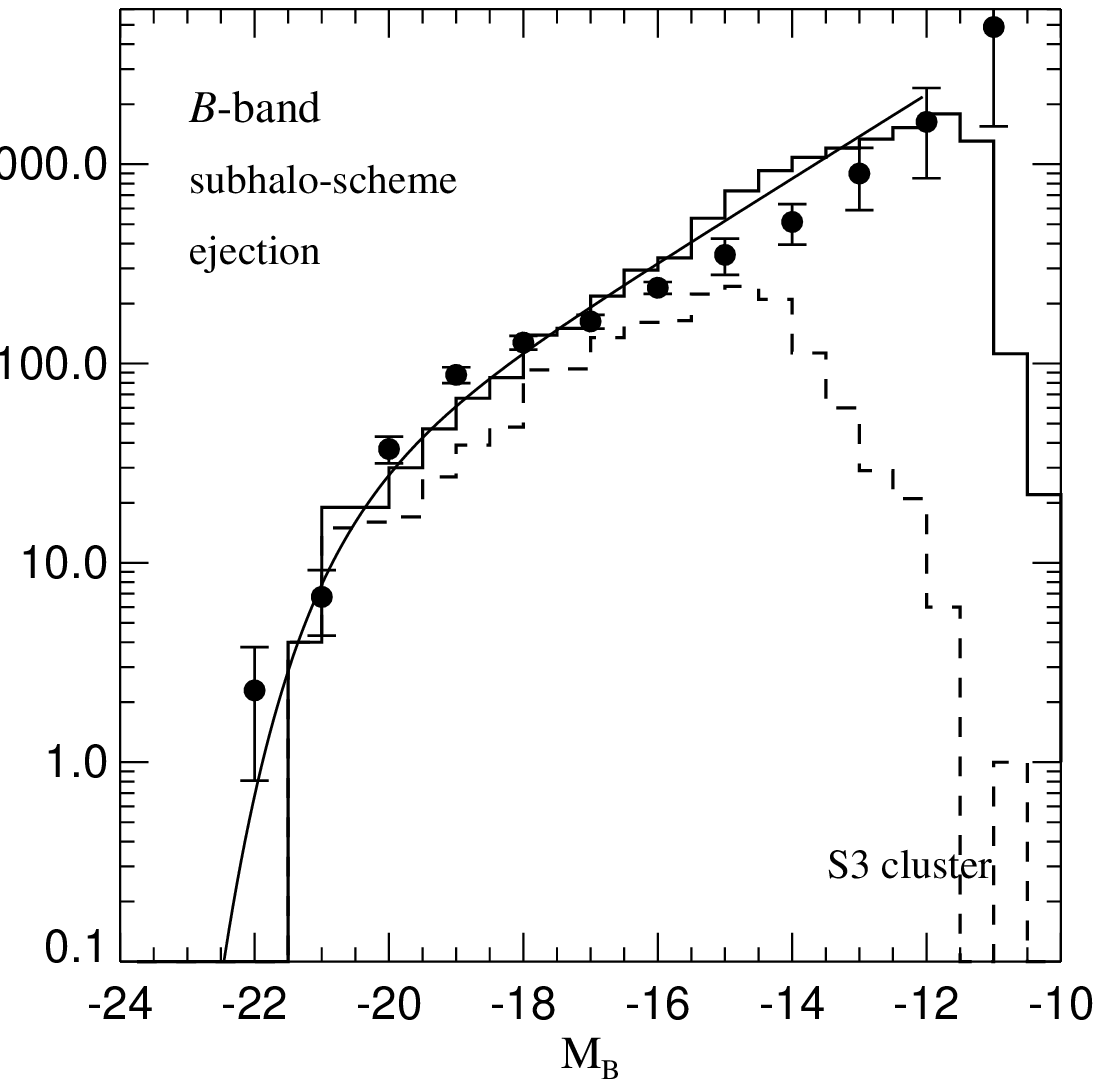}}%
\resizebox{8cm}{!}{\includegraphics{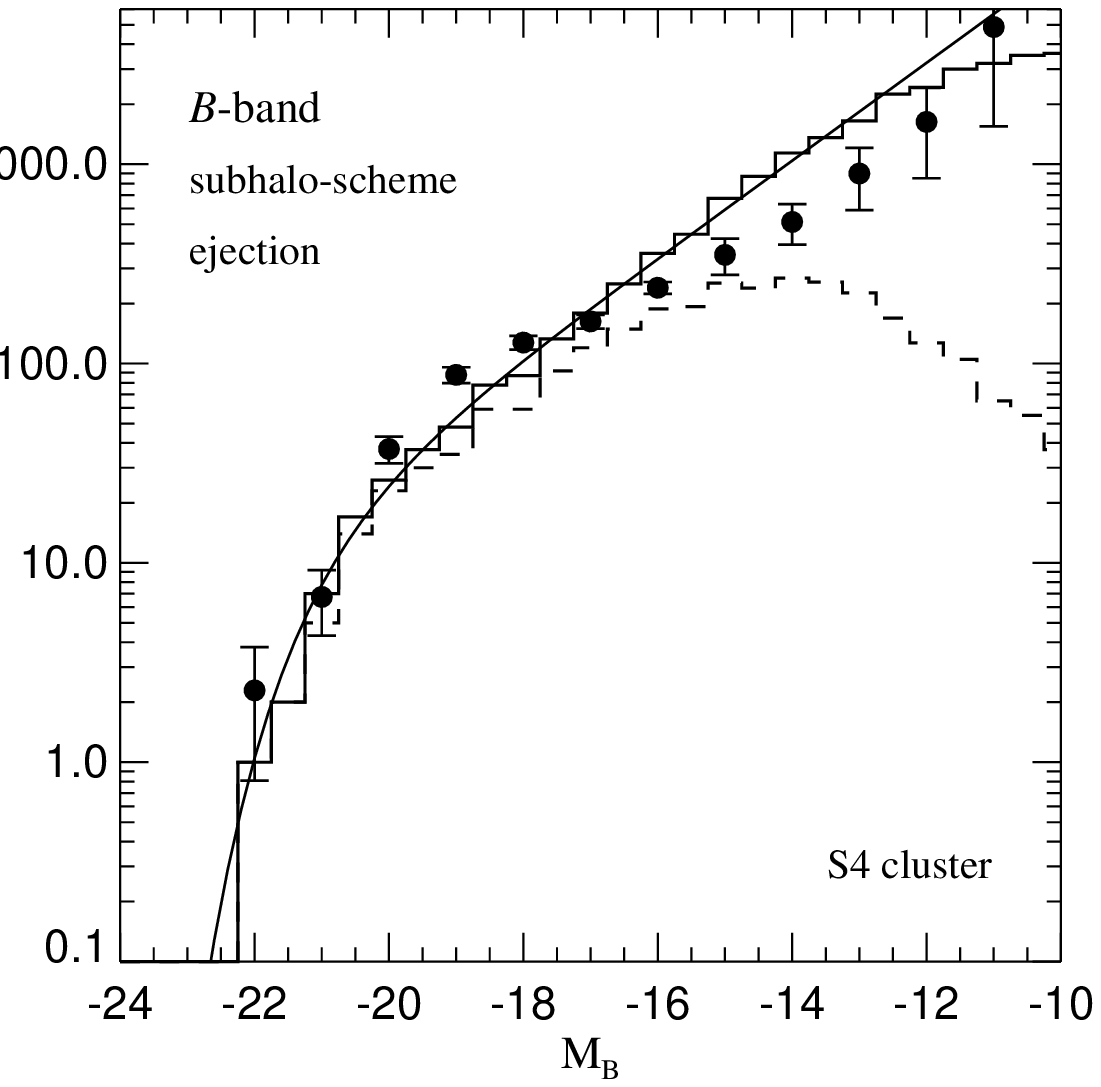}}\\%
\caption{The {\em B}-band cluster luminosity functions obtained for
the S1, S2, S3, and S4 simulations using the subhalo scheme and
ejection feedback.  The solid lines are three-parameter Schechter
function fits to the histograms of bin-size 0.5 mag. The formal values
for faint-end slope and cut-off of these fits are: S1 ($\alpha=-1.22$,
$M_{\star}=-22.8$), S2 ($\alpha=-1.21$, $M_{\star}=-21.8$),
S3 ($\alpha=-1.21$, $M_{\star}=-21.7$), and S4 ($\alpha=-1.24$,
$M_{\star}=-22.0$).
Symbols with error bars
give the composite observational result by \citet{Tre98}. Dashed histograms
give the luminosity functions of subhalo galaxies alone, i.e.~galaxies
that still have an associated dark matter subhalo.
\label{figLumB2}}
\ec
\end{figure*}

Compared to the field, the luminosity functions of clusters tend to be
considerably steeper, and a slope of $-1.31$ can be accommodated
with existing data. However, the standard-scheme also produces
a brightest cluster galaxy with luminosity in excess of most normal cD
galaxies. For example, in the S2 cluster, the standard-scheme produces
a central galaxy with {\em B}-magnitude $-23.9$.

In contrast, the luminosity function obtained with the subhalo
formalism has a more curved shape. There are more
$L_{\star}$-galaxies, resulting in a flatter faint-end slope of
$\alpha=-1.21$, and in a more pronounced ``knee'', which is much
better fit with a Schechter function. In addition, the luminosity of
the central galaxy is reduced.  The overall shape of the resulting
luminosity function is in reasonably good agreement with observed
cluster luminosity functions, as shown by a comparison with the
composite luminosity function of \citet{Tre98}, which is indicated
with symbols in Figure~\ref{figLumB}.

What causes this difference between the subhalo-scheme and the
standard scheme?  Note that the excessive brightness of the central
galaxy in the latter model is unlikely to be caused by an overcooling
problem. We have already cut-off the cooling flow for central galaxies
in haloes with a virial velocity larger than $350\kms$. Furthermore,
the cooling model was essentially the same for both schemes, which is
also reflected in their similar overall mass-to-light ratios.

The most important difference between the `standard' semi-analytic
scheme and the `subhalo'-methodology is the explicit tracking of
subhaloes, which allows  more faithful modeling of the actual merging
rate in any given halo.  Recall that one critical assumption in the
study of KCDW is that the time of survival of a satellite can be
estimated using a simple dynamical-friction formula.  However, this
description is crude and the excessive brightness of the central
galaxies may result from an overestimate of the overall merging rate,
or from merging the `wrong' galaxies. Recall that a second important
assumption of KCDW has been that the position of a satellite galaxy
can be traced by a single particle identified as the most-bound
particle of the satellite's halo just before it was accreted by a
larger system.

Before we investigate these assumptions in more detail, we plot
in Figure~\ref{figLumB2} the cluster luminosity functions for the S1,
S2, S3 and S4 clusters, obtained with the subhalo scheme and ejection
feedback. All four simulations produce luminosity functions which can
be well described by Schechter functions with a well defined cut-off.
Their shape is in fact quite close to the obervational result by
\citet{Tre98}, shown with symbols in each of the four panels.  The
vertical normalization of his composite observational result is
arbitrary, and we have set it to match the total luminosity of our
cluster(s). It is however the same in all four panels.  Agreement of
the luminosity functions between the four simulations is quite
good, with the simulations of higher mass resolution able to probe the
luminosity function to ever fainter magnitudes.  It is interesting to
note that as the resolution of the simulations increases, a larger and
larger fraction of the galaxies have their own dark matter halo (this
is indicated by a dashed line in the plot).  In S4, almost all bright
galaxies still have an associated dark matter substructure, and it
seems likely that in still larger numerical simulations the
correspondence would be perfect.

However, when comparing the luminosity functions with each other,
there is a trend of decreasing brightness of the first ranked
cluster galaxies with increasing resolution.  We think that this is a
reflection of inaccuracies in estimates of merging timescales, which
become more serious in simulations with lower
resolution. This effect may also be responsible for the weak trend in
the zero-points of the Tully-Fisher relations (Figure \ref{figTF2}).

We now test whether this problem is responsible for the difference
between the subhalo scheme and the standard formalism. To this end we
try to answer the question: How many of the galaxies present in
subhaloes of S2 at $z=0$ are prematurely merged with the center in the
standard scheme?

To do this, we follow each subhalo of the
cluster back in time along its merging history until it is a main halo
itself for the first time.  The corresponding FOF-halo will host a
central galaxy in the standard scheme, and the descendant of this
galaxy at $z=0$ should directly correspond to the galaxy of the
originally selected subhalo.

Among the 494 subhaloes identified in the S2-cluster at $z=0$, we find
in this way that 39 of them do not have a directly corresponding
satellite galaxy in the `standard'-formalism.  The satellites that
should correspond to these 39 subhaloes have prematurely disappeared
by merging processes with the central galaxy, because their merging
timescales have been underestimated. Note that the total number of
mergers with the central galaxy in the standard scheme is 116, while
this number is 132 in the subhalo formalism. This suggests that the
overall rate of mergers is not too high in the standard
scheme. However, in the standard scheme the central galaxy accretes 24
galaxies with stellar masses of more than $10^{10}\msunh$, among them
8 galaxies with stellar masses larger than $10^{11}\msunh$. On the
other hand, in the subhalo formalism there is only 1 merger involving
more than $10^{11}\msunh$, and 11 with more than $10^{10}\msunh$. This
means that a larger number of very bright galaxies is merged with the
central galaxy in the standard scheme.  We also note that the galaxies
in the 39 subhaloes that appear to have been merged prematurely in the
standard scheme tend to be quite bright. If we take the results for
the subhalo-model and add the luminosities of the corresponding
halo-galaxies to that of the central galaxy, its brightness increases
from -21.9 to -24.1 mag, quite close to the -23.9 obtained in the
standard-scheme. It thus appears that the excessive brightness of the
central cluster galaxy in the standard scheme is mainly caused by an
underestimate of the merging timescales for some fraction of the
bright galaxies.

We have also tested how well the position of subhaloes corresponds to
those of single particles used to track satellites in the standard
scheme.  We find that a fraction of 86.8\% of the subhaloes in S2
still contain the most-bound particle that is used in the
standard-scheme to track the position of the satellite.  This number
is 86.6\% in S1, 82.2\% in the S3 cluster, and 76.3\% in the S4
cluster. Note that these numbers have been computed for all the
subhalos in each of the simulations.  To the mass resolution limit of
the S1 cluster, the corresponding success rates are 93.6\%, 89.4\% and
96.8\% in S2, S3, and S4, respectively.  Using just a single particle
identified at a time before a structure was accreted onto the cluster
can thus provide a good estimate of the subhalo's position within the
cluster halo at later times.

\begin{figure}
\bc
\resizebox{8cm}{!}{\includegraphics{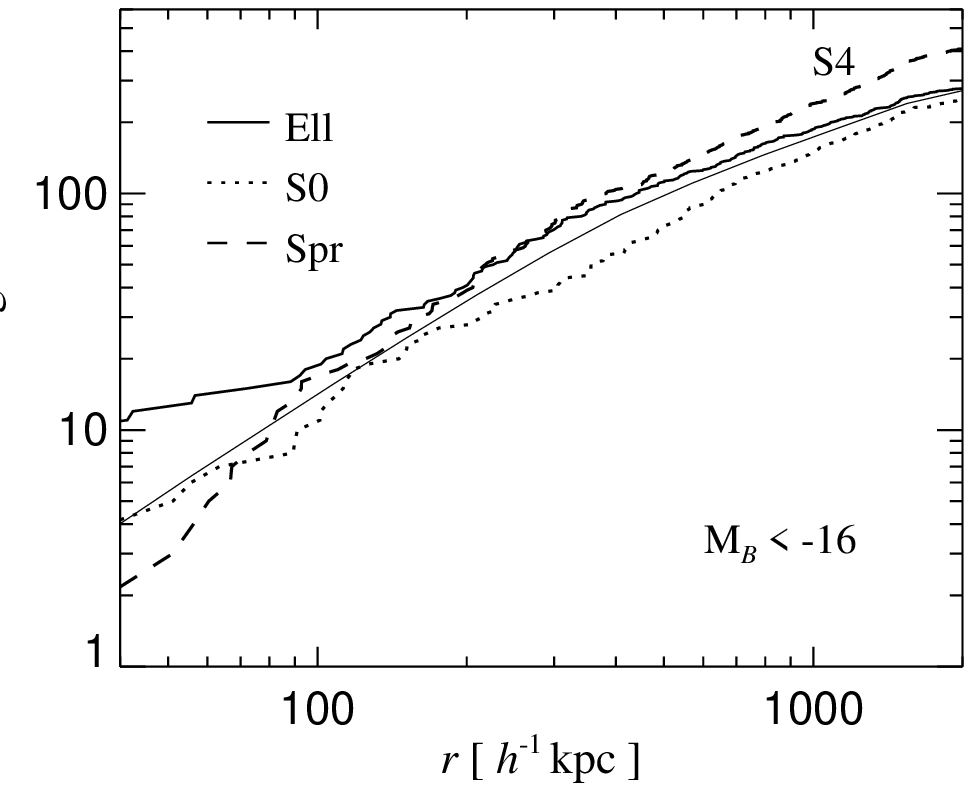}}\\%
\resizebox{8cm}{!}{\includegraphics{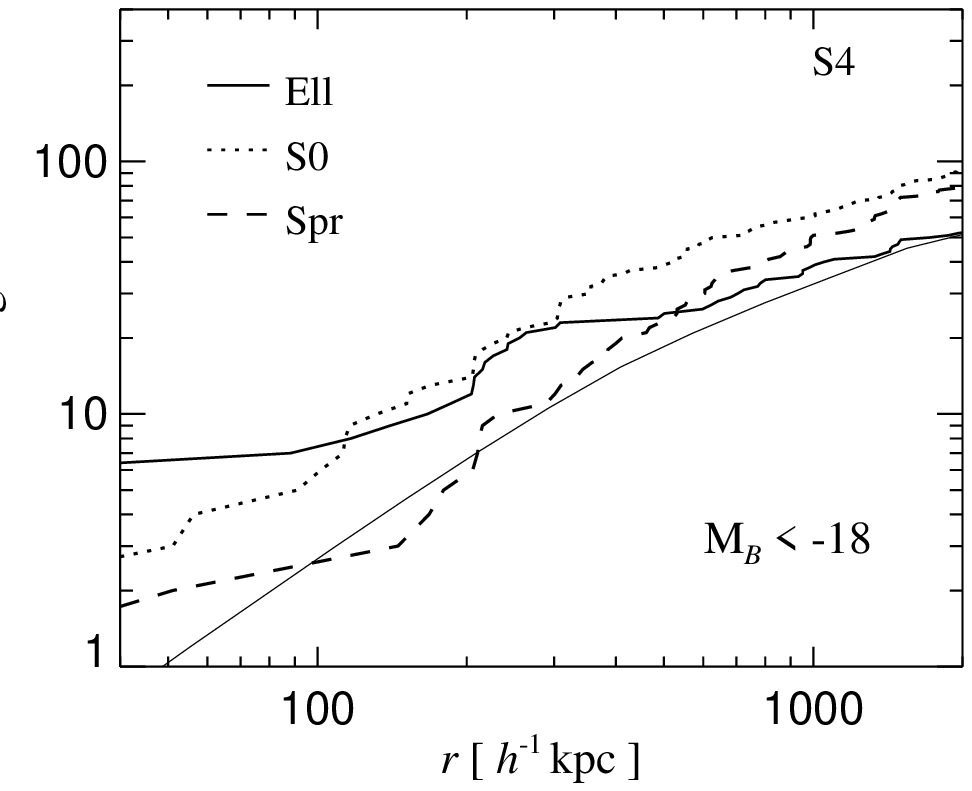}}\\%
\caption{Cumulative number of galaxies in the cluster as a function of
radial distance to the cluster center. In the top panel, we count all
galaxies with $M_{\rm B}< -16$, binned into three classes of different
Hubble type. In the bottom panel, only the bright galaxies with
$M_{\rm B}<-18$ are shown.  In both panels, the shape of the
cumulative dark matter mass distribution is indicated by a thin solid
line, for comparison.  The results shown are for the S4 cluster using
ejection feedback.
\label{figGalRad}}
\ec
\end{figure}

It thus seems clear that the difference between the luminosity
functions of the standard and subhalo schemes is largely caused by
differences in the assumed merger timescales of galaxies that fall
into the cluster. Recall that for the choices we adopted in equation
(\ref{friceq}), the dynamical friction formula is essentially the
current dynamical time of virialized haloes times the mass ratio of
the halo and its infalling satellite. Large estimates of merger
timescales arise when the mass of the satellite is much smaller than
that of the halo.  In fact, small satellites will typically have
merger timescales that are much larger than a Hubble time, while large
infalling haloes are predicted to merge with the central galaxy
quickly. As a result, the standard scheme appears to merge some of the
associated bright galaxies `too early', making the central galaxy
excessively bright. In the subhalo scheme, these galaxies are still
around and populate the `knee' of the luminosity function. Note
however that the dynamical friction estimate in the simple form
applied here does not include the effects of tidal truncation and
disruption, which presumably act to reduce the lifetime of small
satellites, but may increase the lifetime of infalling massive
satellites. In fact, the results of \citet{Tor98} indicate that the
dependence of the merger timescale on the mass ratio, measured just
before the satellite falls in, is weaker than linear.

Presently it is unclear whether an improved parameterization of the
merger timescales in the standard scheme can produce results as good
as those obtained by explicitly following the subhaloes. However, a
more detailed investigation of the probability distribution of actual
merging timescales may well make this possible, and offers the
prospect of implementing more accurate satellite merging in the
`standard' semi-analytic scheme, even when simulations with relatively
poor resolution are used.

\subsection{Morphology density relation}

\citet{Dr80} has shown that the relative frequency of elliptical
galaxies is higher in denser environments. In particular, in the cores
of clusters, spirals are quite rare, while they are the dominant type
in the field, and in the Universe as a whole. \citet{WhiG93} have
argued that this morphology-density relation reflects a more
fundamental morphology-clustercentric relation; the correlation
between morphology and clustercentric separation seems tighter than
that between morphology and projected density. This assertion is still
controversial.

\begin{figure*}
\bc
\resizebox{8cm}{!}{\includegraphics{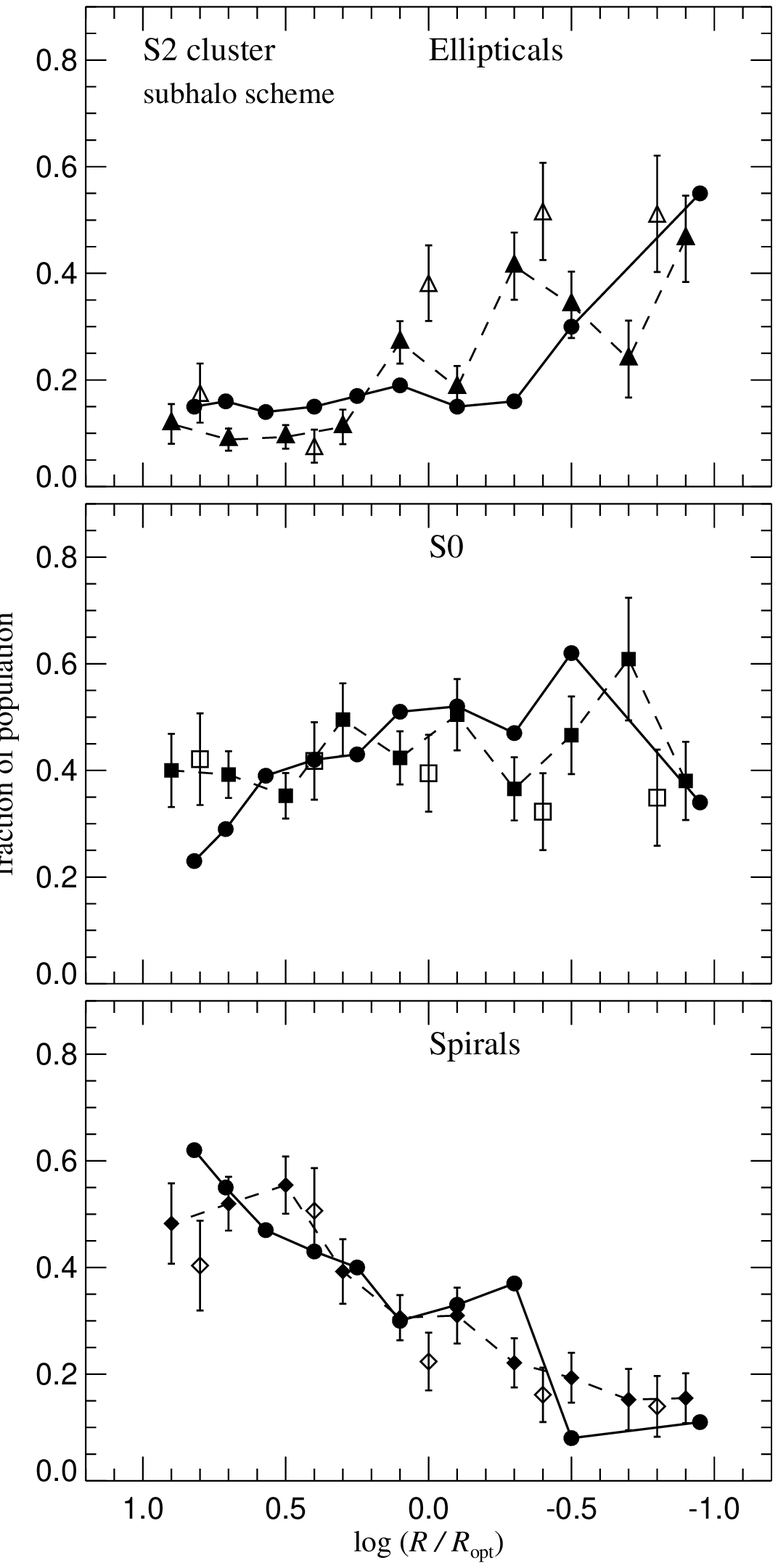}}\hspace{0.3cm}%
\resizebox{8cm}{!}{\includegraphics{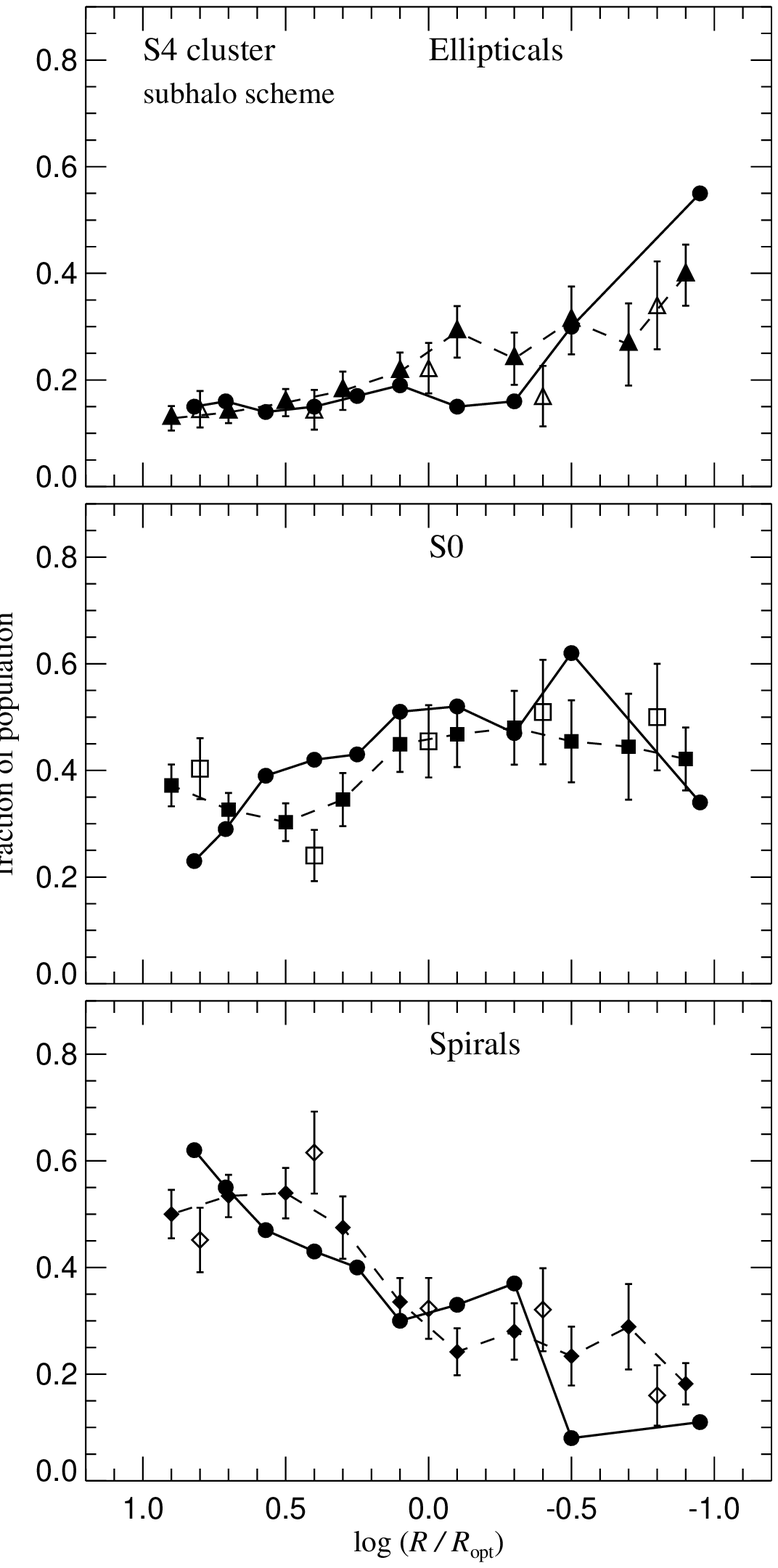}}\\%
\caption{Morphological mix of galaxies as a function of clustercentric
radius. From top to bottom, the symbols in the three panels show the
relative fraction of elliptical, S0, and spiral galaxies in spherical
shells around the cluster center, with error bars given by counting
statistics. Filled symbols are for galaxies brighter than $M_{\rm
B}=-17$, and hollow symbols for a sample selected at $M_{\rm
B}=-18.5$.  Filled circles show the observational results of
\citet{WhiG93}. Note that we here follow these authors in letting the
radius decrease to the right, i.e.~the cluster center is found on the
right side of the diagrams.  The results shown here are for the S2 and
S4 simulations, using the subhalo modeling with ejection feedback.
Results for the other two simulations are similar.
\label{figMorphMix}}
\ec
\end{figure*}

In Figure~\ref{figGalRad}, we show the cumulative number of galaxies
of different Hubble type as a function of distance from the cluster
center.  The top panel gives all the galaxies with $M_{\rm B}<-16$,
while in the bottom panel we just show the bright galaxies with
$M_{\rm B}<-18$.  From the cumulative distribution it can be inferred
that the three-dimensional density profile of the ellipticals has a
steeper slope than that of the spirals, i.e.~they are more
concentrated towards the center. Note that the thin solid line gives
the shape of the cumulative dark matter profile for comparison.  It is
also interesting to note that the bright galaxies in the center are
primarily ellipticals, while their mean Hubble type shifts to later
types towards the outskirts of the cluster.

These trends in the morphological mix of galaxies can be more clearly
seen in Figure~\ref{figMorphMix}, where we show the
morphology-clustercentric relations obtained for the S2 and
S4-clusters, using the subhalo-scheme with ejection feedback.  Note
that the fraction of elliptical galaxies strongly rises towards the
center of the cluster, while that of the spirals declines
accordingly. The quantitative strength of these trends is in
reasonable agreement with the results of \citet{WhiG93}, although
there are small differences in detail.  For the classification of
galaxies, we applied cuts in $T$-space that count galaxies with
$0<T<5$ as S0's, and lower (higher) types as ellipticals (spirals),
respectively. The resulting relative populations of galaxies of
different types are consistent with the observed morphological mix,
and largely independent of the magnitude cut-off for the sample. Note
however that in the standard observational classification a slightly
different cut in $T$-space is used for S0's. This may just reflect
the difficulty to unambiguously define the transitional type S0 in our
coarse morphological classification scheme, or it may mean that
additional physical processes are at work to create real S0 galaxies.

Recall that our morphological modeling is solely based on the merging
history of galaxies. Ellipticals are formed in major mergers, which
occur more frequently in higher density environments. This simple
model suffices to establish a pronounced morphology density relation.
This shows that the morphology density relation is built-in at a very
fundamental level in hierarchical theories of galaxy formation.

In passing we note that in the standard recipe, where satellites are
just traced by single particles once they have fallen into a larger
halo, the morphology-clustercentric relation is also present, although
it is not quite so well defined. The stronger clustering of early type
galaxies was also found by KCDW, \citet{Dia99} and more recently by
\citet{Dia2000}. Using techniqes somewhat similar to ours,
\citet{Oka2000} also found a morphology-density relation. All these
papers argued that additional processes are required to understand the
S0 population.

Figure~\ref{figMorphMix} also shows that the fraction of S0's in our
cluster shows a weaker dependence on cluster-centric distance than the
observations.  This also suggests that S0's experience additional
processes in clusters that are not included in our analysis.  For
example, disk galaxies orbiting in a cluster experience high-speed
encounters with other galaxies. Together with global cluster tides
this can drive a morphological evolution towards spheroids, a process
termed galaxy harassment \citep{Moo96,Moo98}. It has been suggested
that the spiral galaxies seen abundantly in clusters at moderate
redshift are harassed and slowly transformed to S0's at the present
time.  Such an effect might explain our deficit of S0 galaxies inside
the cluster.

\subsection{Cluster mass-to-light ratio}

The observed mass-to-light ratios of clusters of galaxies are known to
be much larger than those of individual galaxies, and this fact has
been recognized early on as strong evidence for the existence of large
amounts of dark matter in clusters.  Typical measured values for the
cluster mass-to-light range from $\Upsilon_{B}= 200\, h\,
\Upsilon_{\odot}$ to $\Upsilon_{B}= 400\, h\, \Upsilon_{\odot}$ in the
{\em B}-band. For the Coma cluster, \citet{Ken82} measured
$\Upsilon_{B}= 360\, h\, \Upsilon_{\odot}$, while X-ray data seem to
point to a lower value.  For example, Cowie et al. (1987) inferred a
mass-to-visual-light ratio for Coma as low as $\Upsilon_{V}= 180\,
h\,\Upsilon_{\odot}$.

\begin{figure}
\bc
\resizebox{8cm}{!}{\includegraphics{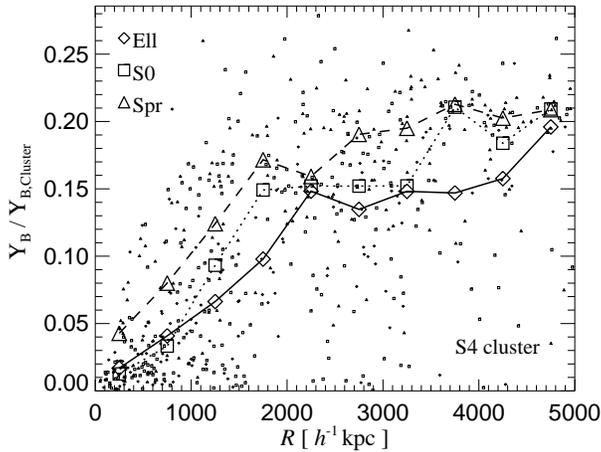}}\\%
\caption{$B$-band mass-to-light ratio of halo galaxies as a function
of distance to the cluster center (small symbols), expressed in units
of the mass-to-light ratio of the cluster as a whole. For the masses
of the galaxies we here simply took the dark matter masses of their
corresponding subhaloes. The large symbols give the median
mass-to-light ratios in radial bins around the cluster center. There
is a clear trend of decreasing $M/L$ towards the cluster center,
largely reflecting the mass loss of subhaloes due to tidal
truncation. Note however that early type galaxies appear to have a
systematically lower mass-to-light ratio than late types.
\label{m_over_l}}
\ec
\end{figure}

The galaxy population constructed for the S4 cluster using the subhalo
model has a total magnitude of $M_{B}=-26.04$. For its total mass of
$8.36\times 10^{14}\msunh$, the cluster mass-to-light ratio is thus
$\Upsilon_{B}= 420\, h\, \Upsilon_{\odot}$. In the {\em V}-band, we
obtain $M_{V}=-26.97$ and $\Upsilon_{V}= 324\, h\, \Upsilon_{\odot}$.
These mass-to-light ratios fall on the high side  compared to the
mean of observational results, although there are also measurements
that are as large, or larger, as our values. For example,
\citet{Ken83} obtained $\Upsilon_{V}= 600\, h\,\Upsilon_{\odot}$ for
the Perseus cluster.

Small changes of the model parameters can, however, enhance the overall
brightness of the cluster, and thus reduce the mass-to-light ratios to
a desired value. One simple possibility is to change the
conversion factor between the virial velocity and the circular
velocity. If the circular velocity of a disk in a dark matter halo of
given virial velocity is on average 15\% larger than we assumed, the
normalization to the \citet{Gio97} Tully-Fisher relation results in an
overall increase in brightness of 0.5 mag, and a corresponding
reduction of the mass-to-light ratio to 63\% of its old value.

Dust obscuration can account for effects of similar size.  It is quite
likely that the observed average brightness of galaxies in the
\citet{Gio97} sample is somewhat reduced by dust, even though the {\em
I}-band is less affected by dust than bluer
wavelengths. Since we do not correct for dust extinction in this work,
our fit to the observed Tully-Fisher relation is expected to
underestimate the stellar masses of these spirals.
Correcting this would produce larger stellar masses for all galaxies
at the present epoch and consequently lower mass-to-light ratios
for the cluster.

We may also ask whether we detect systematic variations of the
mass-to-light ratio of subhalo galaxies as a function of their
position in the cluster. In Figure~\ref{m_over_l} we show the $B$-band
mass-to-light ratio of halo galaxies as a function of radial distance
to the cluster center. Here we took the dark matter masses of
the corresponding subhaloes to compute $M/L$ ratios. The clear trend
of decreasing $M/L$ towards the cluster center is caused primarily by
mass loss through tidal truncation. However, for a
given distance from the cluster center, there is also a systematic
variation of mass-to-light ratio with morphological type.  At
each luminosity, early type galaxies tend to have lower mass halos 
than late types, reflecting their earlier incorporation and more
effective stripping within the cluster. All the galaxies have 
mass-to-light ratios substantially smaller than the cluster as a whole.

\begin{figure}
\bc
\resizebox{8cm}{!}{\includegraphics{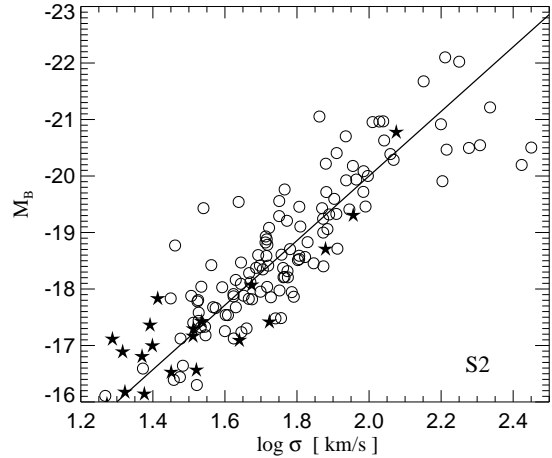}}\\%
\resizebox{8cm}{!}{\includegraphics{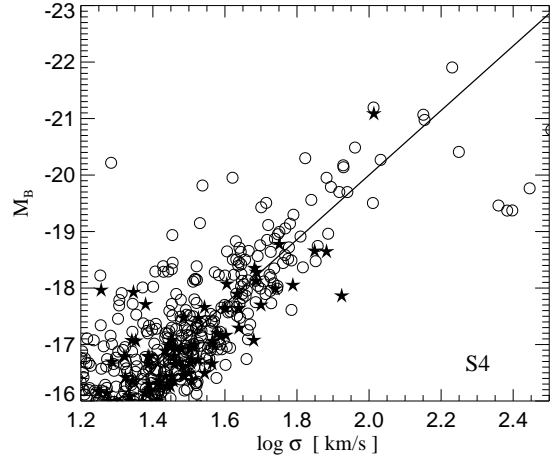}}%
\caption{%
Faber Jackson relations for cluster ellipticals (circles) and
field/group ellipticals (filled stars).  We here plot the $B$-band
luminosity of subhalo-galaxies versus an estimate of the
one-dimensional stellar velocity dispersion.  The latter was taken to
be $0.9/\sqrt{3}$ times the {\em measured} 3D-velocity dispersion of
the galaxies' dark matter subhalo identified in the simulation.  The
factor $1/\sqrt{3}$ converts from 3D to line-of-sight velocity
dispersion, and the factor $0.9$ is needed to bring the zeropoint in
agreement with the observed correlation. The obervational
Faber-Jackson relation is shown as a solid line, and is adopted from
the fit of \citet{Bend96} to data for Coma and Virgo ellipticals.
\label{figfabjack}}
\ec
\end{figure}

\subsection{Faber-Jackson relation}

For elliptical galaxies, there is a well-known scaling relation
between luminosity and central velocity dispersion, the so-called
Faber-Jackson relation.  In our formalism, we explicitly follow dark
matter subhaloes that orbit inside larger group- or cluster-sized
haloes.  While the mass of these subhaloes decreases strongly over
time by tidal truncation, their velocity dispersion remains relatively
stable until they finally disrupt. Thus the velocity dispersion of
a heavily truncated subhalo still reflects that of the original halo just
before it fell into the cluster. It is plausible that the stellar velocity
dispersions of elliptical galaxies correlate with the dark matter
dispersions of their haloes \citep{Rix97}. We may thus hope to
find a Faber-Jackson-like relation if we plot the
luminosity of our elliptical galaxies aginst the dark-matter
velocity dispersion of their surrounding (sub)haloes.

In Figure~\ref{figfabjack} we show the resulting $B$-band
Faber-Jackson relation for the S2 and S4 simulations. We here
converted the measured 3D velocity dispersions of the dark matter of
the subhalos to 1D dispersions, and we multiplied the resulting values
by a factor $0.9$ to obtain an estimated stellar velocity dispersion.
It is remarkable that the zeropoint, slope and scatter are consistent
with the observed relation for cluster ellipticals, which is here
adopted from \citet{Bend96} for the Virgo and Coma clusters and drawn
as a solid line. It is reassuring that the kinematics of the dark
matter substructure has indeed the right properties to explain the
Faber-Jackson relation. Note that some of the bright elliptical
galaxies in this plot have heavily stripped subhaloes of relatively
small mass, but the dark matter velocity dispersion of the residual
subhalo is still large enough to put them at about the right place
according to the Faber-Jackson relation.

\subsection{Luminosity segregation}

\begin{figure}
\bc
\vspace*{-0.5cm}\resizebox{8cm}{!}{\includegraphics{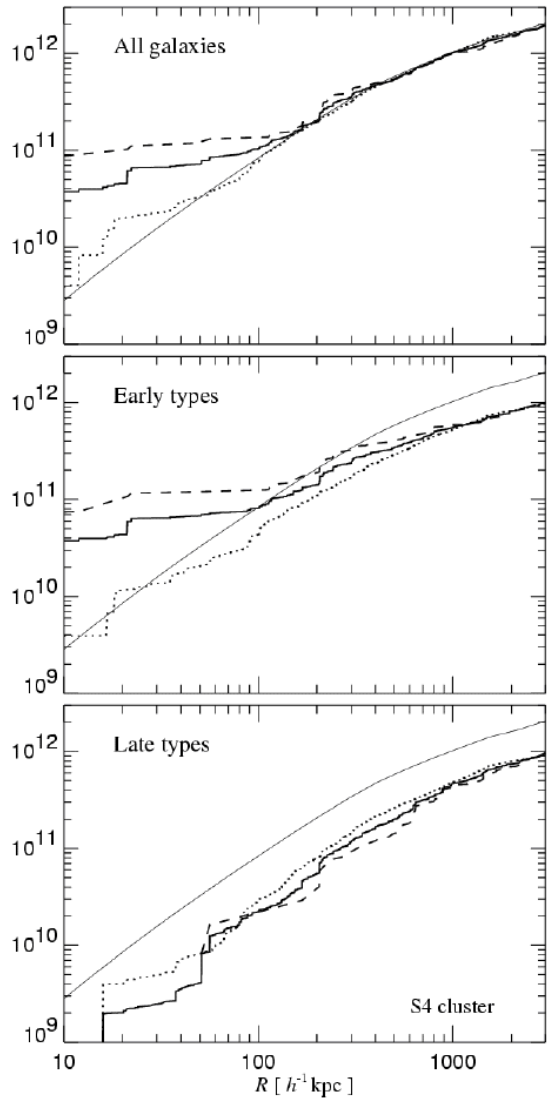}}\vspace*{-0.5cm}%
\caption{Luminosity segregation in the cluster S4.  We here show the
cumulative luminosity profile of $B$-band light of cluster galaxies
(thick solid lines), and compare it to the cumulative dark matter mass
profile (thin sold lines). The latter has been scaled by the overall
mass-to-light ratio of the cluster to put the curves in the three
panels on a common scale. While the top panel shows all the galaxies,
we have devided them by morphological type into two equally luminous
groups in the middle and bottom panels.  In each panel, we further
split the galaxies into a luminous (dashed lines) and a faint (dotted
lines) group such that each of the groups produces just half of the
total luminosity in the panel. Finally, we shifted the total
luminosity profile in each panel by a factor of two to align the
curves at the virial radius. 
\label{figLumprofile}}
\ec
\end{figure}

We have already seen that inside the cluster the number distribution
of luminous elliptical and spiral galaxies does not follow the mass
distribution of the dark matter. Does this also hold for the total
light emitted by the cluster galaxies?  In Figure~\ref{figLumprofile},
we show the cumulative luminosity profile of the cluster, and compare
it to the dark matter mass profile.  The luminosity profile is
shallower than that of the mass, showing that the light is actually
more concentrated than the mass, or phrased differently, the
cumulative mass-to-light ratio decreases towards the cluster center.
Note that the `steps' in the light profile are caused by the finite
number of galaxies and our assumption that they emit as point sources.

A related question is that of luminosity segregation, i.e.~are 
more luminous galaxies more strongly concentrated than less
luminous ones?  Note that clustering strength is a strong function of
circular velocity in CDM models \citep{Wh87}, so some form of
luminosity segregation should perhaps be expected. Since early-type
galaxies tend to be brighter than late-type galaxies, luminosity
segregation may also be seen as a consequence of the morphology
density relation. However, it is then not so clear which of these two
correlations is more fundamental. In fact, they might have a common
origin in the local dynamics of the cluster core with its frequent
mergers.

\begin{figure*}
\bc
\resizebox{8cm}{!}{\includegraphics{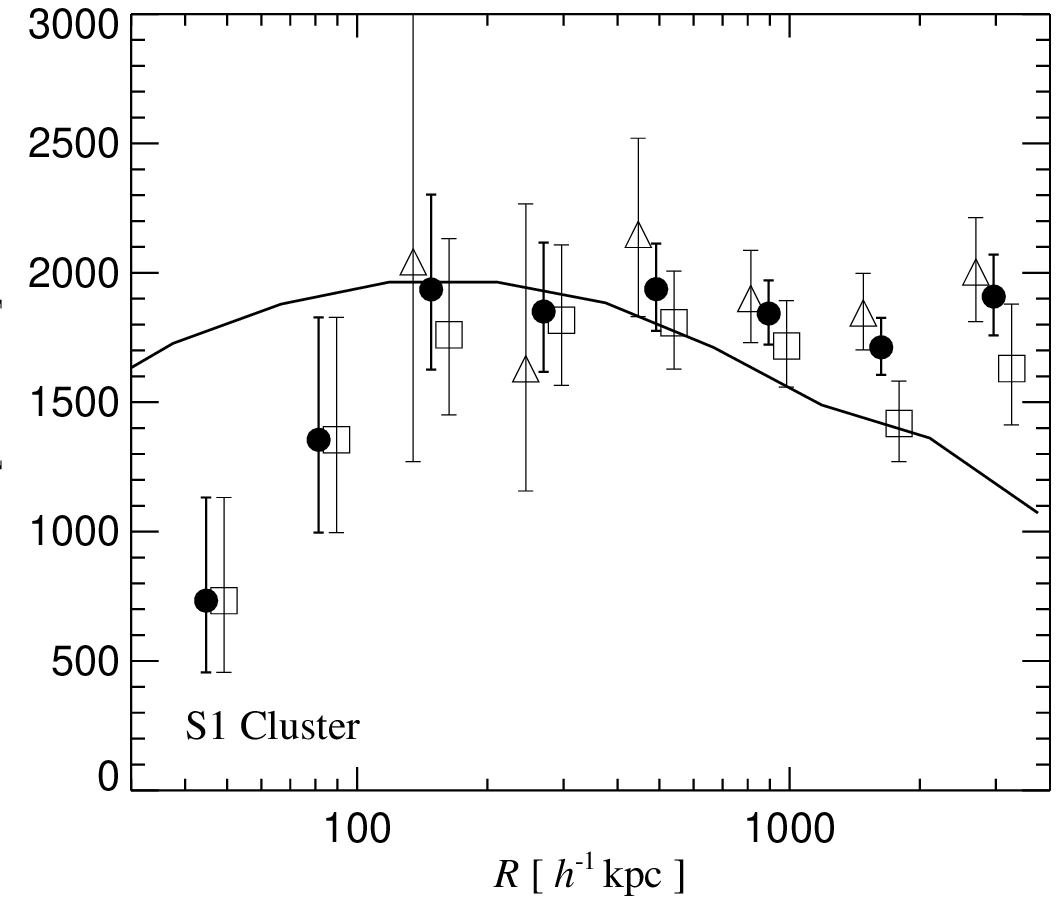}}%
\resizebox{8cm}{!}{\includegraphics{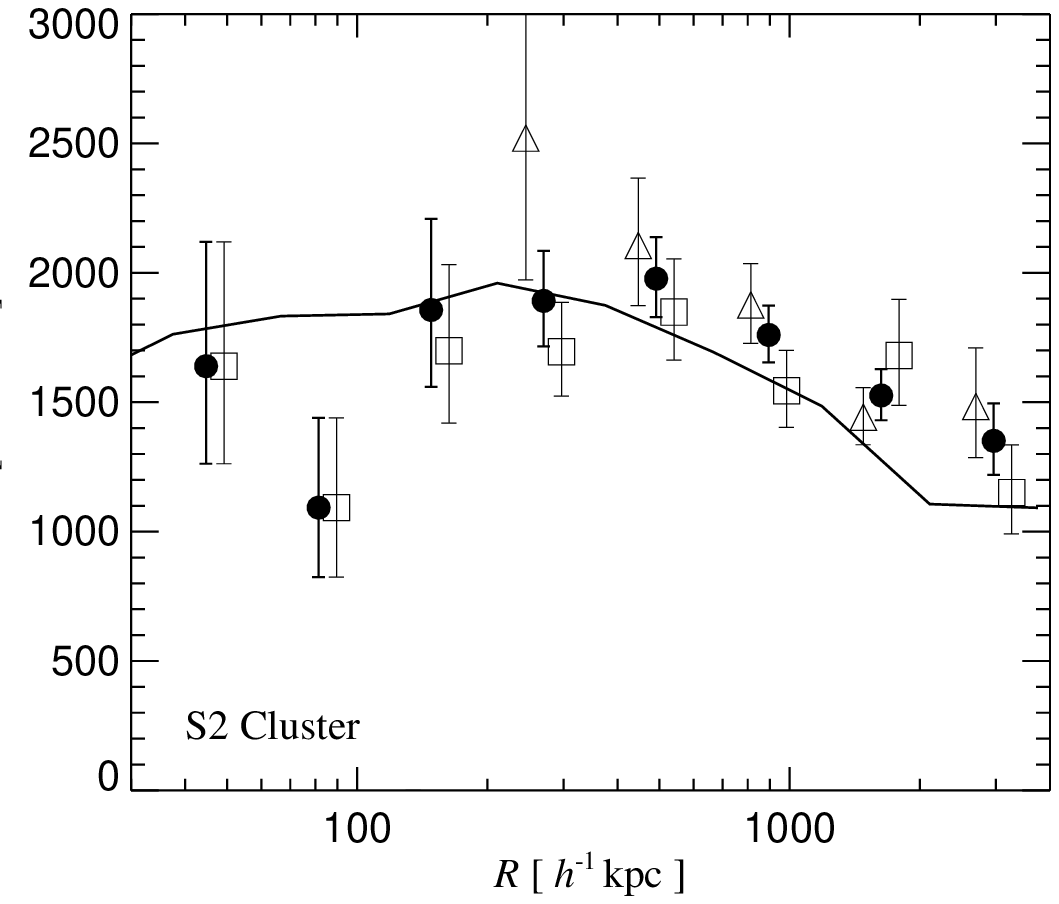}}\\%
\resizebox{8cm}{!}{\includegraphics{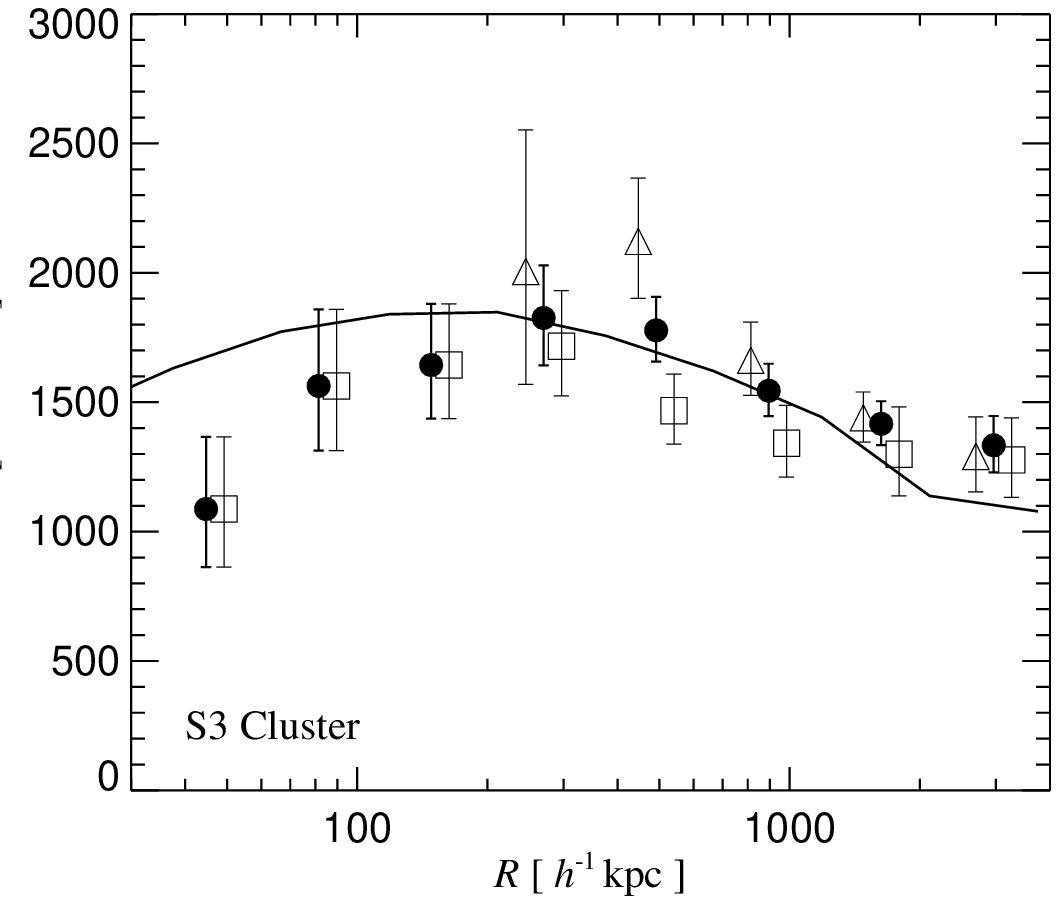}}%
\resizebox{8cm}{!}{\includegraphics{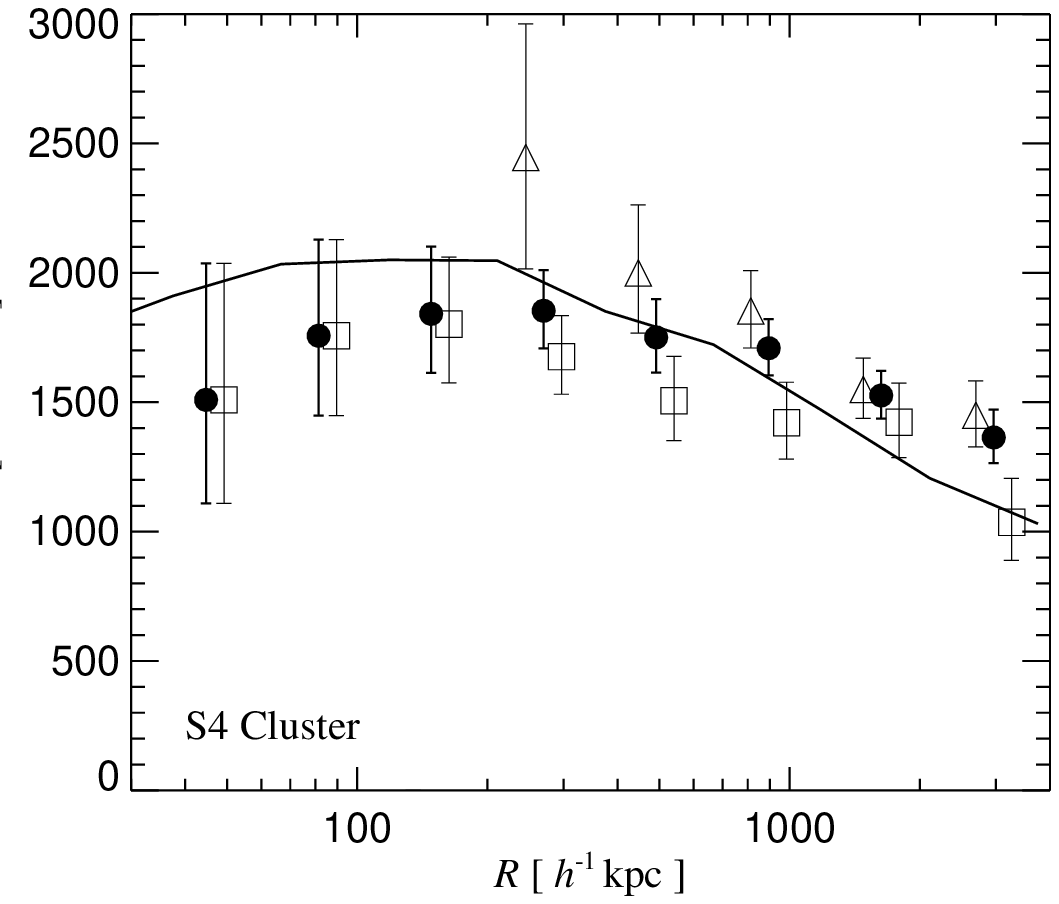}}\\%
\caption{One-point velocity dispersion of galaxies (filled circles)
brighter than $M_{B}=-17$ as a function of distance to the cluster
center.  The velocity dispersion of the dark matter is given by the
solid lines in the four panels.  The error bars mark the 68 per cent
confidence intervals based on counting statistics. We have further
split up the galaxies into a blue and a red sample of equal size,
based on their $B-V$ color.  The blue galaxies are shown as triangles,
the red ones as boxes. Note that there are no very blue galaxies 
near the cluster center, hence triangular symbols are absent at small
radii.
\label{figVelDisp}}
\ec
\end{figure*}

Observational evidence for luminosity segregation is still relatively
sparse and controversial, primarily because of the difficulty in
obtaining sufficiently large samples of faint cluster galaxies.
\citet{Kas98} have recently studied the dependence of luminosity
segregation on morphological type in the Coma cluster.  In order to
disentangle luminosity segregation from the morphology-density
relations it is critical to understand whether it occurs within a
given morphological type. \citet{Kas98} found that galaxies with high
central concentration (early types) show signs of luminosity
segregation, while galaxies of low central concentration (late types)
do not exhibit any strong segregation.

In the two bottom panels of Figure~\ref{figLumprofile}, we investigate
this issue for our cluster galaxies. We have divided the galaxies into
early and late types, and we split each of the groups into a high and
a low luminosity sample, such that the total luminosity in each group
was divided into two equal parts. Interestingly, only the light of the
bright elliptical galaxies is more concentrated than the mass of the
cluster, while the low luminosity early types, and the spiral galaxies
are distributed like the mass. We thus find evidence for luminosity
segregation of early type galaxies. This is in agreement with the
findings of \citet{Kas98}. Notice however that \citet{Dia2000} found no clear
evidence of luminosity segregation in their analysis of the GIF
simulations.

\subsection{Velocity dispersion of galaxies}

One common technique to estimate masses of real clusters of galaxies
is based on the virial theorem, requiring measurements of the
line-of-sight velocity dispersion and of the projected separations of
galaxies \citep{Heis85}. However, the accuracy of this technique can
be compromised if cluster galaxies are {\em biased} tracers of the
dark matter velocity distribution. Such systematic differences in the
velocity fields will usually give rise to errors in the mass estimates
of clusters.  In fact, there have been controversial claims in the
literature whether such a bias exists, and whether it is positive or
negative. Since there is currently no consensus on this issue, we
briefly address it here.  In the following, we will focus on the
one-point velocity bias, defined as the ratio $b_{v}=\sigma_{\rm
gal}/\sigma_{\rm DM}$ of galaxy and dark matter velocity dispersions.
We will refer to values of $b_{v}$ greater (less) than one as
positive (negative) velocity bias.

In the early simulation work of \citet{Car94} and \citet{Fre96}, a
negative velocity bias of up to 20-30\% was reported.  However, using
the `standard' scheme of semi-analytic modeling applied to the
GIF-simulations, \citet{Dia99,Dia2000} found a positive velocity bias in the
outskirts of clusters, a result which can be explained as being due to
the recent infall of star-forming blue galaxies.

Recent numerical studies with sufficient resolution
to resolve substructure have led to somewhat
conflicting results.  \citet{Col99} found a substantial positive
velocity bias, except perhaps in the innermost region of their cluster.
A similar result, albeit of weaker strength, was obtained by
\citet{Oka99}.  On the other hand, \citet{Gh98} did not find any
significant velocity bias in their analysis of cluster subhaloes.
Similarly, \citet{Kly99} found no velocity bias on cluster mass
scales, but a significant anti-bias for small groups of galaxies.  In
a more detailed analysis of this issue, \citet{Ghig99} recently
reported a small positive velocity bias for subhaloes in the innermost
region of their clusters, although the overall signal was considerably
smaller than that found by \citet{Col99}.
 
In light of this theoretical uncertainty, it is interesting to examine
the velocity bias of our cluster simulations.  In
Figure~\ref{figVelDisp}, we plot the velocity dispersion of our model
galaxies 
in the
clusters S1-S4 as a function of radius. We used logarithmic bins and
computed 68\%-confidence intervals under the assumption that the
galaxies in each bin are drawn from an underlying Gaussian velocity
distribution. In this case, the estimated dispersions are expected to
follow $\chi^2_{n-1}$ distributions, where $n$ is the number of
galaxies in each bin.

Upon comparison with the dark matter velocity dispersion (solid lines)
it is seen that we detect evidence for a small negative velocity bias
in the central region of the cluster. The significance of this result
is small for a single bin, but adjacent bins are independent here, and
`noise' between the four realizations is also uncorrelated, giving
rise to a quite significant finding when the individual results are
combined. We thus conclude that our models show evidence for a
negative velocity bias in the innermost $\sim 300\lu$ of the cluster,
while further outside, velocity bias appears to be small, if present
at all.  Such a lower velocity dispersion in the center appears to be
consistent with a number of observational studies
\citep[e.g.][]{Biv92,Ada98} that have found systematically smaller
velocities for the brightest galaxies in the cores of clusters,
although the observational evidence for this has not so far been
fully convincing.

In Figure~\ref{figVelDisp}, we have also split up the galaxies into a
blue and a red sample according to their $B-V$ color.  Interestingly,
it is seen that the blue galaxy sample shows larger velocity
dispersions in the radial range $300-1000\lu$. This is in
agreement with the findings of \citet{Dia99,Dia2000}.  Since these galaxies
experienced relatively recent star formation, they have mostly just
fallen into the cluster. As a result their orbits have systematically
larger apocentres than those of red galaxies in the same region of the cluster.
Note that if one
selects the most massive subhaloes, one also obtains a population that
has just fallen into the cluster and is biased towards its
outskirts. This may explain why studies focusing on subhaloes have
reported signs of positive velocity bias \citep{Col99}.

\subsection{$B-V$ color distribution}

\begin{figure}
\bc
\resizebox{8cm}{!}{\includegraphics{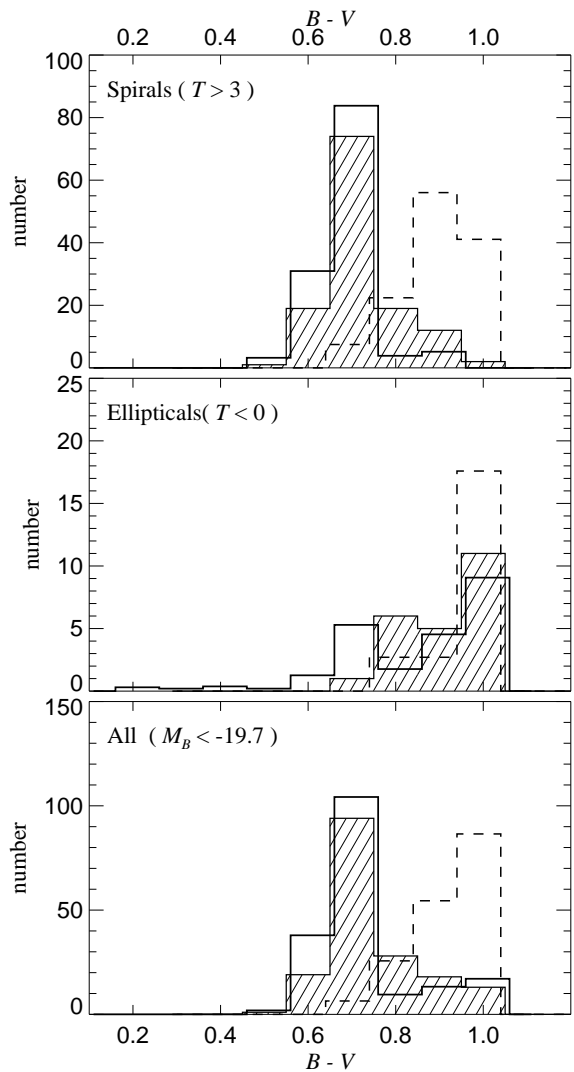}}%
\caption{The $B-V$ color distribution of galaxies brighter than
$M_{B}=-19.7$ mag in the `field' region of the S4 cluster simulation
(shaded histogram).  The dichotomy of the distribution mainly reflects
the difference in the stellar populations of elliptical and spiral
galaxies.  Spirals are significantly bluer due to the their recent
star formation, while ellipticals have older stellar populations,
resulting in redder colors.  The thick histogram gives the observed
color distribution as determined by KCDW from the RC3 catalogue
\citep{Vau91}. Finally, the dashed line outlines the $B-V$ color
distribution of galaxies in the cluster. Clearly, these galaxies tend
to be much redder than the field sample. Even most of the
cluster-galaxies that are classified as spirals are dominated by
relatively old stellar populations.
\label{figBminusV}}
\ec
\end{figure}

In Figure~\ref{figBminusV}, we show the $B-V$ color distribution of
galaxies brighter than $-19.7$ mag in the S2 simulation. A
corresponding plot has been shown by KCDW, and we here obtain similar
result.  The distribution of $B-V$ colors of field galaxies is
bimodal, with two peaks at $B-V\simeq0.7 $ and $B-V\simeq1.0$. By
plotting the color distribution for individual morphological types, it
becomes apparent that this dichotomy mainly reflects the differences
in star formation history between elliptical and spiral galaxies. The
recent star formation in spirals makes them blue, while the older
stellar populations of ellipticals give them redder colors. The color
distribution of field galaxies is in good agreement with the RC3
sample \citep{Vau91} of local galaxies. In Figure~\ref{figBminusV} we
also plot the color distribution of the galaxies in the cluster
(dashed histograms).  These galaxies tend to be much redder,
reflecting their substantially older stellar populations. Note that
since we have not included dust in our models, the apparent agreement
in Figure~\ref{figBminusV} probably means that the populations of 
late-type galaxies in our model are systematically older than those 
of the corresponding galaxies in the RC3.

\section{Discussion}

In this study, we have used cosmological N-body simulations combined
with semi-analytic techniques to construct the galaxy population of a
rich cluster of galaxies.  The very high resolution of our simulations
allowed us to extend the methodology for galaxy formation of KCDW to
the regime of substructure within virialized systems. Our goal in this
work has therefore been twofold. We wanted to develop the necessary
technical machinery for a galaxy formation scheme that works with
subhaloes, and we wanted to compare its results with those obtained
with the `standard' techniques by KCDW.

The detection of subhaloes within haloes is a technically difficult
problem, and has only been addressed very recently. Several working
algorithms have been described in the literature, but none seems ideal.
In this paper, we presented our new subhalo finding algorithm,
\subfind.  It relies on the particle positions and velocities at
a single output time, and it reliably identifies locally overdense,
self-bound particle groups within larger systems.
\subfind\ can also detect hierarchies of `haloes within haloes', a
feature that is not possible with alternative techniques. We find
examples of such a hierarchy in our highest resolution simulation.

The mass spectrum of subhaloes in our cluster appears to be close to a
power-law.  Our set of simulations allowed a direct study of
resolution effects, and it is interesting to note that the
lower-resolution simulations predict the right abundance of subhaloes
for any given mass above their respective resolution limit.

We have shown how the detected subhaloes can be traced from simulation
output to output. Just like KCDW, we analysed 51 simulation snapshots,
logarithmically spaced in expansion factor from $z=20$ to $z=0$. This
large number of output times together with the very high resolution of
our simulations allowed a measurement of the merging history of the
dark matter in unprecedented detail. For example, we detected almost
4700 subhaloes in the final halo of the S4 cluster, and we found
hundreds of mergers between subhaloes orbiting inside the progenitor
halo of the cluster, even though the rate of genuine subhalo-subhalo
mergers is relatively low.

Using a small set of modifications, we have adapted the semi-analytic
scheme of KCDW to include subhaloes in the analysis. In both schemes,
the agreement with the observed Tully-Fisher relation is very
good. However, the inclusion of subhaloes results in a substantial
improvement of the cluster luminosity function of the models.  In the
standard scheme, the first ranked cluster galaxies usually turn out to
be too bright, while this problem does not occur in the
subhalo-scheme.  We have shown that this is mainly due to inaccuracies
in the estimated merging timescales in the standard scheme, where too
many bright galaxies are prematurely merged
with the central galaxy.  The direct tracing of subhaloes until their
eventual disappearance allows a more faithful estimate of the actual
merger rate within haloes.  As a result, the luminosity function
becomes more curved, and develops a well defined knee and a flatter
faint-end slope. In fact, in the highest resolution simulations we
have carried out, essentially all the bright galaxies in the cluster
can still be followed in terms of their individual dark matter
subhaloes.

The subhalo-scheme also provides accurate positions and velocities for
galaxies orbiting in the cluster halo. We have shown that our simple
morphological modeling gives rise to a morphology-clustercentric
relation that is qualitatively in good agreement with
observations. Towards the center of the cluster, the morphological mix
of galaxies becomes gradually dominated by ellipticals, while the
contribution of spirals strongly declines.  Note that the morphology
of our model galaxies is primarily determined by their merging
history. A morphology density relation arises quite
naturally in hierarchical theories of galaxy formation. We have also
found that red galaxies near the cluster center can be expected to have
a negative velocity bias, i.e.~they move more slowly than nearby dark matter
particles. For infalling blue galaxies we detect a
small positive velocity bias.

In summary, we find that our subhalo-model for the galaxy population
of the cluster produces results that are in good agreement with a
variety of data. The cluster-luminosity function has a reasonable
shape, the Tully-Fisher relation of field spirals is well fit, the
cluster mass-to-light ratio has the right size, a morphology density
relation results, the $B-V$ color distribution appears to be well
consistent with observations, we find luminosity segregation for early
type galaxies, and we can reproduce the Faber-Jackson relation for
cluster and field ellipticals. Given the approximate treatment of key
physical processes, we think that these are remarkable successes.
However, it is important to note that changes in some of our model
assumptions can have a strong effect on the results. This makes it
possible to isolate the consequences of specific physical assumptions,
thereby guiding further physical modeling. At this point it is worth
noting that the number of free parameters in our models is actually
quite small, in practice, about the same as in direct hydrodynamical
simulations of galaxy formation.

One powerful strength of semi-analytic models is that they provide the
full history of galaxy formation. Using the present time to normalize
the models, they can make predictions at high redshift, where
observational data can be used to put additional strong constraints on
physical processes. We will discuss the evolution of the galaxy
population of our cluster in a companion paper (in preparation).

Combining high-resolution N-body simulations with semi-analytic galaxy
formation methods is currently the only way to simulate directly the
formation of a rich cluster and its constitutent galaxies in their
proper cosmological context.  We think that this approach can be
fruitfully exploited for study of other aspects of galaxy formation in
hierarchical cosmologies.

\section*{Acknowledgments}

The simulations presented in this paper were carried out on the T3E
supercomputer at the Computing Centre of the Max-Planck-Society in
Garching, Germany.  We are grateful for the hospitality of the
Institute for Theoretical Physics at Santa Barbara, where much of the
final writing of this paper was completed.

\bibliographystyle{mnras}
\bibliography{paper}

\renewcommand{\includegraphics}[1]{\ }

\end{document}